\documentclass[useAMS,usenatbib,usegraphicx]{mn2e}
\usepackage{graphicx}
\usepackage{amsmath}
\usepackage{aas_macros}
\usepackage[british]{babel}
\usepackage{multirow}
\usepackage{url}
\usepackage{color}
\usepackage{hyperref}	
\hypersetup{colorlinks=true,linkcolor=blue,citecolor=blue,filecolor=blue,urlcolor=blue}
\bibpunct{(}{)}{,}{a}{}{,} 
\newcommand{\lsun}{L$_{\sun}$}
\newcommand{\msun}{M$_{\sun}$}

\newcommand{\dydz}{$\Delta Y / \Delta Z$}
\newcommand{\ml}{$\alpha_\mathrm{ML}$}
\newcommand{\sigpr}{$\sigma_\mathrm{FeH}$}
\newcommand{\fehpr}{[Fe/H]$_\mathrm{pr}$}

\title[Calibration of $\alpha_\mathrm{ML}$ and $\Delta Y/\Delta Z$  from MS stars]{Bayesian calibration of the mixing length parameter \ml{} and of the helium-to-metal enrichment ratio \dydz{} with open clusters: the Hyades test-bed.}
\author[E. Tognelli, M. Dell'Omodarme, G. Valle, P. G. Prada Moroni, \& S. Degl'Innocenti]
{E. Tognelli$^{1,2}$\thanks{e-mail: ema.tog$@$gmail.com}, M. Dell'Omodarme$^{2}$, G. Valle$^{2}$, P.G. Prada Moroni$^{2,3}$,S. Degl'Innocenti$^{2,3}$\\
$^{1}$INAF, Osservatorio Astronomico d'Abruzzo, via Mentore Maggini, I-64100, Teramo,  Italy\\
$^{2}$Dipartimento di Fisica `E.Fermi', Universit\'a di Pisa, Largo Bruno Pontecorvo 3, I-56127, Pisa, Italy\\
$^{3}$INFN, Sezione di Pisa, Largo Bruno Pontecorvo 3, I-56127, Pisa, Italy
}
\begin{document}
\date{Accepted 2020 November 23. Received 2020 November 21; in original form 2020 September 25}
\pagerange{\pageref{firstpage}--\pageref{lastpage}} \pubyear{2020}
\maketitle
\label{firstpage}
%
\begin{abstract}
We tested the capability of a Bayesian procedure to calibrate both the helium abundance and the mixing length parameter (\ml), using precise photometric data for main-sequence (MS) stars in a cluster with negligible reddening and well-determined distance. The method has been applied first to a mock data set generated to mimic Hyades MS stars and then to the real Hyades cluster. We tested the impact on the results of varying the number of stars in the sample, the photometric errors, and the estimated [Fe/H]. The analysis of the synthetic data set shows that \ml{} is recovered with a very good precision in all the analysed cases (with an error of few percent), while [Fe/H] and the helium-to-metal enrichment ratio \dydz{} are more problematic. If spectroscopic determinations of [Fe/H] are not available and thus [Fe/H] has to be recovered alongside with \dydz{} and \ml, the well-known degeneracy between [Fe/H]-\dydz-\ml{} could result in a large uncertainty on the recovered parameters, depending on the portion of the MS used for the analysis. On the other hand, the prior knowledge of an accurate [Fe/H] value puts a strong constraint on the models, leading to a more precise parameters recovery. Using the current set of PISA models, the most recent [Fe/H] value and the {\it Gaia} photometry and parallaxes for the Hyades cluster, we obtained the average values $<$\ml$>=2.01\pm0.05$ and $<$\dydz$>=2.03\pm0.33$, sensitively reducing the uncertainty in these important parameters.
\end{abstract}

\begin{keywords}
methods: numerical -- methods: statistical -- stars: abundances -- stars: evolution -- stars: fundamental parameters -- stars: low-mass
\end{keywords}
\maketitle
%
\section{Introduction}
\label{sec:intro}
Theoretical stellar models are essential tools to infer the properties and characteristics of single or multiple stars in clusters/associations. Being theoretical computations, the accuracy of the models depends on the treatment of physical processes active in stars, which in turn strongly depends on the adopted input physics \citep[e.g. equation of state, opacities, nuclear reaction rates, and outer boundary conditions,][]{tognelli11, Valle13a, Stancliffe2015}. A crucial point in stellar modelling is the efficiency of convection in external superadiabatic regions. Despite the attempts to find a consistent description of the motion of matter in convective regions in stars using detailed hydrodynamical simulations, at the present it is not possible to follow the entire evolution of a star. Due to their complexity and long computational time, such simulations can be applied only to limited portions of a star, generally to boxes that cover a region of about $10^6$~m in the atmosphere (about 1-10~percent of the atmosphere, depending on effective temperature and surface gravity), then considered as representative of the whole atmospheric region. Moreover, such simulations have been obtained only for few selected evolutionary models (i.e. at selected stellar ages/evolutionary stages). To overcome these problems, simplified treatments of convection are adopted in stellar computations. The most used is the mixing length theory \citep{bohm58}, where the heat transport occurs in a region with a scale length $\ell$ proportional to the local pressure scale $H_P$, namely $\ell =$\ml$H_P$. The free parameter \ml{} is the mixing length parameter, which has to be calibrated to reproduce real data. A method to estimate \ml{} consists in reproducing the radius/colour index/effective temperatures of stars with a superadiabatic convective envelope. To do this it is convenient to select stars in the mass range  [0.6, 1.0]~\msun, which corresponds to models strongly affected by a variation of \ml. 

Besides the input physics, stellar models are severely affected by the adopted chemical composition, especially the content of helium $Y$ and the total metallicity $Z$ \citep[see e.g.][]{castellani99,valcarce12}. Unfortunately, helium cannot be observed in the spectra of cold low-mass stars, but only in stars hotter than about 20\,000~K, while the metallicity can be obtained -- at least in principle -- measuring the equivalent width of several metals absorption lines in stellar spectra. 

For disk stars, assuming a relative metal abundance similar to that of the Sun, it is possible to simply connect the total metallicity and the helium content to the iron abundance measured disc
in the spectra (i.e. [Fe/H]), using the following relation \citep{gennaro10,tognelli13},
\begin{equation}
[\mathrm{Fe}/\mathrm{H}]= \log \frac{Z}{1-Y-Z} - \log \bigg(\frac{Z}{X}\bigg)_{\sun}
\label{eq:feh_zy}
\end{equation}
where $(Z/X)_{\sun}$ is the present photospheric total-metallicity-to-hydrogen ratio that can be measured in the Sun \citep[see e.g.][]{asplund09}. However, to convert [Fe/H] into mass fractional abundances it is necessary to have a relation between $Y$ and $Z$. A commonly adopted method to simply relate these two quantities is to adopt linear relation between $Y$ and $Z$, as originally proposed by \citet{peimbert74}. Thus, by introducing the so-called helium-to-metal enrichment ratio \dydz, it is possible to write a simple linear relation $Y$ vs $Z$,
\begin{equation}
Y = Y_p + \frac{\Delta Y}{\Delta Z} Z
\label{eq:yz}
\end{equation}
In this expression, $Y_\mathrm{p}$ is the helium produced in the Big Bang nucleosyntehsis \citep[see e.g.][]{peimbert07,coc17,pitrou18,fields20}. Using eqs.~(\ref{eq:feh_zy}) and (\ref{eq:yz}) it is possible to derive both $Y$ and $Z$ knowing [Fe/H], \dydz{} and $(Z/X)_{\sun}$. 

The value of \dydz{} has been largely investigated in the past years. One of the possible method of constraining its value consists in comparing isochrones and stellar models in the HR diagram. Pioneering works involving the comparison between models and data of low-mass stars in the HR diagram was conducted by \citet{faulkner67} and \citet{perrin77}, who found respectively \dydz=3.5 and 5. \citet{pagel98} used low-mass stars in the Hipparcos catalogue to obtain \dydz=$3\pm2$. \citet{jimenez03} re-analysed a sample of K dwarfs in the Hipparcos catalogue and they derive \dydz=$2.1\pm0.4$. A similar result was found by \citet{casagrande07a} who obtained \dydz=$2.1\pm0.9$. \citet{gennaro10} performed a similar investigation on Hipparcos stars finding a much larger value \dydz=$5.3\pm1.4$. Moreover, they showed that the adopted age for the sample of stars deeply affects the results, leading to a systematic uncertainty on \dydz{} of about 2. 

Another method to derive \dydz{} involves the construction of a standard solar model, where the present luminosity, radius and surface $(Z/X)_{\sun}$ of the Sun has to be reproduced. This method gives a calibration of the \ml{} parameter, and of the initial $Y_{\mathrm{ini},\sun}$ and $Z_{\mathrm{ini},\sun}$, from which \dydz{} for the Sun can be obtained. Such a value strongly depends on the input physics adopted to compute the solar model, and also on the observed $(Z/X)_{\sun}$. In particular, this last quantity has been largely revised in the last decades, changing more than 40~percent \citep[see e.g.][]{grevesse93,grevesse98,asplund05,asplund09,caffau11}. The consequence of such uncertainty on both $(Z/X)_{\sun}$ and the adopted input physics in solar models leads to values of \dydz{} that can vary between about 0.4 and 1.3 \citep[see e.g][]{gennaro10,serenelli10,tognelli11,valcarce12}.

There are other methods of determining \dydz, such as considering more evolved stars  \citep[horizontal branch and red giant stars, see ][]{renzini94,valcarce16}: in this case, values of \dydz{} between 2 and 3 are obtained. Another possibility is to constrain \dydz{} using galactic and extragalactic HII regions \citep{peimbert74,peimbert76,pagel92a,peimbert00,fukugita06,peimbert08,melendezdelgado20} or planetary nebula \citep{dodorico76,peimbert80,maciel01}. These methods yield values of \dydz{} between 1 and 6 \citep[see e.g. section 9 in ][ for a detailed discussion]{gennaro10}.

The Hyades cluster is an ideal cluster to tightly constrain \dydz, because it is very close and it is not affected by extinction/reddening effects. \citet{perryman98} obtained a clean Hyades sequence using the early release of the Hipparcos parallaxes and used a sample of 40 main-sequence (MS) stars to constrain the cluster helium content. They found a helium abundance $Y=0.26\pm 0.02$, where the central values was slightly smaller than their solar reference value. However, the large uncertainty they estimated (mainly due to the large photometric and distance uncertainty) did not rule out the possibility of having a larger helium content, as expected for such a metal-rich cluster. \citet{lebreton01} analysed a double-lined eclipsing binary system in the Hyades \citep[vB22,][]{peterson88}, which consist of a primary component with $M_1=1.0591\pm0.0062~$\msun, and a secondary of about $M_2=0.7605\pm0.0062$~\msun{} \citep{torres02}: they found that only stellar models with subsolar helium content, namely $Y=0.255\pm0.009$ ($Y_{\mathrm{ini},\sun}=0.2674$), were compatible with the data. The authors  concluded that this is the helium abundance in the Hyades cluster; however, the isochrone with subsolar helium content obtained from their fit starts to deviate from the observed Hyades MS at $M_V \ga 4$-4.5~mag, as the clearly stated. A similar analysis was conducted by \citet{pinsonneault03}: they derived a helium content for the vB22 binary system similar to their solar value. In the same years, \citet{castellani02} using the improved parallaxes provided by \citet{madsen02} obtained a very good fit of the Hyades sequence down to $M_V \sim 5.5$~mag, where the models, and in particular the synthetic photometry used to convert the luminosity into magnitudes, suffer some problems. They found that models with a supersolar helium content ($Y=0.278$), obtained using \dydz$=2$ \citep{castellani99}, well matched the observed sequence. \citet{casagrande07a} re-derived the helium content in the vB22 binary system confirming a slightly subsolar value. More recently, \citet{gennaro10} used faint stars in the Hyades cluster ($M_V \ge 5.2$~mag) to infer the helium content of the cluster: from the analysis they obtained \dydz$=4.75\pm0.35$, which leads to supersolar helium content.

The aim of this paper is to apply a statistical method to analyse theoretically if it is possible to precisely calibrate stellar models on a portion of MS stars extracted from a cluster to derive the best value of \dydz{} and \ml. To do this, we generated a set of synthetic stars to test the capability of the method to simultaneously recover the parameters used to generate the mock data set (i.e. [Fe/H], \dydz{} and \ml), and to identify possible systematic uncertainty sources. Then, we applied our method to the real data of MS stars in the Hyades cluster. This cluster is a good candidate for the present analysis because it is very close (about 47~pc), thus with a negligible reddening (E(B-V)$\la 0.001$~mag) and a very good precision on parallax and photometry. Moreover, it is old enough (500-700~Myr) that the location in the colour-magnitude diagram (CMD) of MS stars is insensitive to the age.
All these characteristics helps in reducing the possible systematic uncertainties coming from unknown reddening, errors in distance, and age estimation.

The paper is structured as it follows. In Section~2 we describe the mock and real data sets  and in Section~3, we describe the Bayesian code used for the analysis. The grid of models employed for the recovery is described in Section~4. In Section~5 we discuss the effects on the models of varying [Fe/H], \dydz{} and \ml. In Section~6 we present the results of the recovery procedure applied to the mock data sets. In Section~7 the same procedure is applied to the real Hyades data set. The conclusions and the results are then presented in Section~8.

\section{The synthetic and Hyades data sample}
\label{sec:met}
With the aim to apply our method to Hyades cluster, we first tested it on a mock data set of synthetic MS stars, to evidence the possible biases and the minimal uncertainties of the procedure. To generate the synthetic stellar sample as close as possible to the real Hyades stars, we took as reference stars the {\it Gaia} DR2 Hyades data \citep{gaia1,gaia2}, corrected for the distance modulus obtained from the parallaxes given by \citet{babusiaux18}. We selected for our analysis the ($G-Rp$, $B_P$) plane, due to the very small uncertainties in these photometric bands \citep[mean errors in magnitudes of about 0.002~mag, in the selected magnitude range ][]{gaia2}. We mention that {\it Gaia} DR2 data still suffers some systematics in both photometry and parallax zero-point definition (depending on the magnitude), estimated to be in the worst cases of about 0.01~mag in photometry \citep{evans18} and of about 0.03-0.05~mas in parallax \citep{lindegren18,zinn19,schonrich19}. We investigated the possible effect of such errors in the synthetic data sets, as discussed below.

The first step was to select the MS portion to be used in the procedure, according to the following requirements: 1) independence of the age, 2) good agreement between theoretical models and data and 3) dependence on helium-to-metal enrichment ratio, \dydz. An additional requirement is that the selected region is unaffected by not well-constrained parameters used in stellar models, such as the core overshooting parameter \citep[see e.g.][]{saslaw1965,shaviv1973}.  
\begin{figure}
\centering
\includegraphics[width=0.98\columnwidth]{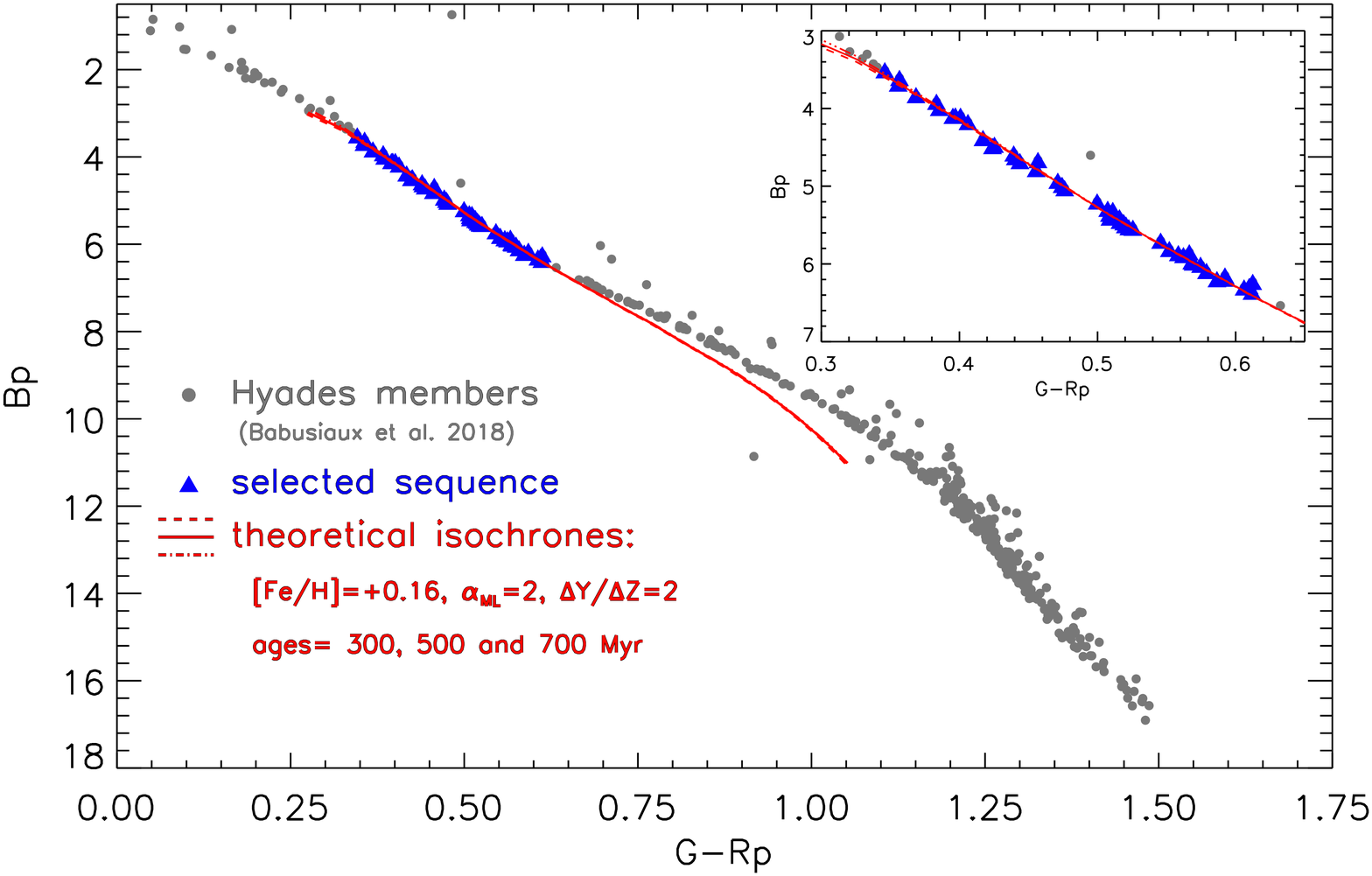}
\caption{Hyades sequence in the {\it Gaia} CMD, data from \citet{babusiaux18}, compared to our partial isochrones with [Fe/H]$=+0.16$ (\dydz=2, and \ml=2) and three different ages, namely 300 (dashed line), 500 (solid line) and 700~Myr (dashed-dotted line): the isochrones at different ages overlap almost perfectly along the whole sequence. The part of the sequence selected for the analysis is shown as blue-filled triangles.}
\label{fig:hyades}
\end{figure}

Figure~\ref{fig:hyades} shows a qualitative comparison between the {\it Gaia} Hyades sequence and our partial isochrones in the selected mass range [0.4, 1.5]~\msun that satisfies the three conditions discussed above. Notice in particular that the location of the MS in this mass interval is almost independent of the age\footnote{The isochrones showed in figure have been computed assuming [Fe/H]=$+0.16$, a helium to metal enrichment ratio of 2.0 and a mixing length parameter of 2 (see next sections), although for this preliminary and qualitative analysis the adopted value of \dydz{} and \ml{} are not essential.}, within the current uncertainty, as proven by the superposition of the three isochrones with ages 300, 500, and 700~Myr. Thus, to avoid age dependence we have adopted a cut-off magnitude of $B_P=3.5$~mag (about 1.4~\msun).

From the same figure it is also evident that the agreement between data and models is very good for $B_P$ magnitudes below 6-7~mag, while for fainter stars the observed sequence significantly deviates from the theoretical one (reaching a maximum difference in colours of about 0.06~mag). To address the problem of the origin of this discrepancy, which has been observed by different authors using several theoretical models and which is already discussed in the literature \citep[see e.g.][]{gagne18} is out of the purposes of this paper. It is common opinion that part of the disagreement is caused by the atmosphere models used to obtain the synthetic spectra, which still have problems in correctly predicting the flux in the optical/blue part of the spectra \citep[e.g.][]{kucinskas05,casagrande14}. Indeed, such a wrong behaviour, that is the bend of theoretical models towards bluer colours, does not occur when infrared colours -- such as 2MASS magnitudes -- are used, down to 0.1-0.2~\msun{} \citep{kopytova16}. However, the observations in infrared colours do not achieve the high photometric precision required for our analysis. To use the {\it Gaia} DR2 data we were forced to limit the analysis to a smaller portion of the observed MS. In particular we restricted to $B_P\le  6.5~$mag (about 0.8~\msun). In this magnitude range we also removed from the real cluster 3 stars that are obviously out of the MS sequence (suspected unresolved binaries). Thus, we have a sample of observed 49 stars and we can built the synthetic data set with the same number of stars. 

Having selected the portion of MS to be used, we tested the capability of the statistical method to constrain the parameters we are interested in ([Fe/H], \dydz, and \ml) from the ($G-Rp$, $B_P$) plane taking into account the uncertainties on the observable quantities, using different mock data sets of artificial stars with characteristics similar to  those of the Hyades. In particular we explored the effects on the result of different choices on (1) the photometry/distance errors, (2) number of stars, and (3) portion of MS used in the recovery. We define the sets of synthetic data as described in the following:
\begin{enumerate}
\item {\bf TYPE\_0}: the set used only to test the recovery algorithm and to evaluate possible systematic biases. It consists of a set of 50 synthetic stars with an unrealistic small error in magnitudes, that is, 0.00005~mag, in the selected $B_P$ range [3.5, 6.5]~mag, which corresponds to populate the isochrone in the mass range [0.83, 1.35]~\msun. The synthetic data set is calculated for [Fe/H]=$+0.16$, \dydz=2, \ml=2, and an age of 500~Myr. This set is designed to evidence biases due to the recovery procedure, since the simulated observational errors are extremely small\footnote{We adopted an error $\sigma =0.00005$~mag, different from zero in the TYPE\_0 case, because the dispersion $\sigma$ appears at the denominator of the quantities used to evaluate the likelihood (see Sect.~\ref{sec:rec}) and $\sigma = 0.0$ would produce a divided-by-zero number in the likelihood evaluation. The value of 0.0005 has been checked to be small enough to produce simultaneously a delta-like function and a to avoid numerical problems.}.

\item {\bf TYPE\_1}: we adopted an uncertainty of 0.002~mag on the colour index ($G-R_P$) and 0.011~mag on the $B_P$ absolute magnitude. The uncertainty on the colour index is designed to mimic the mean value of the errors of the {\it Gaia} colours for the Hyades in the adopted $B_P$ magnitude range. The error on the absolute magnitude $B_P$ corresponds to the uncertainty on the magnitude quoted by the {\it Gaia} Collaboration (0.001~mag) plus an uncertainty of about 0.01~mag on the distance modulus. This last quantity was evaluated using the  estimated parallax uncertainty, $\pi=21.15\pm0.04$~mas \citep{babusiaux18} and a systematic parallax error of 0.1~mas \citep{brown18}. We populated a theoretical isochrone in the mass range [0.83, 1.35]~\msun, which corresponds to an absolute magnitude range $B_P \in [3.5, 6.5]$~mag. We simulated $N_s=50$ stars, a number similar to that found for the real Hyades cluster in the selected magnitude range. This case is designed to be as close as possible to the real Hyades data.
\item {\bf TYPE\_2}: we used the same errors on photometry as in TYPE\_1, but we reduced the systematic error on the parallaxes to 0.01~mas. Thus, the adopted uncertainty on $B_P$ mag is 0.005~mag.
\item {\bf TYPE\_3}: as in TYPE\_1, but for a larger sample of stars that extends down to $B_P = 10$~mag. We adopted the following parameters, $N_s=100$ in the magnitude range $B_P \in $[3.5, 10]~mag, which corresponds to a mass range [0.5, 1.35]~\msun.
\end{enumerate}

\section{Recovery procedure}
\label{sec:rec}
To obtain the most probable set of parameters from a given set of data we used a Bayesian technique, similar to that described in \citet{randich18} \citep[see also ][]{gennaro12}. In the following we will refer to \emph{parameters} as the set of quantities that are inferred from our analysis, which in our specific case are [Fe/H] (hereafter $\zeta$), the helium-to-metal enrichment ratio \dydz{} (hereafter $\xi$) and the mixing length parameter (hereafter \ml). The parameters are obtained through the comparison between theory and observations in the ($G-Rp$, $B_P$) plane using our grid of theoretical models. In particular, given a data set and a set of models, we can define a posterior probability, which -- by Bayes theorem -- is proportional to the likelihood. If we indicate with $\mathbf{\hat{q}}_i\equiv(\hat{q}_{i,1},\hat{q}_{i,2},...,\hat{q}_{i,j},...)$  where $j=1,...,N_\mathrm{obs}$ the vector of observed quantities for the $i$-th star and with $\mathbf{q}\equiv(q_1(\mathbf{p}),q_2(\mathbf{p}),...,q_j(\mathbf{p}),...)$ the corresponding quantity obtained from theoretical models  which depend on the set of parameters $\mathbf{p}\equiv(\zeta,\xi,\alpha_\mathrm{ML})$, we can define the likelihood for the $i$-th star as,
\begin{eqnarray}
\mathcal{L}_i(\mathbf{\hat{q}}_i|\mathbf{p}) \equiv &&\prod_{j=1}^{N_\mathrm{obs}} \frac{1}{\sqrt{2\pi}\sigma_{i,j}}\times \nonumber\\
&& \times \exp{\bigg\{-\frac{1}{2} \frac{[\hat{q}_{i,j} - q_j(\zeta,\xi,\alpha_\mathrm{ML})]^2}{\sigma_{i,j}^2} \bigg\}}
\end{eqnarray}
where $\sigma_{i,j}$ is the observational uncertainty. Then, the total likelihood of the cluster is given by the product of the likelihood of the single stars, 
\begin{equation}
\mathcal{L}(\mathbf{\hat{q}}|\mathbf{p}) = \prod_{i=1}^{N_s} \mathcal{L}_i(\mathbf{\hat{q}}_i|\mathbf{p})
\end{equation}
The posterior probability can also account for the presence of a prior that is, any additional information about the studied case. If we indicate by $f_p(\zeta,\xi,\alpha_\mathrm{ML})$ a prior distribution, then the posterior probability is given by
\begin{equation}
P(\mathbf{p}|\mathbf{\hat{q}}) = C f_p(\zeta,\xi,\alpha_\mathrm{ML}) \mathcal{L}(\mathbf{\hat{q}}|\mathbf{p})
\end{equation}
where $C$ is the normalisation constant, inconsequential in the present analysis. In the following we use only a prior on the cluster metallicity, $f_p(\zeta)$ for which information from stellar spectroscopy is available. We adopted two distributions for the metallicity prior: 1) a flat distribution $f_p(\zeta) = 1$ and 2) a Gaussian prior, defined as:
\begin{equation}
f_p(\zeta) \equiv \exp\bigg[ -0.5\frac{(\zeta - [\mathrm{Fe/H}]_\mathrm{pr})^2}{\sigma_{[\mathrm{Fe/H}]}^2} \bigg]
\end{equation}
 That is a prior centred in the observed \fehpr{} value with its uncertainty assumed as half width at half-maximum, \sigpr, of the Gaussian distribution.

The posterior probability $P(\mathbf{p}|\mathbf{\hat{q}})$ depends on the three parameters to infer. To reduce it to a monodimensional distribution -- thus defining the most probable value of  the three parameters and their corresponding credible intervals (hereafter CI) -- we used the following procedure. We marginalized (integrated) over all the parameters but that we are interested in, that is,
\begin{eqnarray}
&&F(\zeta) = \int \mathrm{d}\xi\int  P((\zeta,\xi,\alpha_\mathrm{ML})|\mathbf{\hat{q}}) \mathrm{d}\alpha_\mathrm{ML}\\
&&G(\xi) = \int \mathrm{d}\zeta\int  P((\zeta,\xi,\alpha_\mathrm{ML})|\mathbf{\hat{q}}) \mathrm{d}\alpha_\mathrm{ML}\\
&&H(\alpha_\mathrm{ML}) = \int \mathrm{d}\zeta\int  P((\zeta,\xi,\alpha_\mathrm{ML})|\mathbf{\hat{q}}) \mathrm{d}\xi
\end{eqnarray}
$F(\zeta)$, $G(\xi)$, and $H(\alpha_\mathrm{ML})$ are thus the distributions from which one derives the most probable values of $\zeta$ ([Fe/H]), $\xi$ (\dydz), and \ml{} (mixing length), respectively.

The most probable value of the marginalized distribution is defined as the maximum of the distribution itself, that is, the mode. Then, the CI is defined as the region that contains $68\%$ of the area under the distribution, by removing $16\%$ of the area contained in the wings from the left and from the right. In other words, if $g(x)$ is a marginalized distribution defined over the interval $[x_\mathrm{min},x_\mathrm{max}]$, the CI $[x_1,x_2]$ is defined as:
\begin{eqnarray}
&& 0.16\mathcal{A} = \int_{x_\mathrm{min}}^{x_1} g(\tau) \mathrm{d}\tau\\
&& 0.84\mathcal{A} = \int_{x_\mathrm{min}}^{x_2} g(\tau) \mathrm{d}\tau
\end{eqnarray} 
where $\mathcal{A}$ is the area under the curve.

We also tested the dependence of the results on the adopted criterion for the most probable value (a different estimator) and CI. In particular we tested three different choices on the estimator and on the CI: (1) the mode with a CI defined using the 16th and 84th percentile (our reference choice), (2) the median of the distribution with a CI defined using the 16th and 84th percentile,  and (3) the mean of the distribution with a CI defined using the root mean square deviation (hereafter RMSD). In the latter case (the mean as estimator), we used the following equations to define the mean $\mu$ and the RMSD $\sigma_\mu$:
\begin{eqnarray}
&&  \mu \equiv \frac{\int_{x_\mathrm{min}}^{x_\mathrm{max}} \tau\,g(\tau) \mathrm{d}\tau}{\int_{x_\mathrm{min}}^{x_\mathrm{max}}g(\tau) \mathrm{d}\tau}\\
&& \sigma_\mu^2\equiv \frac{\int_{x_\mathrm{min}}^{x_\mathrm{max}} (\tau-\mu)^2\,g(\tau) \mathrm{d}\tau}{\int_{x_\mathrm{min}}^{x_\mathrm{max}}g(\tau) \mathrm{d}\tau}
\end{eqnarray} 
We verified that in all the cases presented in this work the difference in the estimated parameters due to the use of the mode, the median or mean as estimator of the best value is smaller than the estimated CI. In particular we found that in most cases the three estimators produce essentially the same result: we discuss this point in more details in Section~\ref{sec:feh}.

\section{The grid of models}
\label{sec:models}
We constructed a grid of stellar models using the PISA stellar evolutionary code \citep{deglinnocenti08,dellomodarme12} in the same configuration described in previous papers \citep{tognelli15b,tognelli18}. We briefly mention the main input parameters/physics that are relevant for the present analysis, which are the convection efficiency in superadiabatic regions, the outer boundary conditions (atmospheric models) and the solar mixture (metal relative abundance in the Sun). For the last quantity, we adopted the \citet{asplund09} metal distribution. Concerning the superadiabatic convection efficiency, we used the mixing length theory \citep{bohm58} which is characterized by the dependence on a free parameter (the mixing length scale), \ml. Such a quantity is a free parameter that has to be calibrated. A common way of calibrating \ml{} consists in constructing a standard solar model, which is a 1~\msun{} model that at the age of the Sun has to reproduce simultaneously the solar radius, luminosity and $(Z/X)_{\sun}$ within a given tolerance. These three quantities depends on the initial values of helium $Y_{\mathrm{ini},\sun}$, metallicity $Z_{\mathrm{ini},\sun}$ and \ml: from the solar calibration we obtained \ml$_{\sun} = 2.0016$, $Y_{\mathrm{ini},\sun}=0.2624$ and $Z_{\mathrm{ini},\sun}=0.01524$. Another key ingredient for the computation of models with a convective envelope (in our case for MS stars with $M\la 1$~\msun)  is the adopted outer boundary conditions (i.e. atmospheric structure) needed to specify the pressure and temperature at the bottom of the atmosphere to close the system of differential equations that describes the stellar structure. We adopted the \citet{allard11} pre-computed models, obtained using input physics very similar to that we used for the computation of the interior structure of the star. The consistency between atmospheric and interior structure, in particular the treatment of convection, is a crucial point to avoid the inclusion of uncontrollable systematic effects. We emphasize that the value of \ml{} we obtained with the solar calibration is basically the same used for the computation of the atmospheric structure (i.e. \ml$_\mathrm{,atm} = 2.0$), which results in a very similar convection treatment in the interior and in the atmosphere, thus giving to the models a high level of internal consistency.

The required mass range for the model grid to be compared with the Hyades sequence is $0.8\la M/$\msun$\la 1.4$. However for some of the test cases we needed a wider mass range. So, we computed 18 stellar models in the mass range $M\in$[0.4, 1.5]~\msun  with a variable spacing (0.05~\msun{} for $M\le 1.0$~\msun ~and $0.1$~\msun for larger masses), each for 11 [Fe/H] values in the range [$+0.08$,~$+0.28$]~dex with a spacing of $0.02$~dex, to covers the range of recent spectroscopic determinations with the related errors. For each value of [Fe/H] we adopted 11 values of \dydz{} in the range [1,~3] with a spacing of $0.2$. Moreover, each model set specified by a given value of mass, [Fe/H] and \dydz{} has been calculated for five different values of \ml{} in the range [1.8,~2.2] with a spacing of $0.1$. The selected \ml{} values are chosen to cover a symmetric interval about our solar calibrated mixing length parameter (namely \ml=2). In total, we computed 605 sets of stellar models. 

The parameters estimation algorithm relies on the set of pre-computed isochrones (to reduce the computational time of the recovery procedure), and no interpolation is performed on the fly. Therefore, a dense grid is needed to achieve a high accuracy in the results and to avoid noisy behaviours \citep[see also][]{Bazot2016}. To do this, the displacement of a model at a fixed mass in the CMD due to the variation of each parameter in the grid (i.e. [Fe/H], \dydz, and \ml) must be smaller than the observational uncertainty. To this regard, the average smallest error is about 0.002~mag, and it comes from the $G-Rp$, colour index. We verified that to achieve a good likelihood in the recovery (smooth and without spurious peaks) we need a variation of about one-half/one-third of such an uncertainty. The grid spacing (in each parameter) has been consequently tuned to satisfy this requirement, by interpolating the models on a finer grid before running the recovery procedure: we adopted a spacing of $0.005$~dex in [Fe/H], $0.05$ in \dydz, and $0.02$ in \ml, for a total of  $35\,301$ sets of stellar models. We checked (extracting some cases from the grid) that the interpolation produces models in a very good agreement with those obtained by a direct computation by means the stellar evolutionary code (negligible differences).

Having constructed the grid of models, we generated the isochrones in range of age [300,~700]~Myr and we converted them to {\it Gaia} magnitudes using the bolometric corrections obtained from the MARCS2008 synthetic spectra \citep[hereinafter M08,][]{gustafsson08}, and the magnitude zero points given in \citet{evans18}. 

\section{Effect of changing \ml, [Fe/H] and $\Delta Y/\Delta Z$.}
\label{sec:qual_test}
\begin{figure*}
\centering
\includegraphics[width=0.45\linewidth]{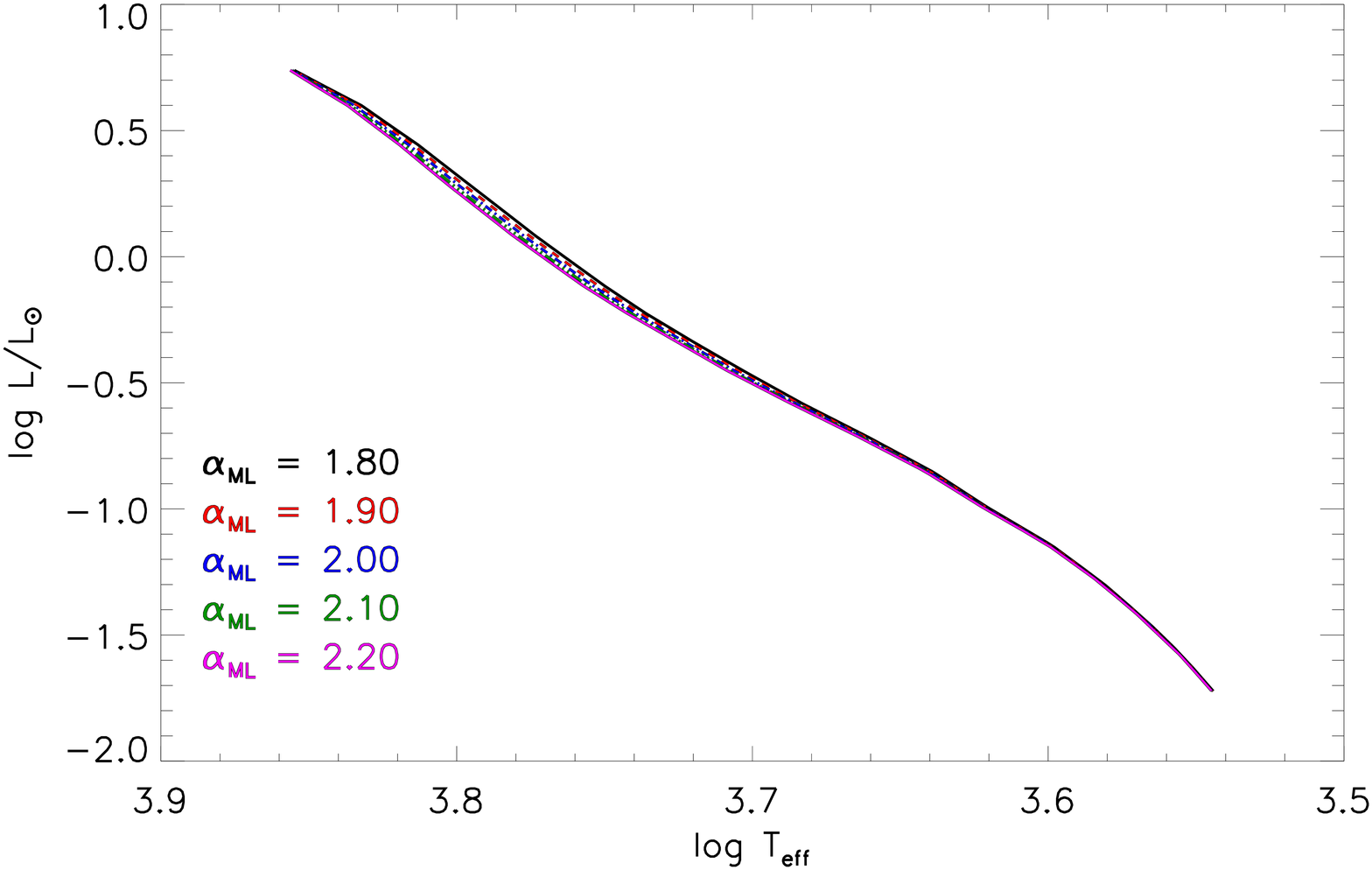} 
\includegraphics[width=0.45\linewidth]{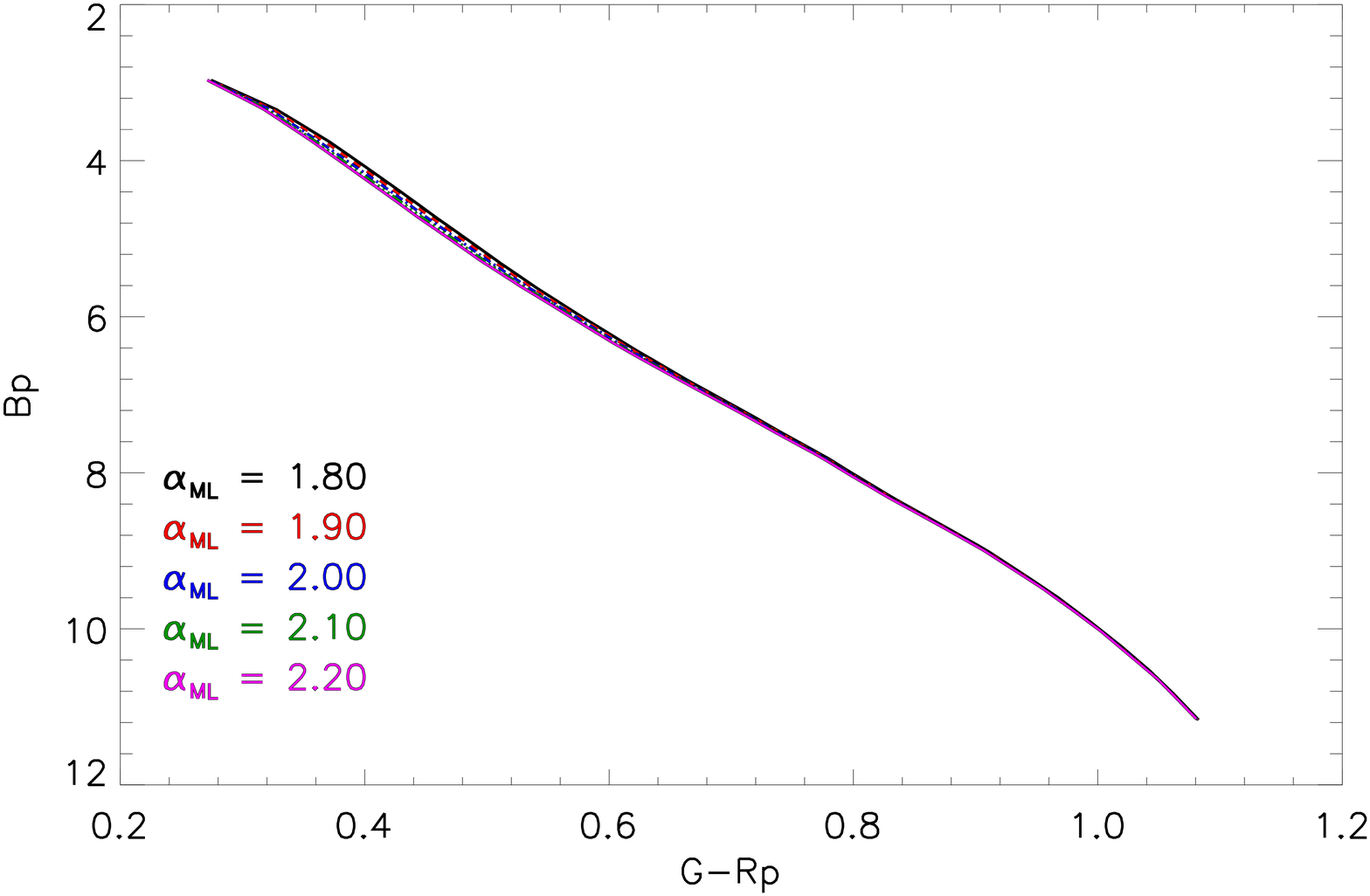}
\includegraphics[width=0.45\linewidth]{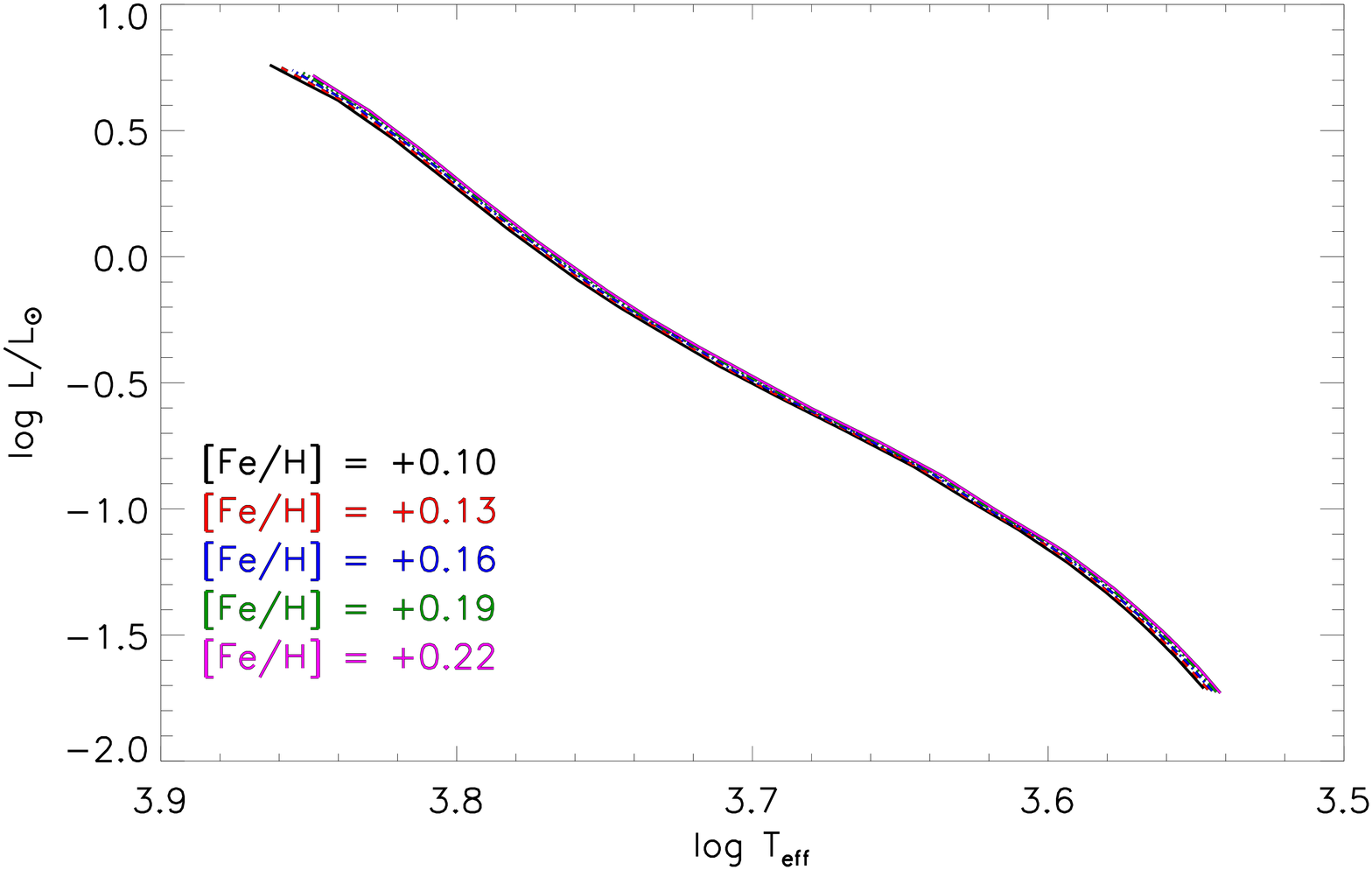}
\includegraphics[width=0.45\linewidth]{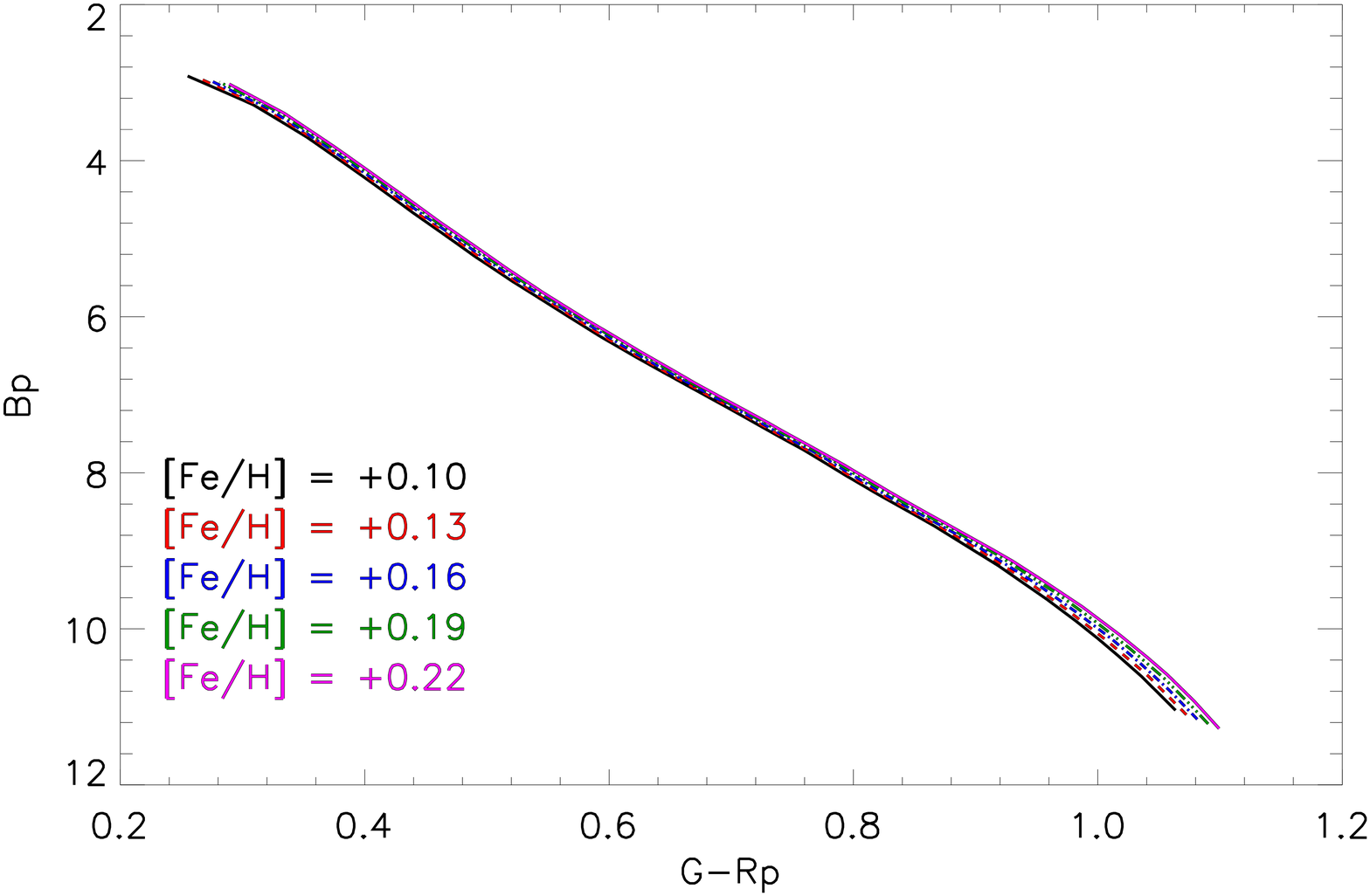}
\includegraphics[width=0.45\linewidth]{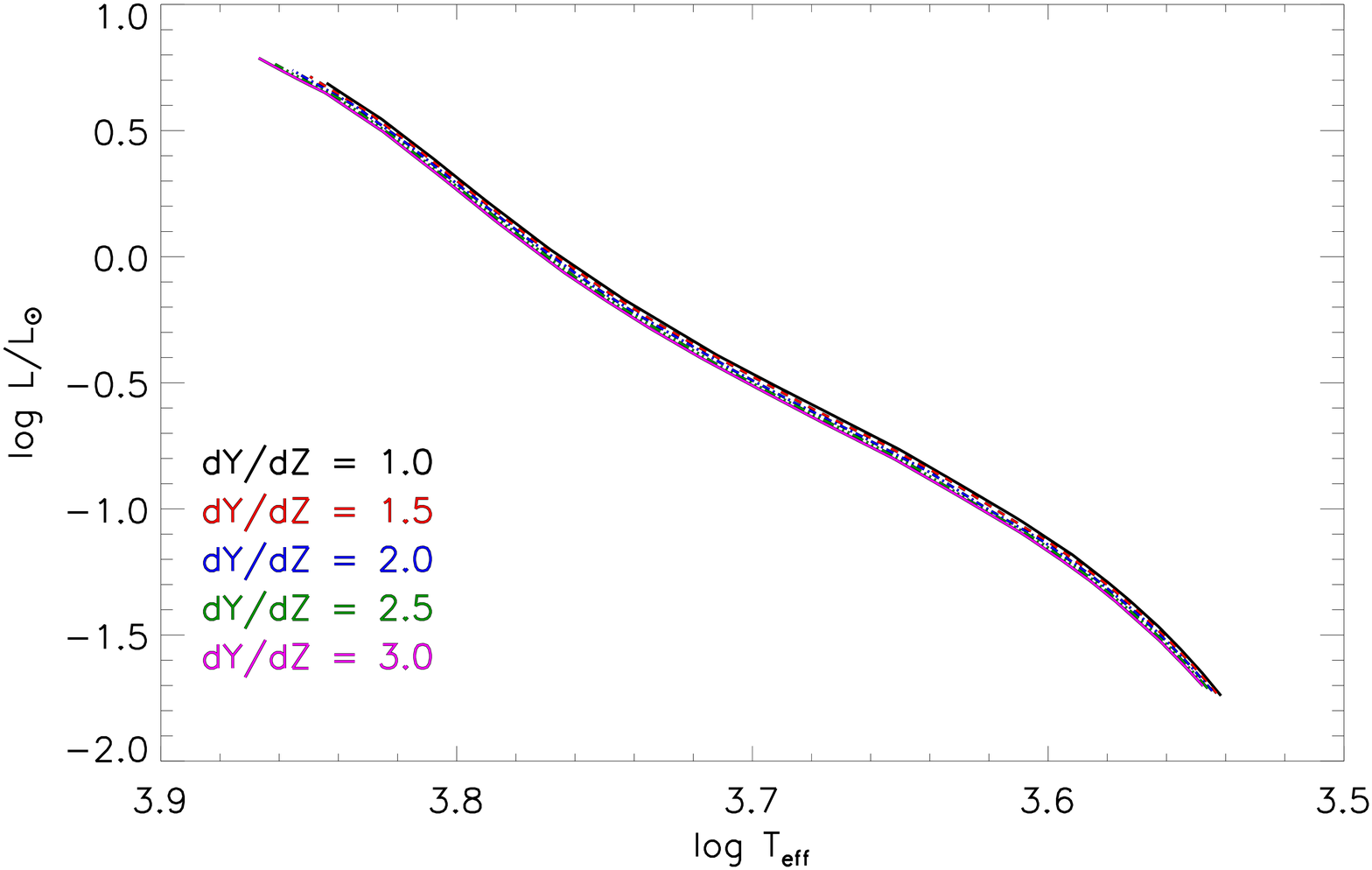}
\includegraphics[width=0.45\linewidth]{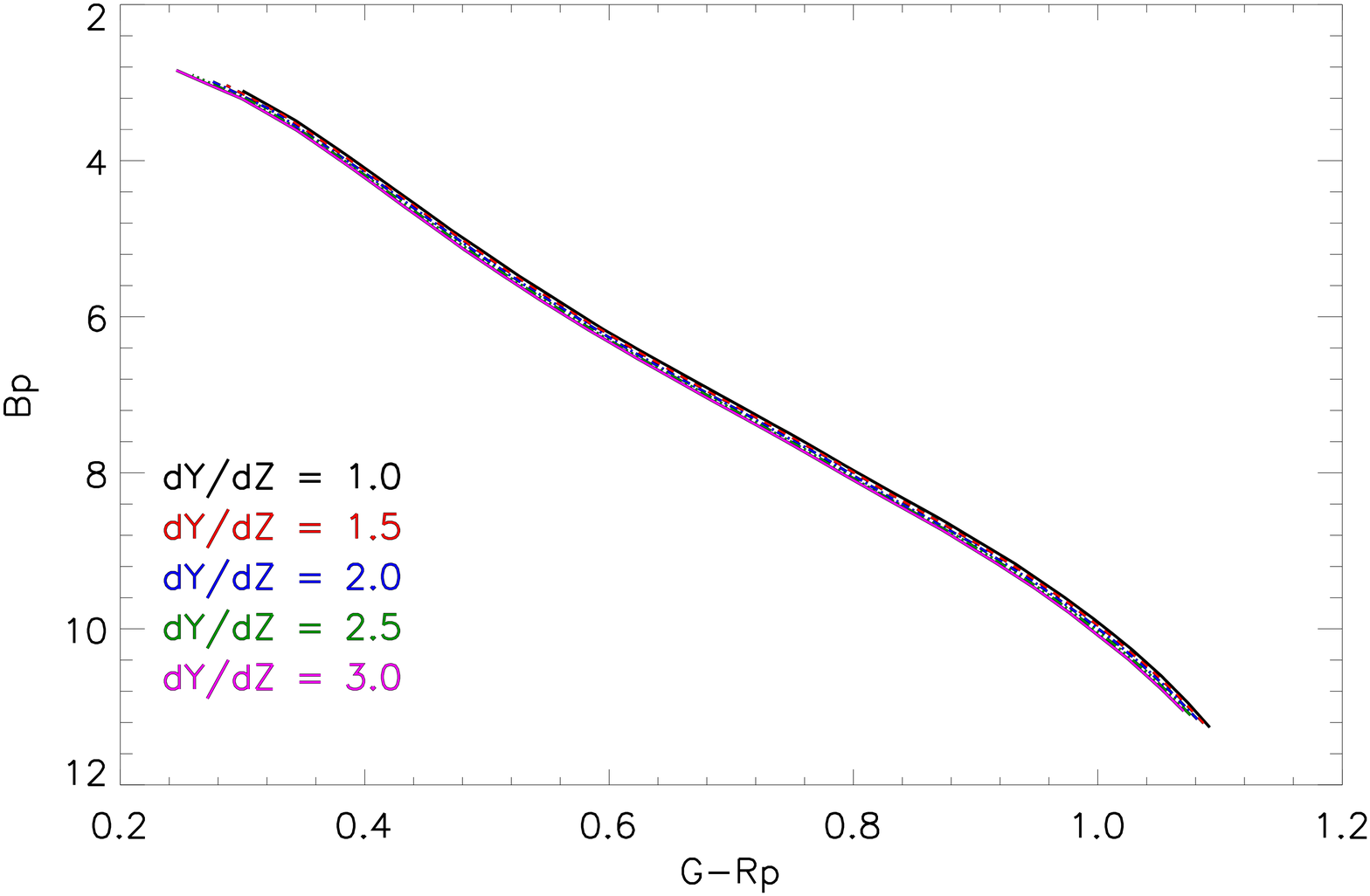}
\caption{Change of the position of a 500~Myr isochrone in the HR (left-hand column) and {\it Gaia} CMD (right-hand column) caused by a variation of the mixing length parameter (top panels), [Fe/H] value (middle panels), and \dydz{} (bottom panels).}
\label{fig:eff}
\end{figure*}

Before starting the analysis, it is useful to remind the effect of [Fe/H], \ml and \dydz{} on theoretical isochrones. Figure~\ref{fig:eff} shows the effect of varying separately \ml{}, [Fe/H], and \dydz{} within the selected range on a 500~Myr isochrone both in the HR and the CMD. 

Concerning the mixing length (top panels), it is well known that the effect strongly depends on the stellar mass. The $T_\mathrm{eff}$ and $R$ of stars with $M \la 0.4$-$0.5$~\msun{} (i.e. $B_P \ga 10$) in MS are independent of the the value of \ml{} because the structure is dense enough to have an almost adiabatic temperature gradient even in external regions of the star. On the other hand, increasing the stellar mass, the star gets progressively more and more superadiabatic in the external convective layers, showing a progressively stronger dependence on \ml. However, the convective envelope gets thinner and thinner as the mass increases, consequently the sensitivity on \ml{} first increases and then decreases and eventually vanishes for $M\ga$1.2-1.4~\msun ($B_P \la 3.5$~mag). This means that the impact of \ml{} on the CMD location of an isochrone is not rigid but depends on the stellar mass (or equivalently on the magnitude): at fixed $B_P$ magnitude a variation of \ml{} of $\pm 0.25$ results in a colour change of about 0.01~mag for $B_P \in [3.7,~5]~$mag which progressively (almost linearly) decreases to about 0.005~mag at $B_P \approx 6$-6.5~mag.

Middle panels of Fig.~\ref{fig:eff} show the effect of varying the initial value of [Fe/H]. The a variation of [Fe/H] results in a change of both $Y$ and $Z$, although the net variation of $Z$ is much larger than the variation of $Y$. As an example, moving from [Fe/H] = $+0.16$ to $+0.22$ results in a change of about $14$~percent in $Z$ and about $2$~percent in $Y$. The effect on the models of a variation of [Fe/H] in the range [$+0.10$, $+0.22$] is smaller than that due to the change of \ml, and it is almost the same for all the models along the isochrone. Only low-mass stars tail ($\log L$/\lsun$\la -1.2$  or $Bp \ga 9~$mag which corresponds to $M\la 0.6$~\msun) show a slightly different dependence on [Fe/H]. An increase of [Fe/H] of $+0.06$~dex produces a rigid shift to the right part of the diagram (cooler models), at fixed $B_P$ magnitude, of about 0.005-0.006~mag in the colour index (in the selected magnitude interval $B_P \in [3,~7]~$mag). Low-mass star tail ($M \la 0.6$~\msun) of the isochrone exhibits a slightly larger dependence on [Fe/H] than the region populated by more massive stars. 

Bottom panels of Fig.~\ref{fig:eff} show the effect of a variation of \dydz. A variation of \dydz{} at fixed [Fe/H] affects mainly the helium content. Similarly to [Fe/H], the variation of the initial helium content produces an almost rigid shift of the isochrone in the HR/CMD. In particular, an increase of \dydz{} of $1$ moves the models towards the left-hand part of the plane (hotter models) of about 0.005-0.007~mag in the colour index at a fixed $B_P$ magnitude.
\begin{figure*}
\centering
\includegraphics[width=0.45\linewidth]{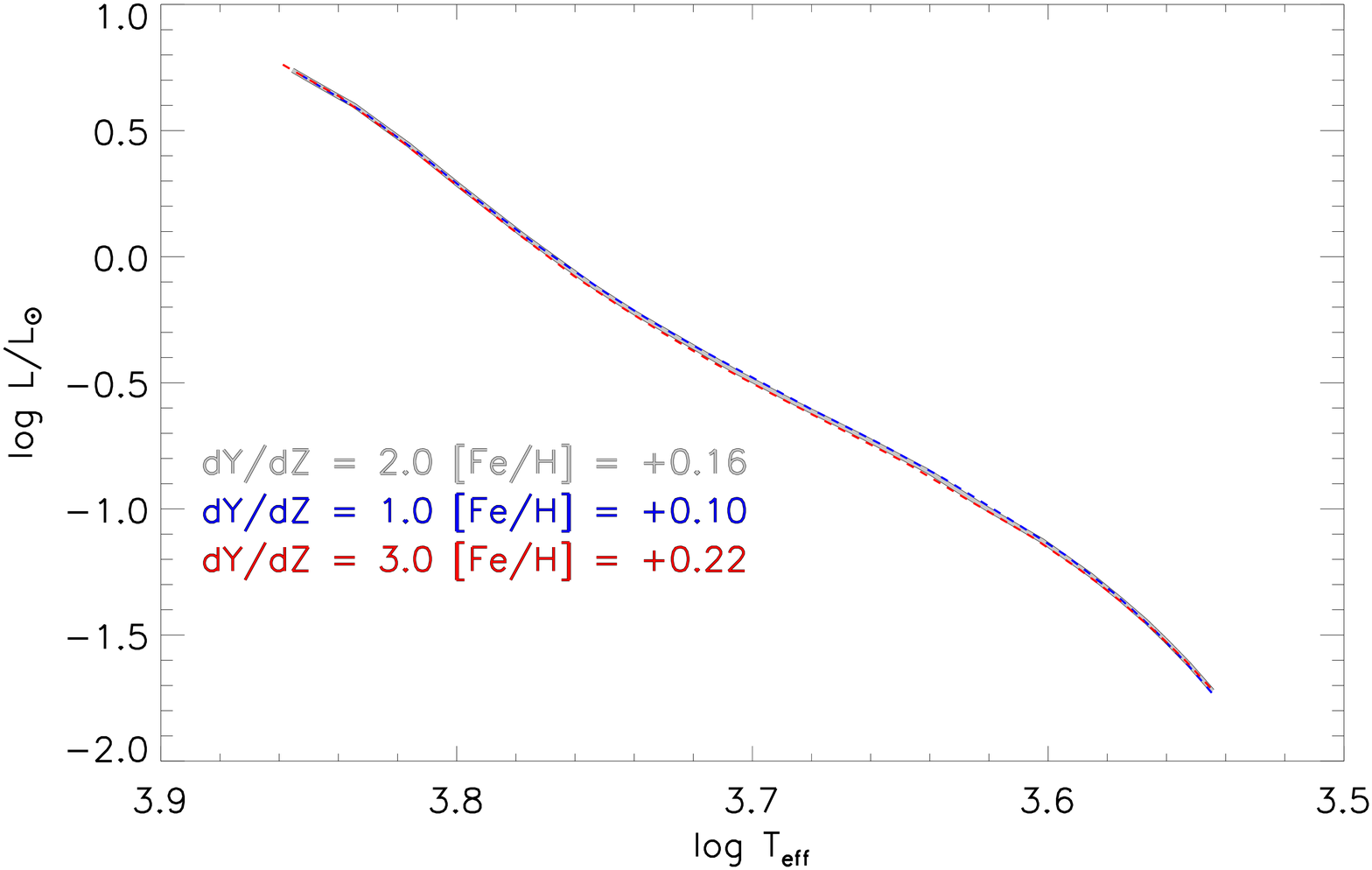}
\includegraphics[width=0.45\linewidth]{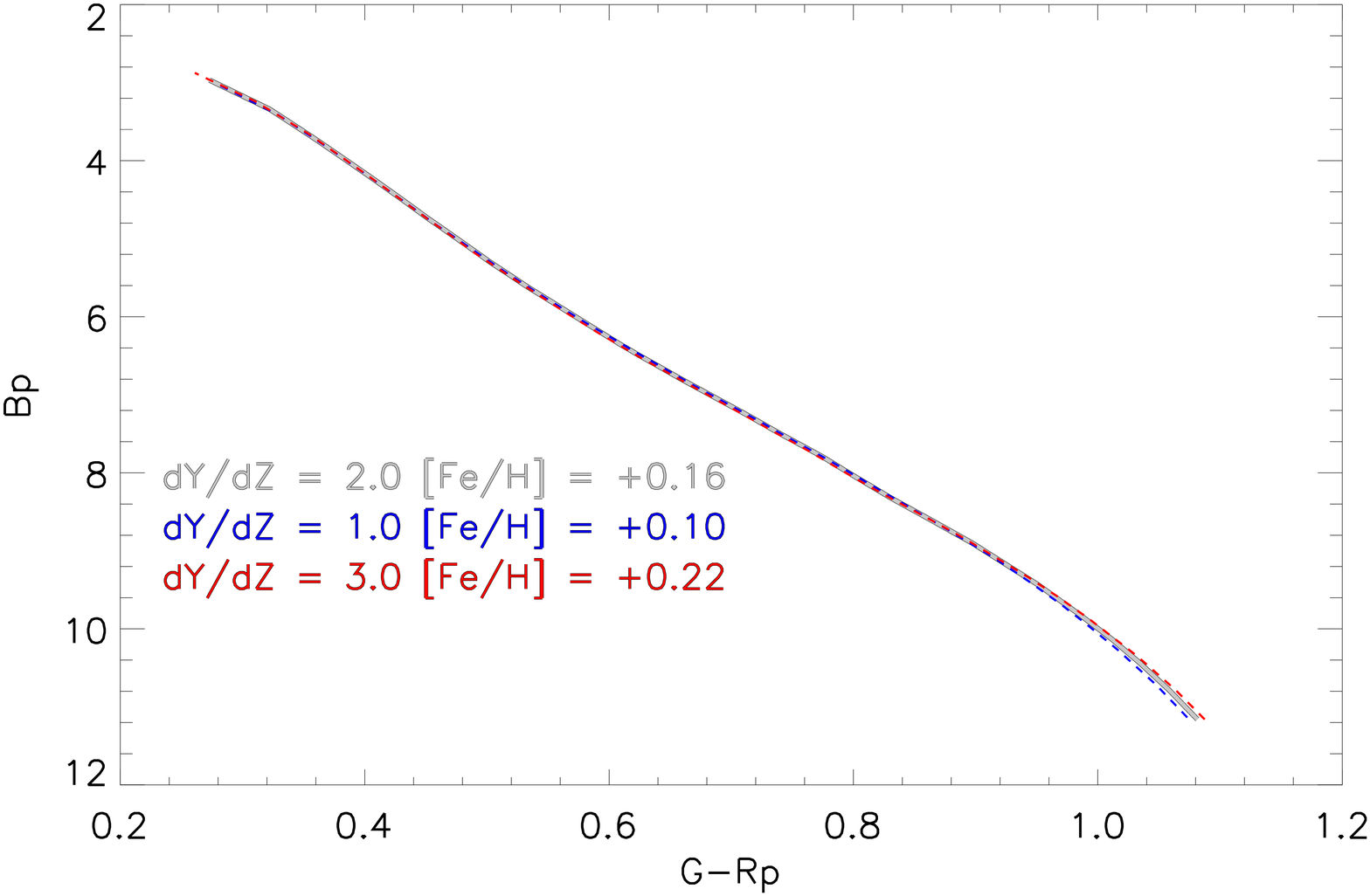}
\caption{Effect on the position of a 500~Myr isochrone in the HR and {\it Gaia} CMD of the simultaneous variation of \dydz{} and [Fe/H] with respect to the reference chemical composition. The reference isochrone (\dydz=2 and [Fe/H]=$+0.16$, grey line) is compared to isochrones with \dydz=1 and [Fe/H]=$+0.10$ (blue dashed line) and \dydz=3 and [Fe/H]=$+0.22$ (red dashed line).}
\label{fig:bal}
\end{figure*}

An increase of \dydz{} has an effect opposite to that caused by an increase of [Fe/H]. Thus, depending on the adopted \dydz{} and [Fe/H] it is possible that the two effects  balance each other. An example is shown in Fig.~\ref{fig:bal} where we plotted a 500~Myr isochrone with the reference chemical composition ([Fe/H]=$+0.16$ and \dydz=2) together with two other 500~Myr isochrones, one with [Fe/H]=$+0.10$ and \dydz=1 and the other with [Fe/H]=$+0.22$ and \dydz=3. The three models are very close both in the HR and {\it Gaia} CMD. There is only a slight difference in HR diagram in the middle part of the isochrone, while for the CMD only at the faint tail of the isochrone (out of the region selected in the present analysis) very small differences are visible. This figure clearly shows that, as well known in the literature, it exists a certain level of degeneracy between [Fe/H] and \dydz{} in a large part of the MS; in particular a variation of $\pm 0.06$~dex in [Fe/H] can be almost entirely balanced by a variation (in the same direction) of $\pm 1$ in \dydz{} \citep[see e.g.][]{castellani02} both in the CMD and in the HR diagram. However, this is not completely true for the low-mass tail of the isochrone ($B_P \ga 9$-10~mag, thus for $M\la 0.6$~\msun): the presence of such a low-mass tail could mitigate the \dydz-[Fe/H] degeneracy. 

This qualitative analysis suggests that, in this work, due to the restrictions on the MS portion suitable for the analysis, it might be difficult to reach an high precision in the simultaneous derivation of \dydz{} and [Fe/H] using MS stars. This problem could be overcame if a reliable spectroscopic estimate of [Fe/H] is available\footnote{From the observational point of view, the measurements of [Fe/H] in K or M dwarfs is difficult and uncertain; however, if such stars belong to the same cluster the mean [Fe/H] of the cluster can be safely used.}. In this case, the adoption of a proper prior on [Fe/H], could be used to constrains the interval of reasonable [Fe/H] values and consequently to better constrain the \dydz{} value.

\section{Testing the performance of the recovery procedure on synthetic data sets}
Before performing an analysis on real Hyades data, it is mandatory to test the capability of the method on a synthetic data set to highlight the bias and the uncertainties related to the recovery of the selected parameters. Thus, the first step is to prove the capability of the algorithm to recover the true parameters, that is, those used to generate the synthetic data set of artificial stars sampled from the same grid of theoretical models used in the recovery itself. More in detail, the mock data have been obtained by taking one of the isochrone available in the grid (with a given set of [Fe/H], \dydz, and \ml), populating it with a given initial mass function (not important for the present analysis), and then by applying a Gaussian perturbation (error) on the magnitudes of each synthetic star to simulate the observational uncertainty. 

It is worth noticing that such a preliminary test corresponds to the best configuration, because the physics adopted in the models of the recovery grid is the same used to construct the data set. In the real world one expects some differences between the physics of observations and of the models which can lead to additional biases and uncertainties.

\subsection{TYPE\_0 data set: ideal data set}
The first test was performed on a mock data set with negligible observational errors (TYPE\_0, 0.00005~mag on the magnitude), which means that the simulated MS stars are not scattered with respect to their original position on the MS. This step is necessary to evaluate the presence of a possible systematic hidden inside the procedure. We generated 50 stars in the $B_P$ range [3.5,~6.5]~mag at 500~Myr adopting as reference values [Fe/H]$=+0.16$, \dydz=2, and \ml=2. 

We applied the recovery procedure to the artificial stars, for two different choices about the adopted prior for the [Fe/H] value: (1) no prior, all the [Fe/H] values are equally probable, and (2) a Gaussian distributed [Fe/H] centred on the value adopted for the synthetic data set. Independently of the adopted prior on [Fe/H], in all the cases, the most probable values of [Fe/H], \dydz, and \ml{} obtained from the recovery procedure are exactly the same used to generate the data sample. The CI (credible interval of each recovered parameter) is much smaller than the step on the grid of models. Thus, we can conclude that, on a set of synthetic data with a negligible observational errors, the method does not show any systematic or bias\footnote{We do not show the plots corresponding to TYPE\_0 models because they are essentially plots of delta-functions centred in [Fe/H], \dydz{} and \ml{} equal to those used to generate the mock data set.}.
 
\subsection{TYPE\_1 data set: Hyades-like data set}
After the validation of the method, we applied the analysis to the TYPE\_1 data set, which was generated to obtain a data set as close as possible to the Hyades. We adopted the same chemical composition and \ml{} of the TYPE\_0 data set. We selected the interval of magnitude/colours, errors and number of stars, as described in Sec.~\ref{sec:met}: we simulated 50 stars in the $B_P$ range [3.5,~6.5]~mag, with a Gaussian distributed error on the $G-R_P$ colour of 0.002 and 0.012~mag on the $B_P$ magnitude. As said before, the relatively large error on $B_P$ -- if compared to that on the colour -- comes from the systematic parallax uncertainty of 0.1~mas (i.e. an error in the distance modulus of about 0.01~mag). 
\begin{figure*}
\centering
\includegraphics[width=0.32\linewidth]{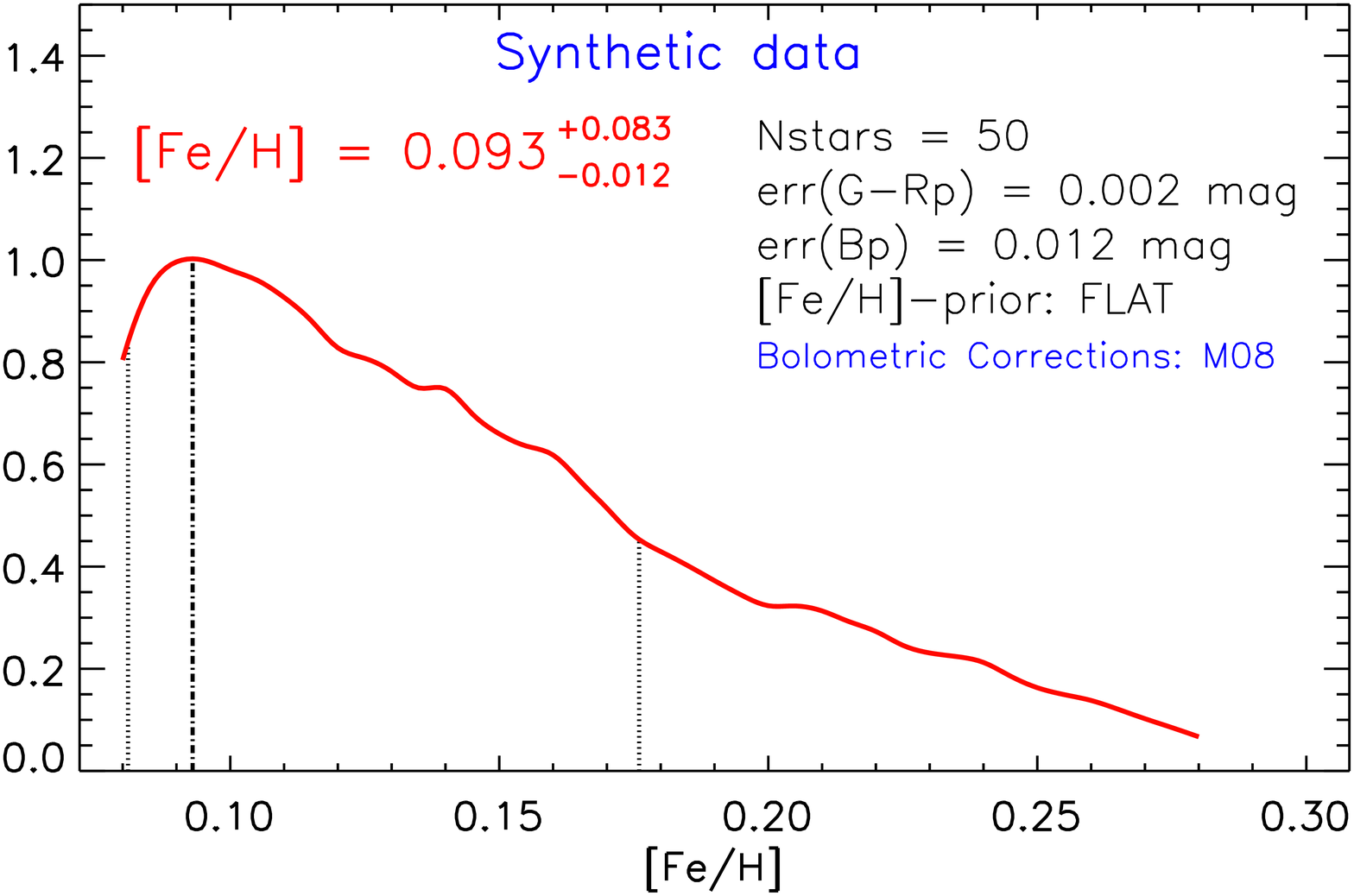}
\includegraphics[width=0.32\linewidth]{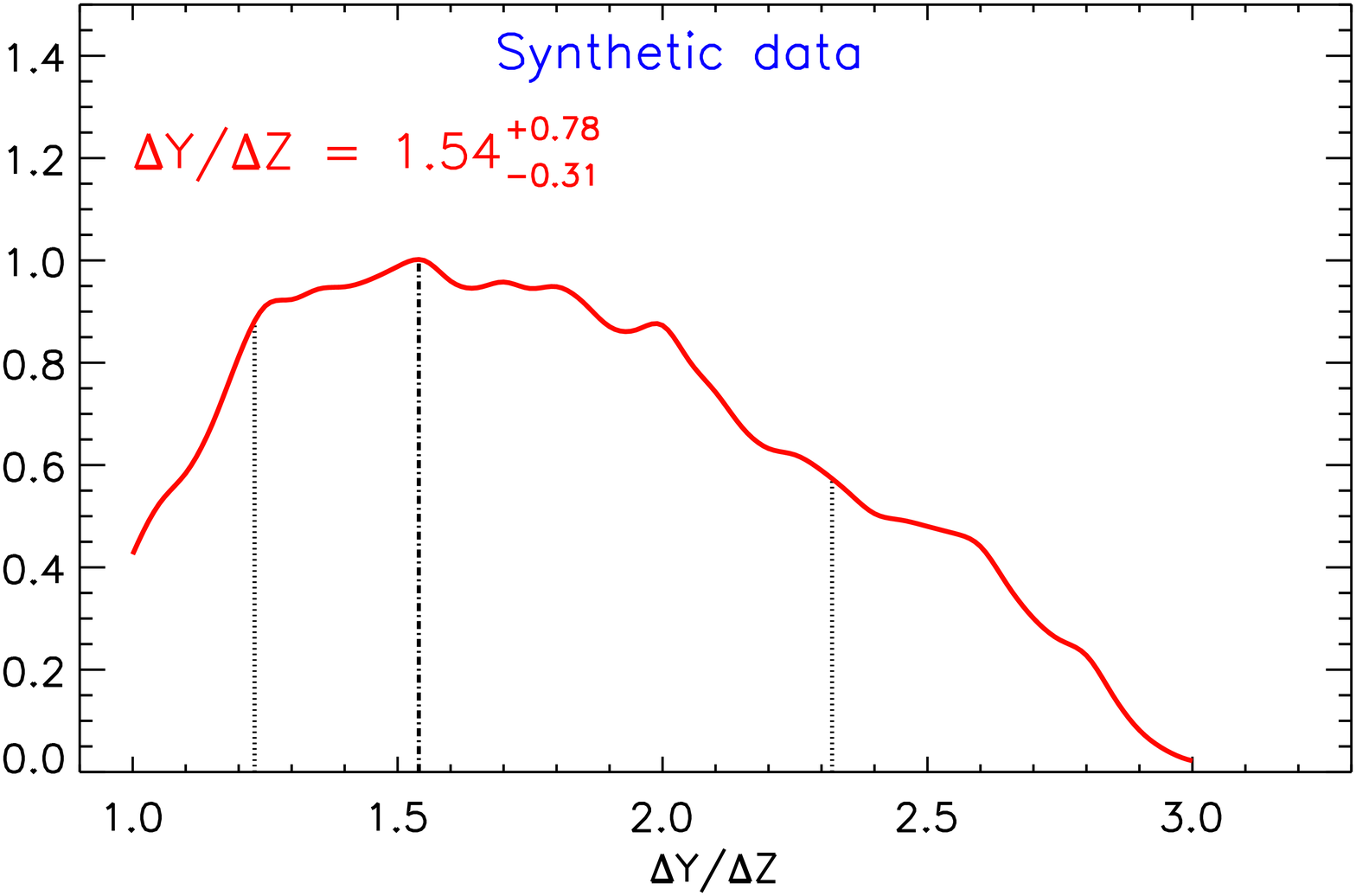}
\includegraphics[width=0.32\linewidth]{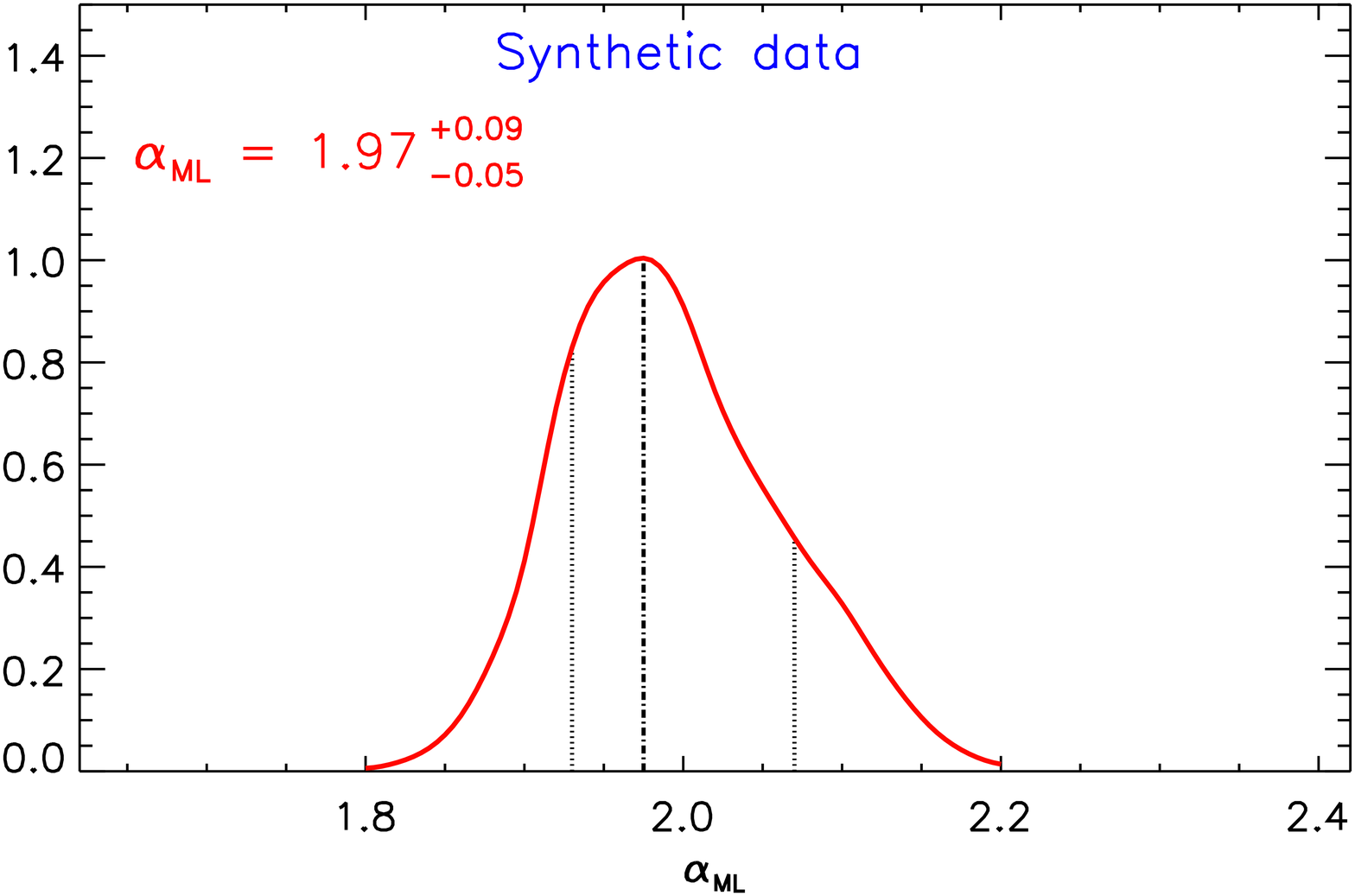}\\
\includegraphics[width=0.32\linewidth]{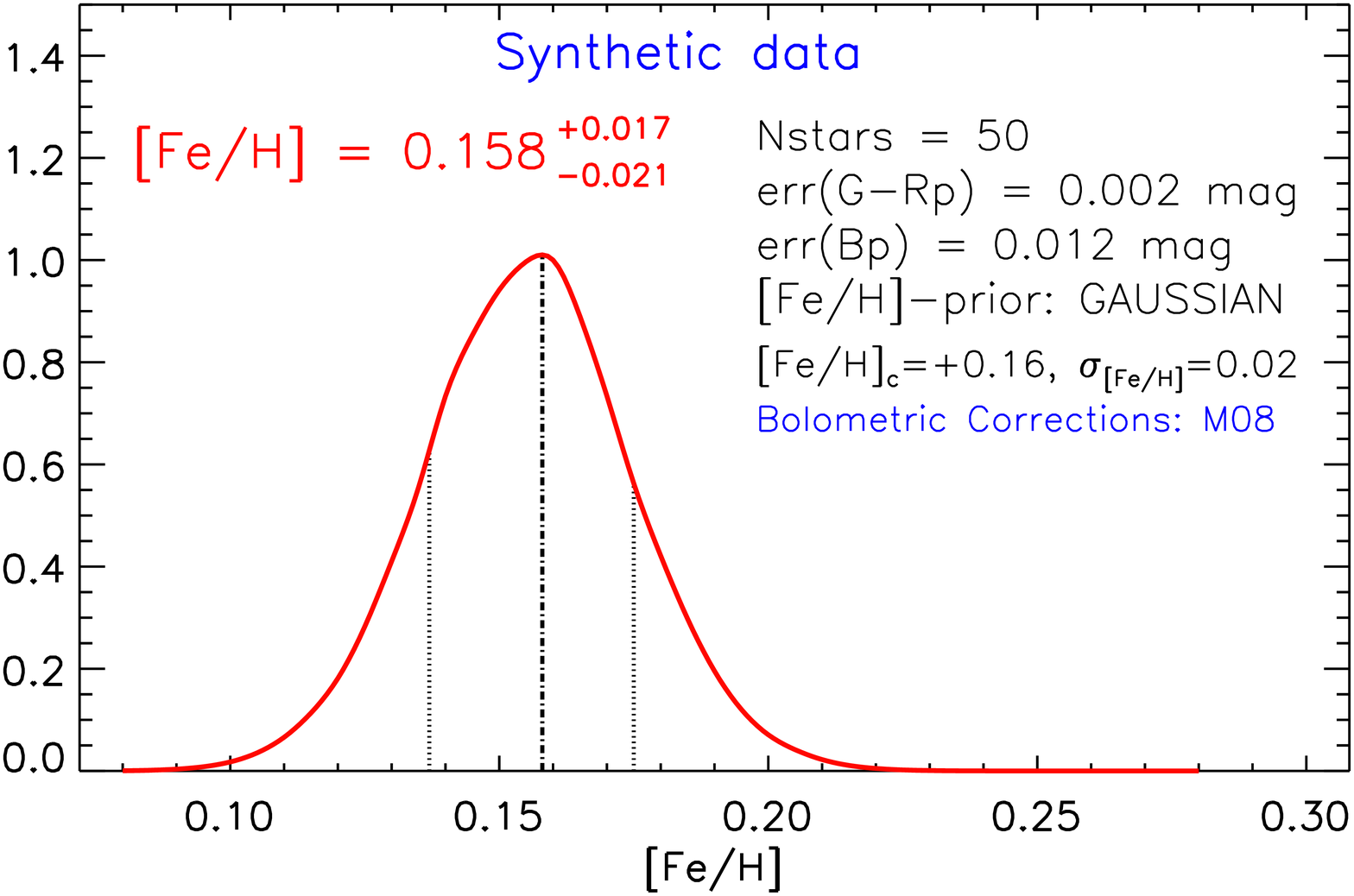}
\includegraphics[width=0.32\linewidth]{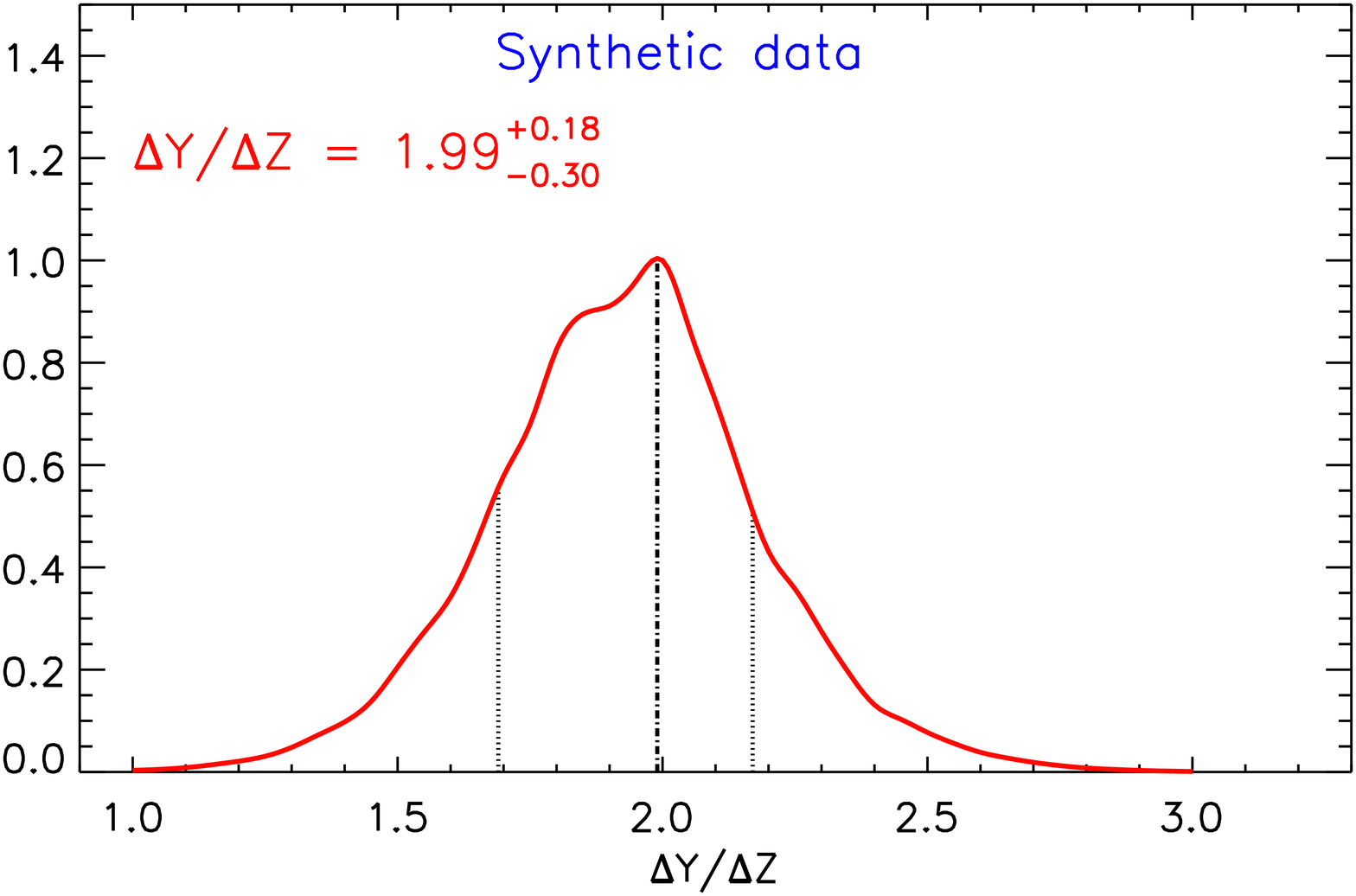}
\includegraphics[width=0.32\linewidth]{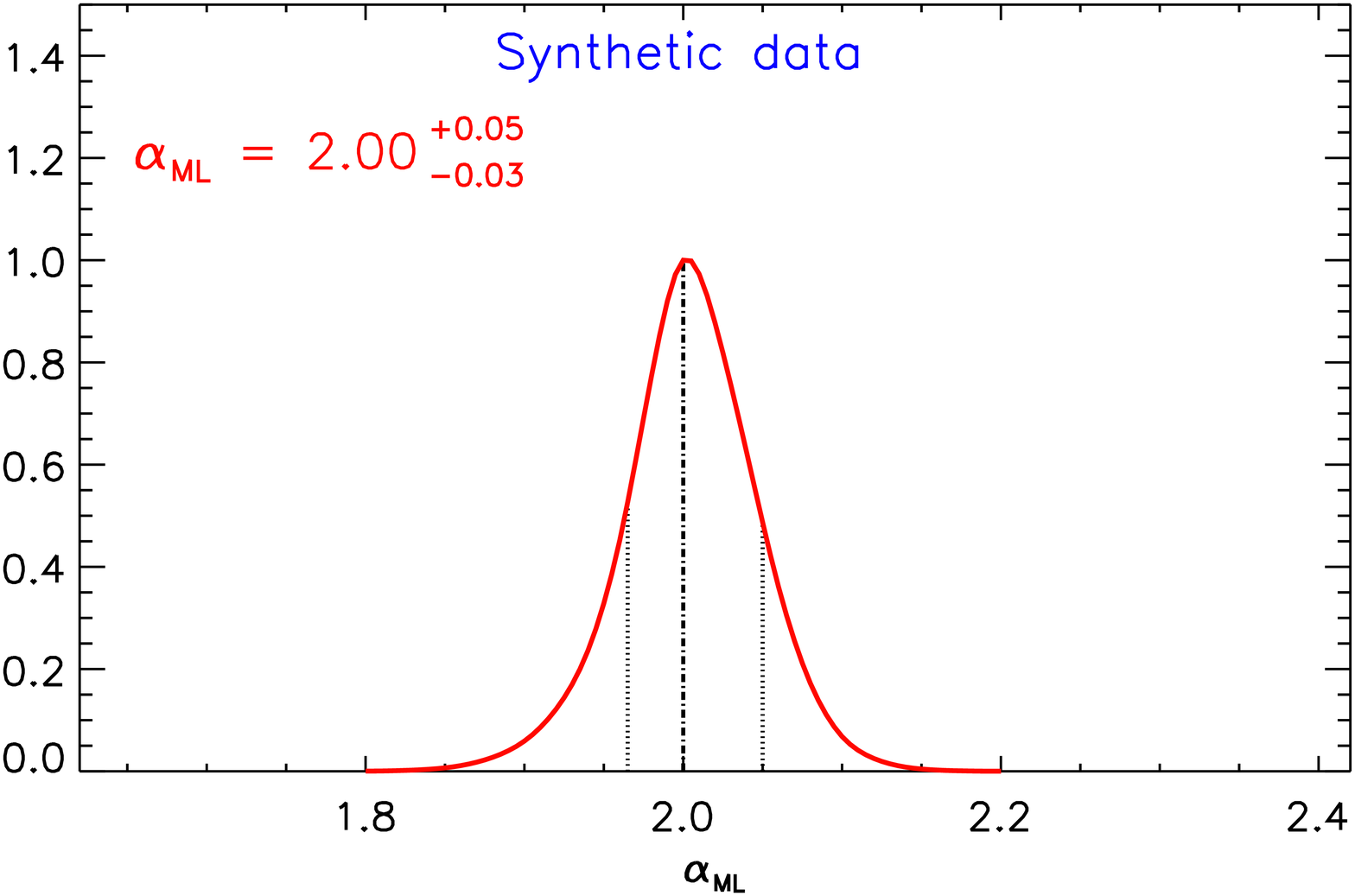}\\
\caption{Posterior marginalized distributions for TYPE\_1 data set, for [Fe/H] (left-hand panels), \dydz{} (middle panels), and \ml{} (right-hand panels), for two choices of the prior on the metallicity, namely flat (top panels) and Gaussian with \fehpr=$+0.16$~dex and \sigpr=0.02~dex (bottom panel).}
\label{fig:type1_lkl}
\end{figure*} 

We ran the recovery procedure to simultaneously derive the most probable values of \dydz, \ml, and [Fe/H]. The top panels of Fig.~\ref{fig:type1_lkl} show the marginalized likelihood for [Fe/H], \dydz, and \ml{} for a flat prior on [Fe/H]. [Fe/H] is badly recovered from the analysis. The most probable value (peak of the distribution) is at about [Fe/H]$=+0.09$, whereas the true one is $+0.16$, and the derived CI on [Fe/H] is very broad, from about [Fe/H]=$+0.08$ to about $+0.18$. The shape of the likelihood shown in figure deserves a discussion. The peak at [Fe/H]=$+0.09$~dex, very close to the edge of the grid, shows that in this first test the value of [Fe/H] is badly recovered by the procedure. The mixing of [Fe/H], ML and \dydz{} on the grid is so strong that the method fails in identifying the true value of [Fe/H] used to simulate the data in this mock data set. 

Similarly to [Fe/H] also \dydz{} is badly recovered and, as shown in figure, the peak is at \dydz$\approx 1.5$, thus underestimating the value of about 0.5, with a large CI, \dydz=[1.2,~2.3]. The large CI in \dydz, comes essentially from the large CI in [Fe/H]. 

The distribution of the \ml{} values looks better. Indeed, as showed in the previous sections, a change in \ml{} affects the models along the isochrone in a differential way, thus producing a different slope of the isochrone in the plane used for the analysis. Moreover, for $B_P \in [4,~6]$~mag the isochrones are very sensitive to the adopted \ml{} and even a small change in \ml{} produces a relatively large effect on the models. This allows a better identification of the most probable value (i.e. the peak and the width of the marginalized likelihood). The recovered value is only 1.5~percent lower than the true one, and the corresponding CI is not very large, \ml$\in[1.92,~2.06]$. 

From the analysis we can conclude that, as expected, the unavailability of independent [Fe/H] determinations drastically complicates the problem of deriving \dydz{} and/or [Fe/H] values with this method even when synthetic data are used. More importantly, the derived values of [Fe/H] and \dydz{} are affected by a large bias. This result is expected and it is the consequence of the degeneracy between [Fe/H] and \dydz{} values, and of the weak dependence of the models on [Fe/H] and \dydz.

We tested whether the availability of an independent estimate of [Fe/H] might be used to improve the goodness of the recovery. To do this we assumed a Gaussian prior on [Fe/H]. This  prior was centred at \fehpr=$+0.16$~dex (the value used to simulate the data), with a dispersion of \sigpr=0.02~dex. The results are shown in the bottom panel of Fig.~\ref{fig:type1_lkl}. The posterior distribution of [Fe/H] -- obtained from the analysis -- is close to a Gaussian, and the most probable value of [Fe/H] (i.e. $+0.158$) and the dispersion (CI) [$+0.017$,~$-0.021$] are very similar to those used in the prior. This result points out an important fact, which is that in presence of a prior distribution well peaked at a certain value of [Fe/H], the posterior distribution is strongly influenced by the prior. This is clear if one looks at the posterior distribution of [Fe/H] (also of \dydz) when no prior is adopted (top left panel of Fig.~\ref{fig:type1_lkl}). The distribution is very broad and the inclusion of a Gaussian prior overcomes any information available from the original distribution. The recovered [Fe/H] value is basically defined by the prior.  

Also the \dydz{} distribution looks better when the Gaussian prior on [Fe/H] is adopted. The distribution of \dydz{} shows a clear peak (at variance with the result for the flat-prior case). The most probable value of \dydz{} is 1.99 (basically equal to the true one), with a narrow CI (i.e. between 1.69 and 2.17). Concerning the mixing length parameter,  the distribution of \ml{} is almost independent of the prior on [Fe/H], and the most probable value of \ml{} is equal to the simulated one with a very small CI, namely [1.97,~2.05]. Such a good accuracy on \dydz{} is a consequence of the prior effect on [Fe/H] distribution: indeed, given the degeneracy [Fe/H]-\dydz{} in isochrones, once [Fe/H] is derived with a good accuracy, also \dydz{} can be obtained with a relative high accuracy.

As a final comment, this first data set clearly shows that only the mixing length parameter can be recovered (independently) without any additional information (i.e. prior) from this set of data. The other quantities, [Fe/H] and \dydz{} are badly recovered if no prior on [Fe/H] is specified. On the other hand, the inclusion of a Gaussian prior allows to derive values for the [Fe/H] prior, \dydz{} close to that used to generate the data set. However, the values of the parameters and their CI might be affected by the adopted prior on [Fe/H], that is, \fehpr{} and \sigpr. The dependence of the results on the adopted [Fe/H] value in the prior must be taken into account when the method is applied to the real data, as we will discuss it in the next sections.

\subsection{TYPE\_2 data set: Hyades-like data set with artificially reduced systematic parallax errors}
As a next step, we checked whether the results are affected or not by a realistic future reduction of the observational uncertainties, in particular we want to analyse the effect of a possible reduction of the uncertainty in the parallax. We recall that the assumed uncertainty on the $B_P$ mag in the TYPE\_1 set was dominated by the uncertainty in the distance, a systematic uncertainty of 0.1~mas. Such an uncertainty is expected to be reduced at the end of the {\it Gaia} mission, so it is worth to evaluate the impact of it on the derived parameters. To do this we generated another sample of synthetic data, with the same characteristics of the TYPE\_1 data set,  but with a uncertainty on the $B_P$ mag reduced by a factor of 2, that is, error$(B_P)=0.005~$mag (TYPE\_2). We  applied the method to this data set and we found that the reduction of the distance uncertainty does not produce any appreciable effect on the recovered parameters or on the derived CI. 

\subsection{TYPE\_3 data set: extended Hyades-like main-sequence data set}
We performed another test in which we increased the extension of the observed sequence used to compare models and data, TYPE\_3. In this case, we included faint stars in our sample, down to $M\sim 0.5~$\msun, which corresponds to stars with $B_P\in$~[3.5,~10]~mag. Having increased the magnitude interval, we also increased the number of simulated stars to 100 (as in the real Hyades cluster for the same magnitude interval). The errors are the same used in TYPE\_1 case (namely err($B_P$)=0.012~mag and err($G-R_P$)=0.002~mag).
\begin{figure*}
\centering
\includegraphics[width=0.32\linewidth]{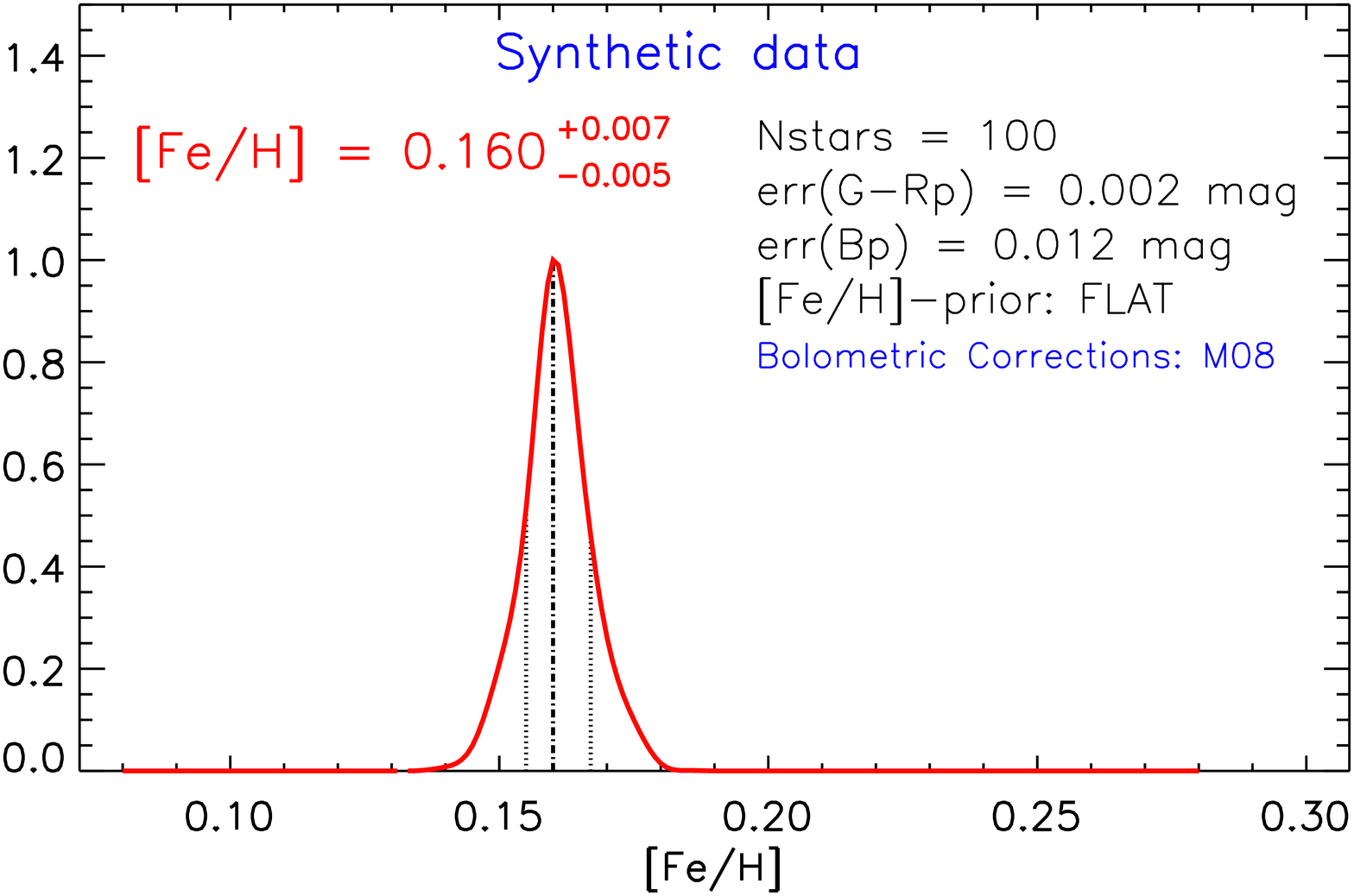}
\includegraphics[width=0.32\linewidth]{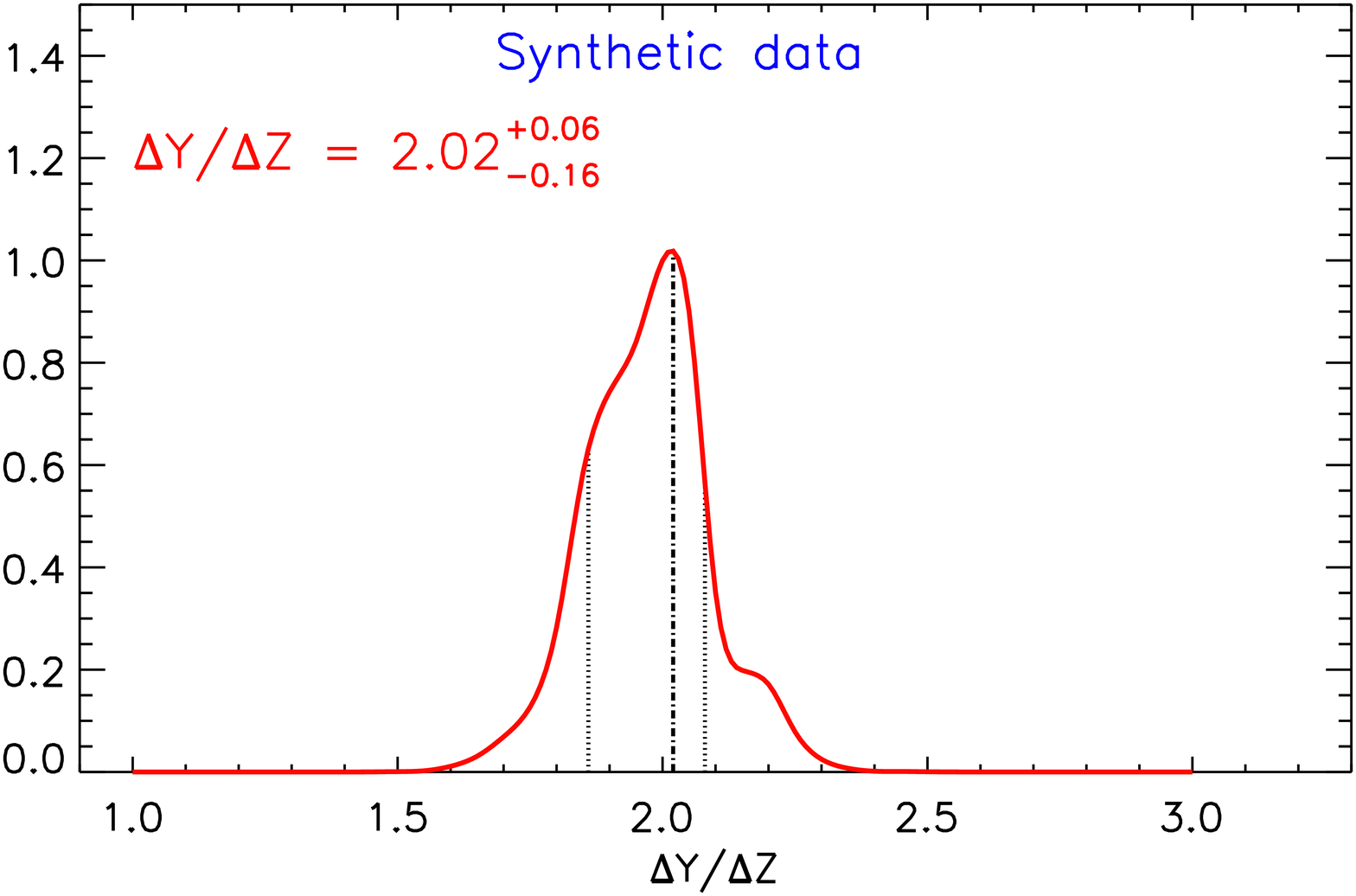}
\includegraphics[width=0.32\linewidth]{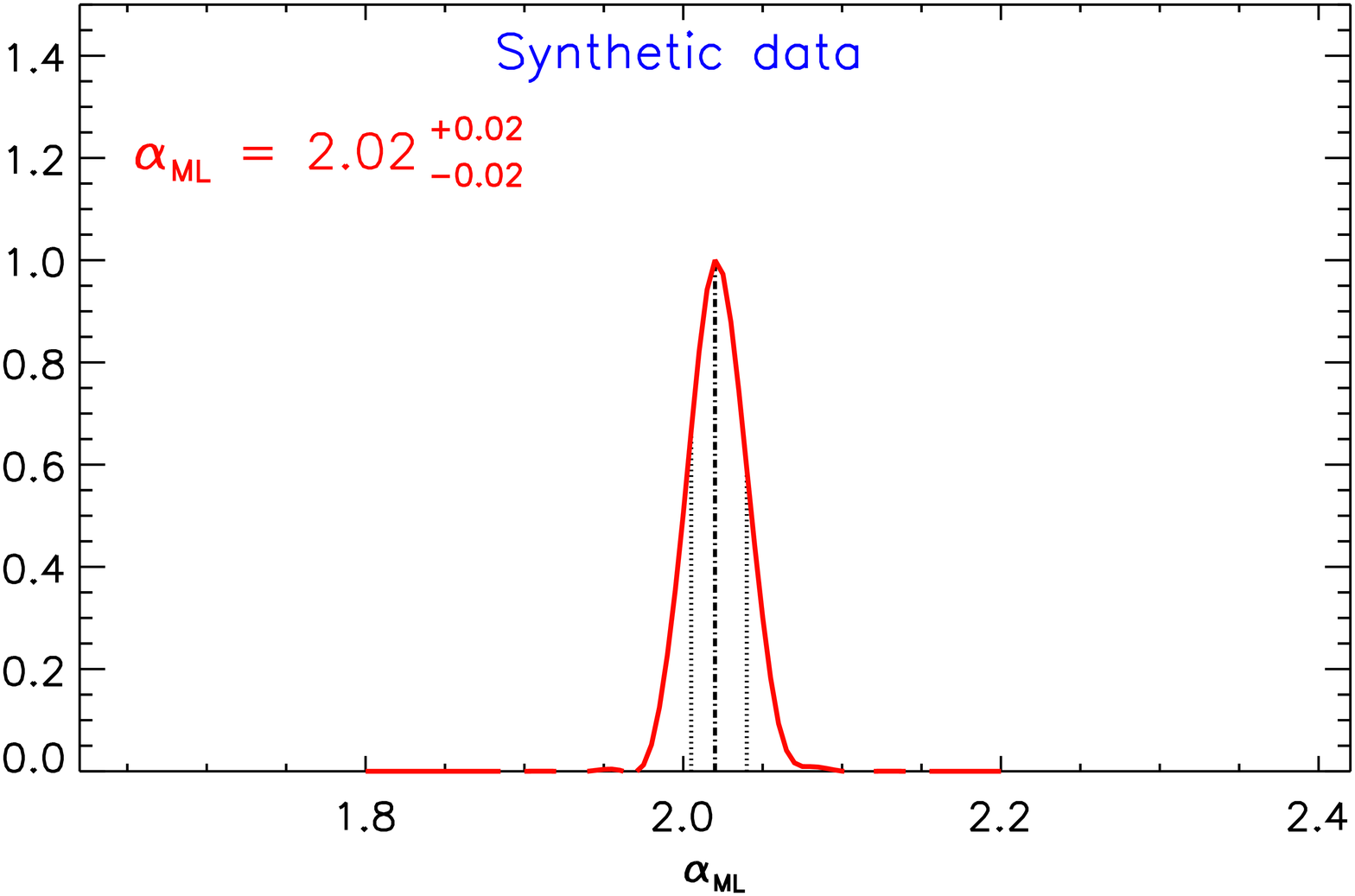}\\
\includegraphics[width=0.32\linewidth]{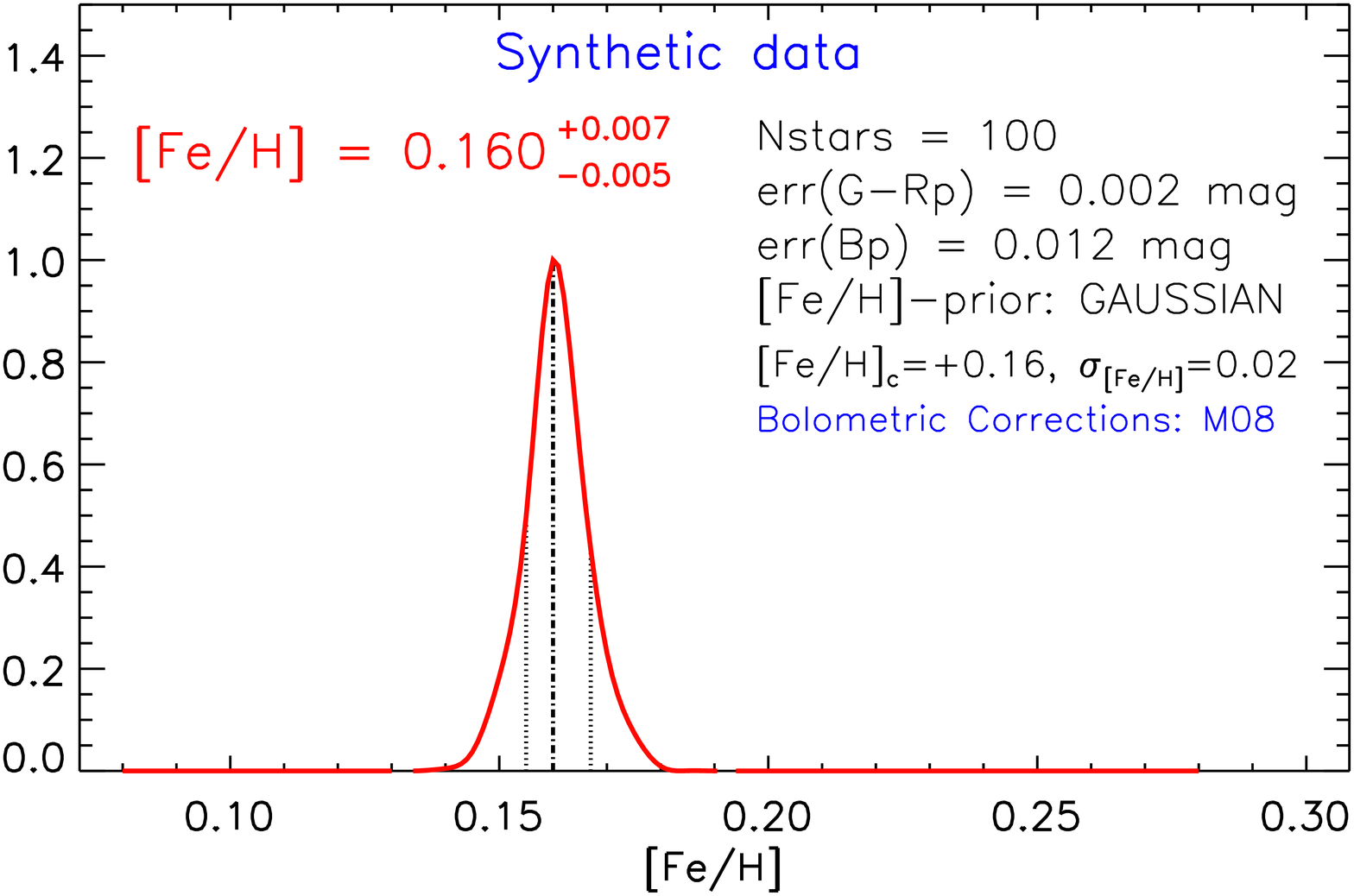}
\includegraphics[width=0.32\linewidth]{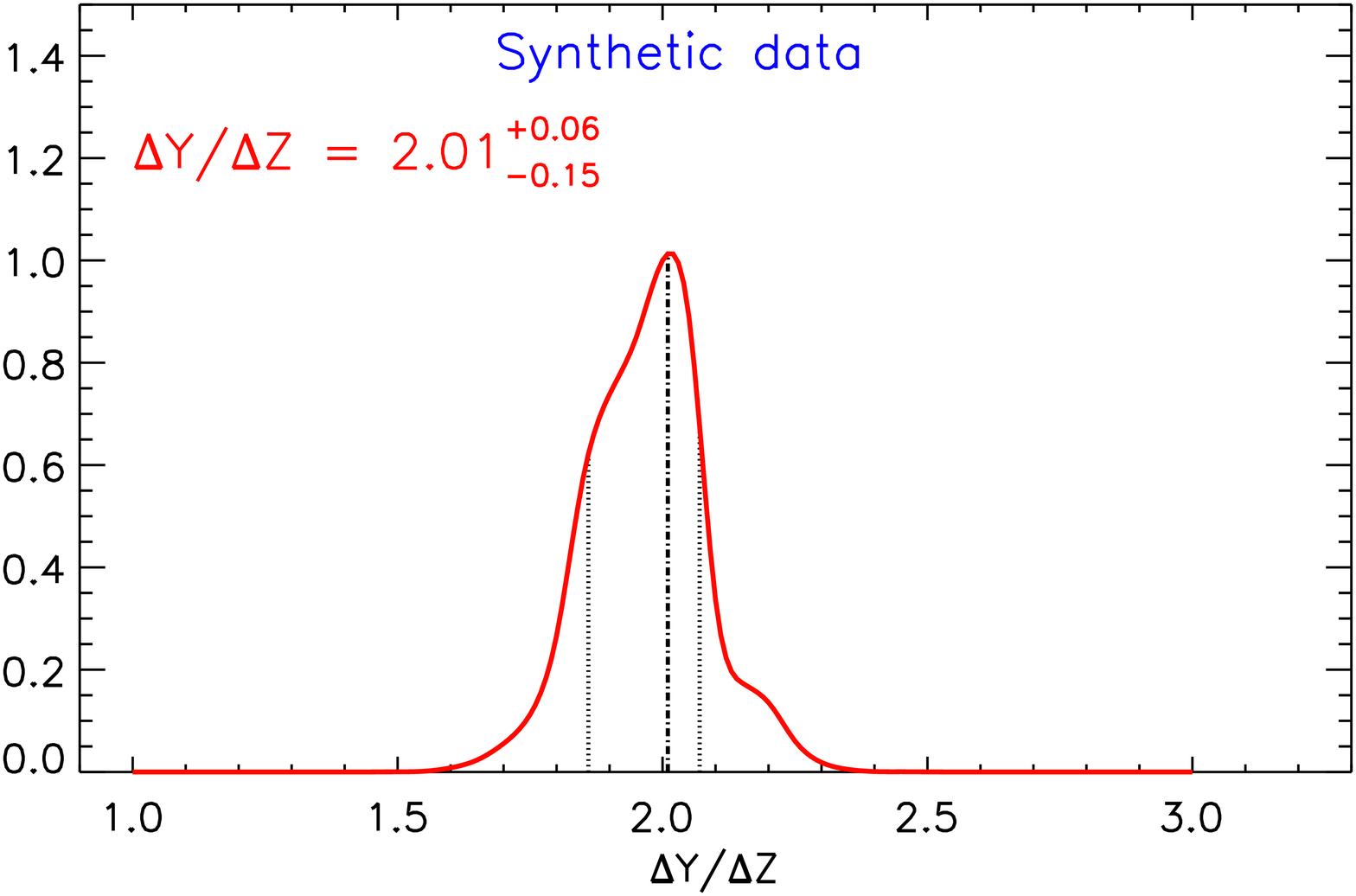}
\includegraphics[width=0.32\linewidth]{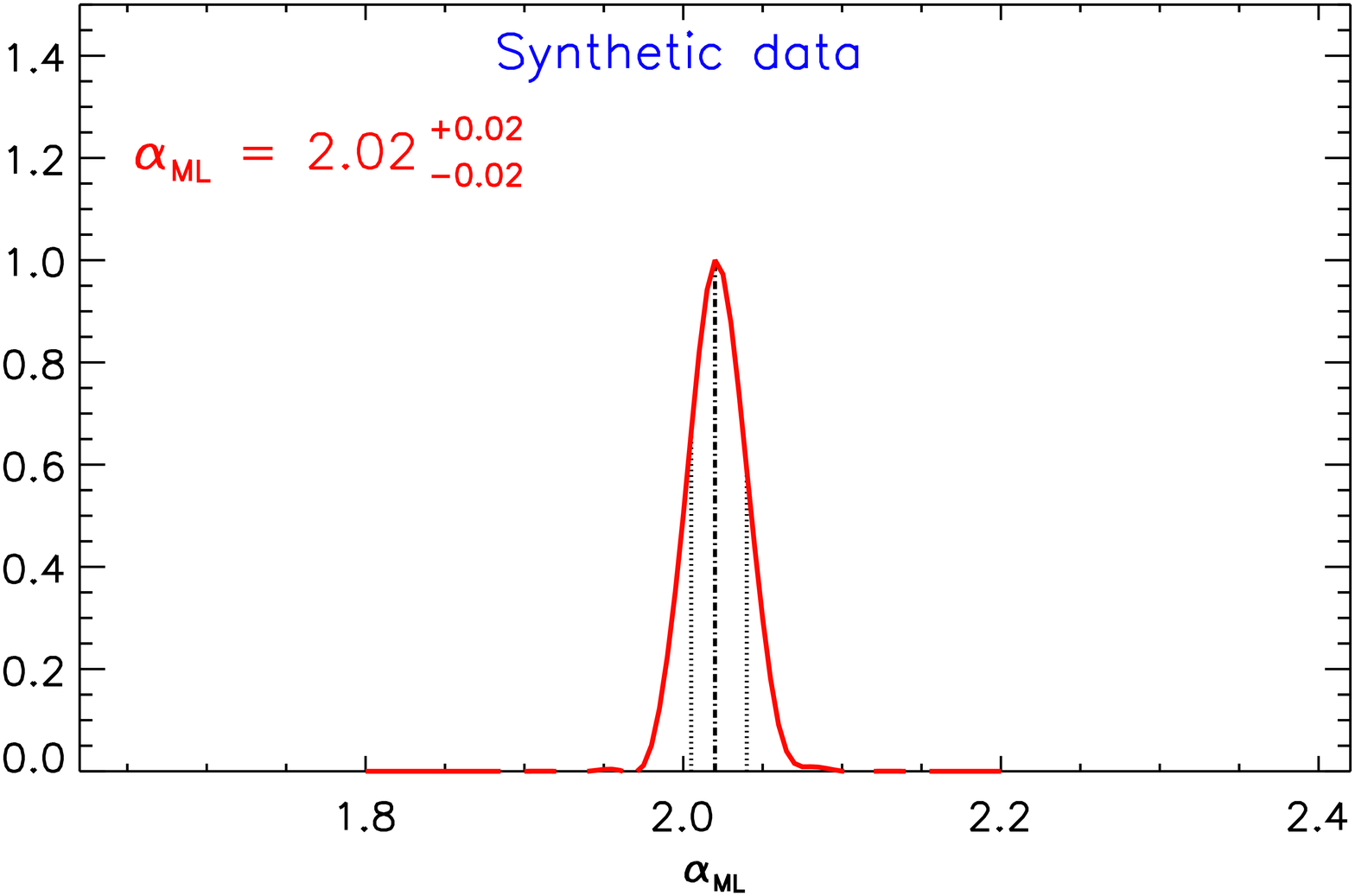}\\
\caption{Posterior marginalized distributions as in Fig.~\ref{fig:type1_lkl}, but for TYPE\_3 data set.}
\label{fig:type3_lkl}
\end{figure*}

The use of a more extended sequence leads to the best results, both with and without a Gaussian prior on the metallicity. The reason for such an improvement is double. On one hand, the inclusion of fainter stars (i.e. $B_P\ga 9.5$-10~mag) partially removes the degeneracy [Fe/H]-\dydz, as we discussed in Section~\ref{sec:qual_test}. Such stars are affected by the helium and [Fe/H] variation in a different way with respect to brighter ones. The same couple of [Fe/H] and \dydz{} that produces (almost) overlapping MS for brighter stars does not lead to the same overlap for fainter stars. This gives the possibility to obtain more precise results. On the other hand, having extended the sequence (i.e. the magnitude interval) we have more stars, which in our case is twice those in TYPE\_1, thus reducing the statistical fluctuations.

Figure~\ref{fig:type3_lkl} shows the result of the recovery for TYPE\_3 data set for the two choices on the prior on the metallicity, namely flat (top panels) and Gaussian centred in \fehpr=$+0.16$~dex with \sigpr=0.02~dex (bottom panels). The most interesting difference with respect to TYPE\_1 data set is achieved when a flat prior on [Fe/H] is adopted. Even in this case, the most probable values of [Fe/H], \dydz, and \ml{} are extremely close to the true ones, that is, those used to generate the synthetic data sample, and the corresponding CI obtained in the TYPE\_3 flat prior case are very small if compared to those obtained in TYPE\_1 flat prior test. It is also interesting to notice that the quality of the results is not improved if the  Gaussian prior on the metallicity is adopted (bottom panels). This is due to the fact that the selected sequence is  extended enough to clearly disentangle the effect of a variation of [Fe/H], \dydz, and \ml. So, this data set would correspond to an ideal configuration to derive the parameters, in the sense that we would expect the results to be more precise than those obtained in the other cases. Unfortunately, as already discussed, the whole MS $B_P \in  [3.5,\,10]$~mag,  cannot be used for comparison with real data, due to the large current disagreement between predictions and observations for $B_P \la 6.5$-7~mag. 

\subsection{Effect of a systematic error in [Fe/H]}
\label{sec:feh}
In the previous section we showed that the estimate of stellar parameters is affected by the functional form, flat or Gaussian, of the assumed prior on [Fe/H] even if the prior distribution if centred on the true value. It is thus natural to expect a stronger impact on the recovered parameters of the adopted prior on [Fe/H] whenever a systematic bias in its determination is present. When dealing with real stellar cluster rather than synthetic ones, it is common to have quite different spectroscopic estimates. This section is then devoted to analyse the effect of systematic bias in the adopted value of \fehpr{} and \sigpr{} on the  estimate of stellar parameters. For what concerns real data, the value of [Fe/H] measured by several authors for the same cluster might vary more than the quoted uncertainties (because of the use of a sample of different stars, or a different analysis technique, or the adoption of different calibration for the [Fe/H] scale). If this is the case, it is not so straightforward to choose the best value of [Fe/H] to be used in the prior and its dispersion. Thus, one should be aware that \fehpr{} and \sigpr{} can alter the estimate of the parameters at some level that we want to quantify.

\begin{figure*}
\centering
\includegraphics[width=0.33\linewidth]{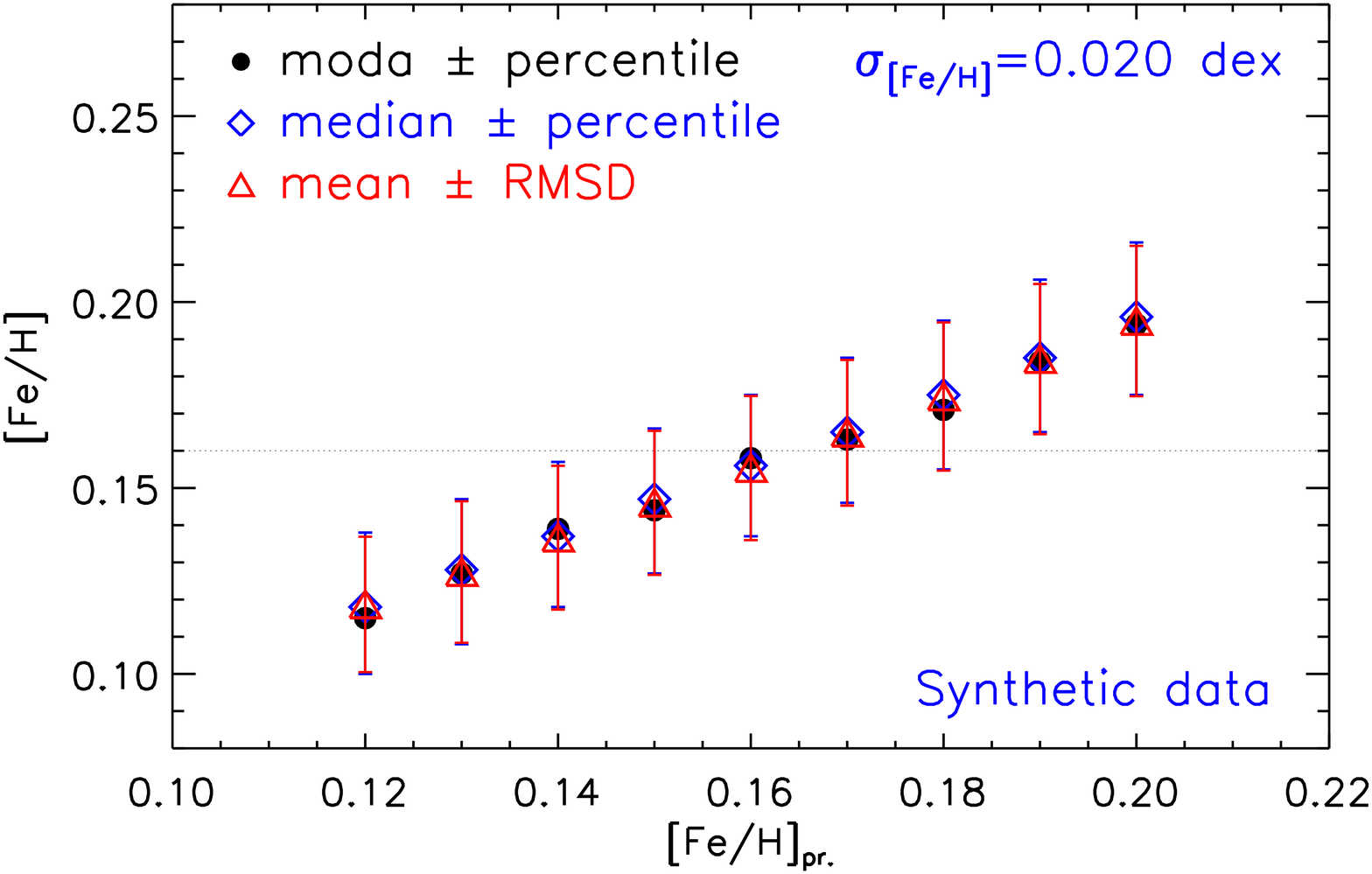}
\includegraphics[width=0.33\linewidth]{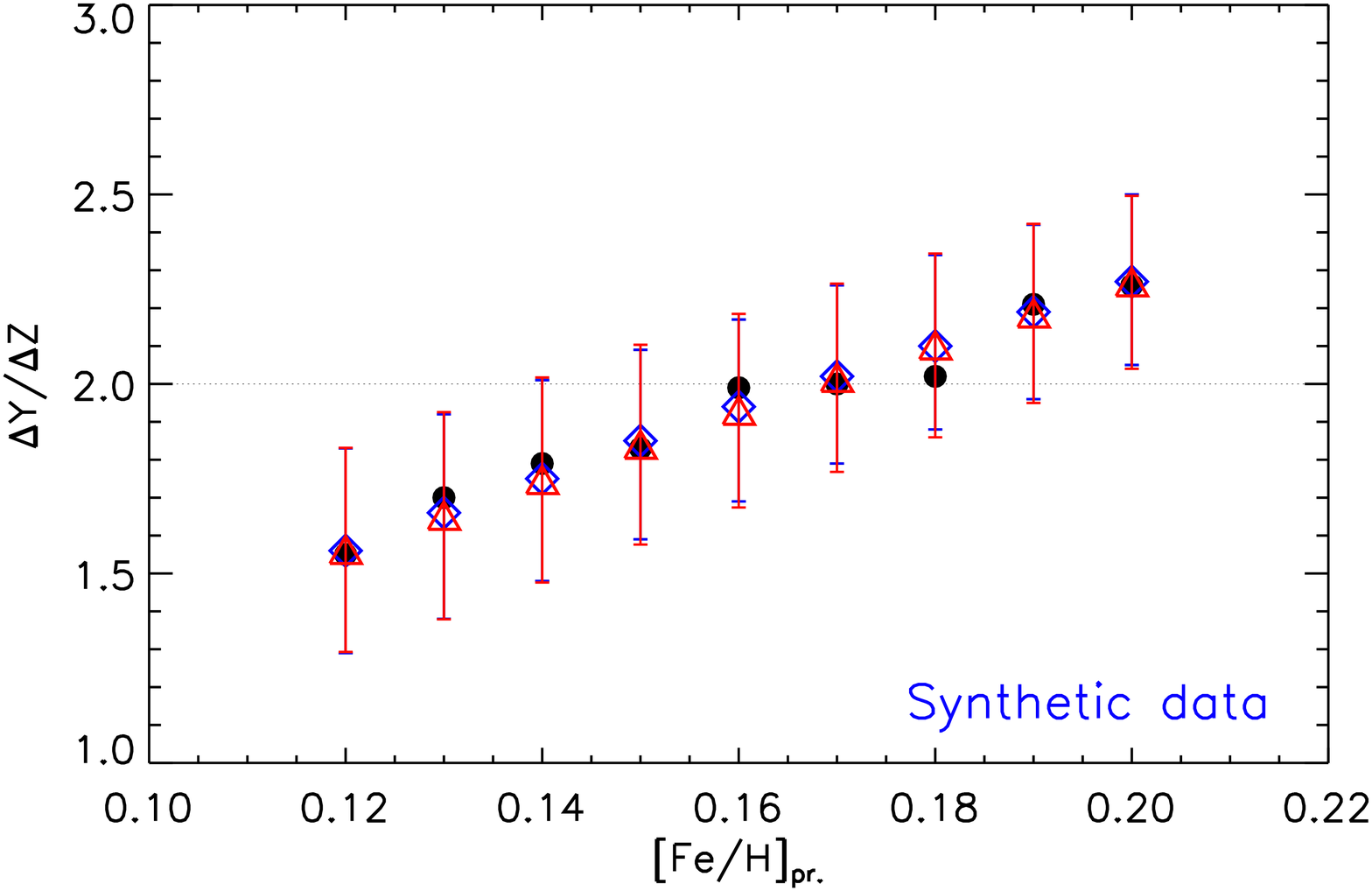}
\includegraphics[width=0.33\linewidth]{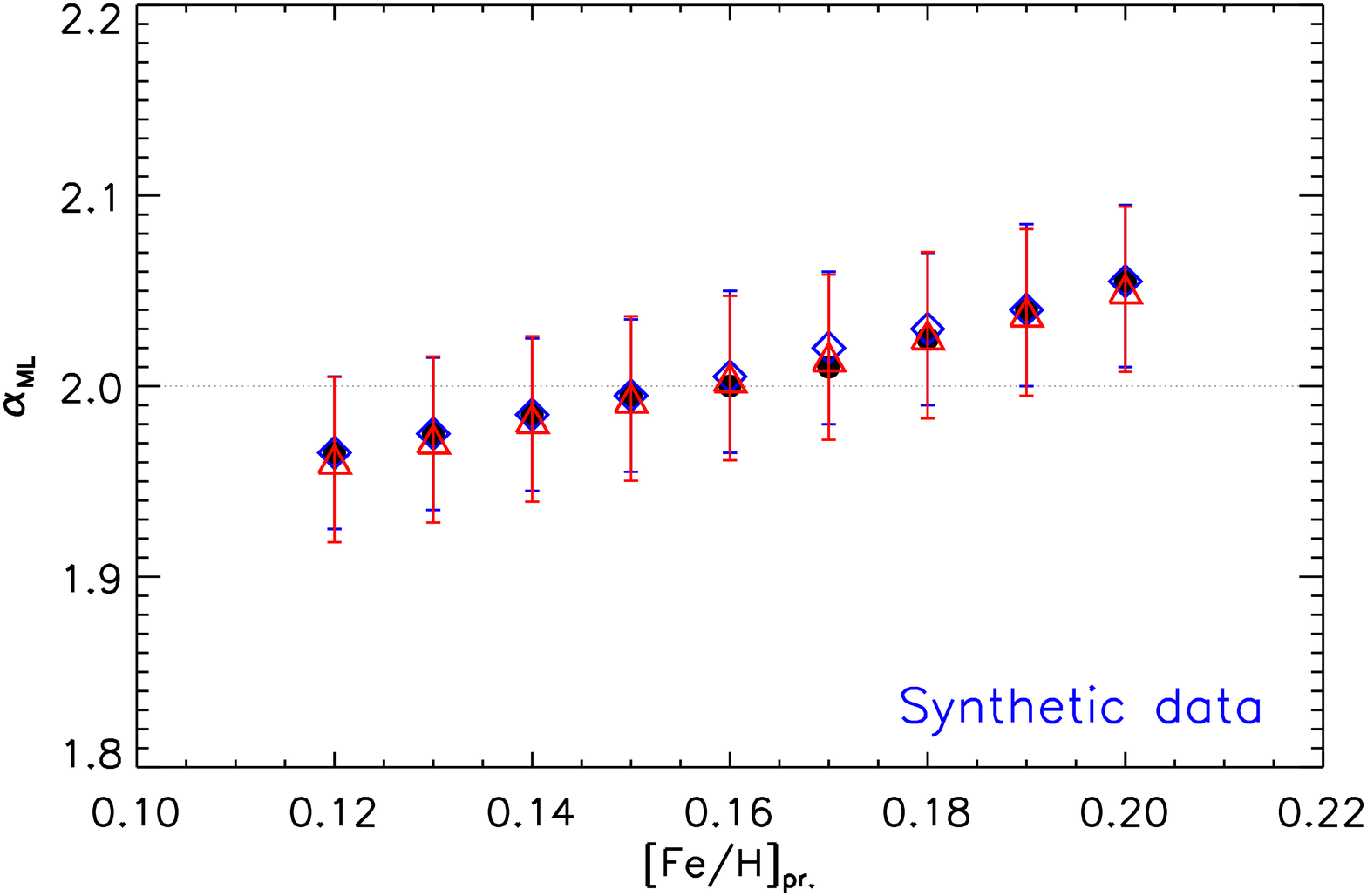}\\
\includegraphics[width=0.33\linewidth]{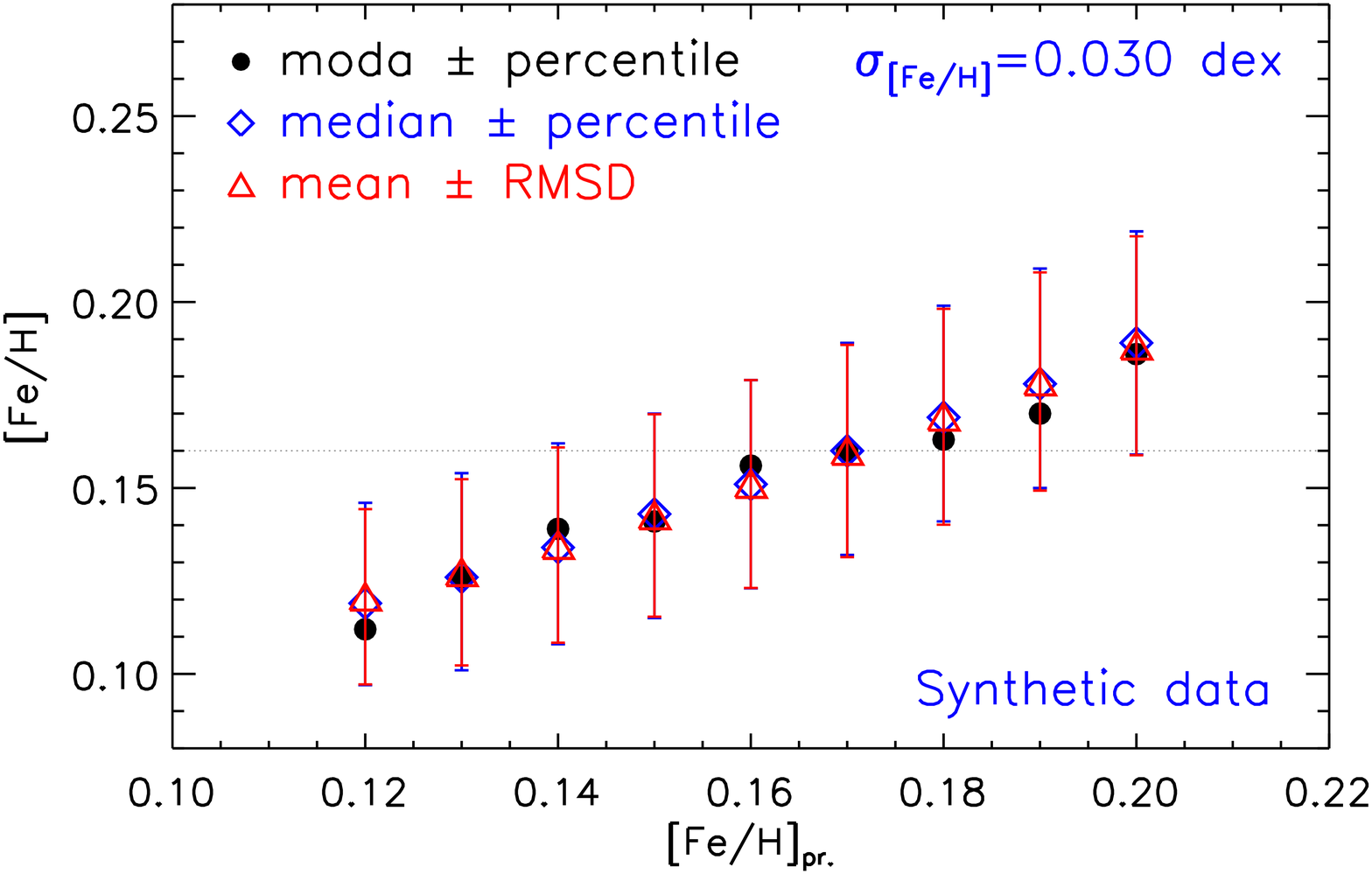}
\includegraphics[width=0.33\linewidth]{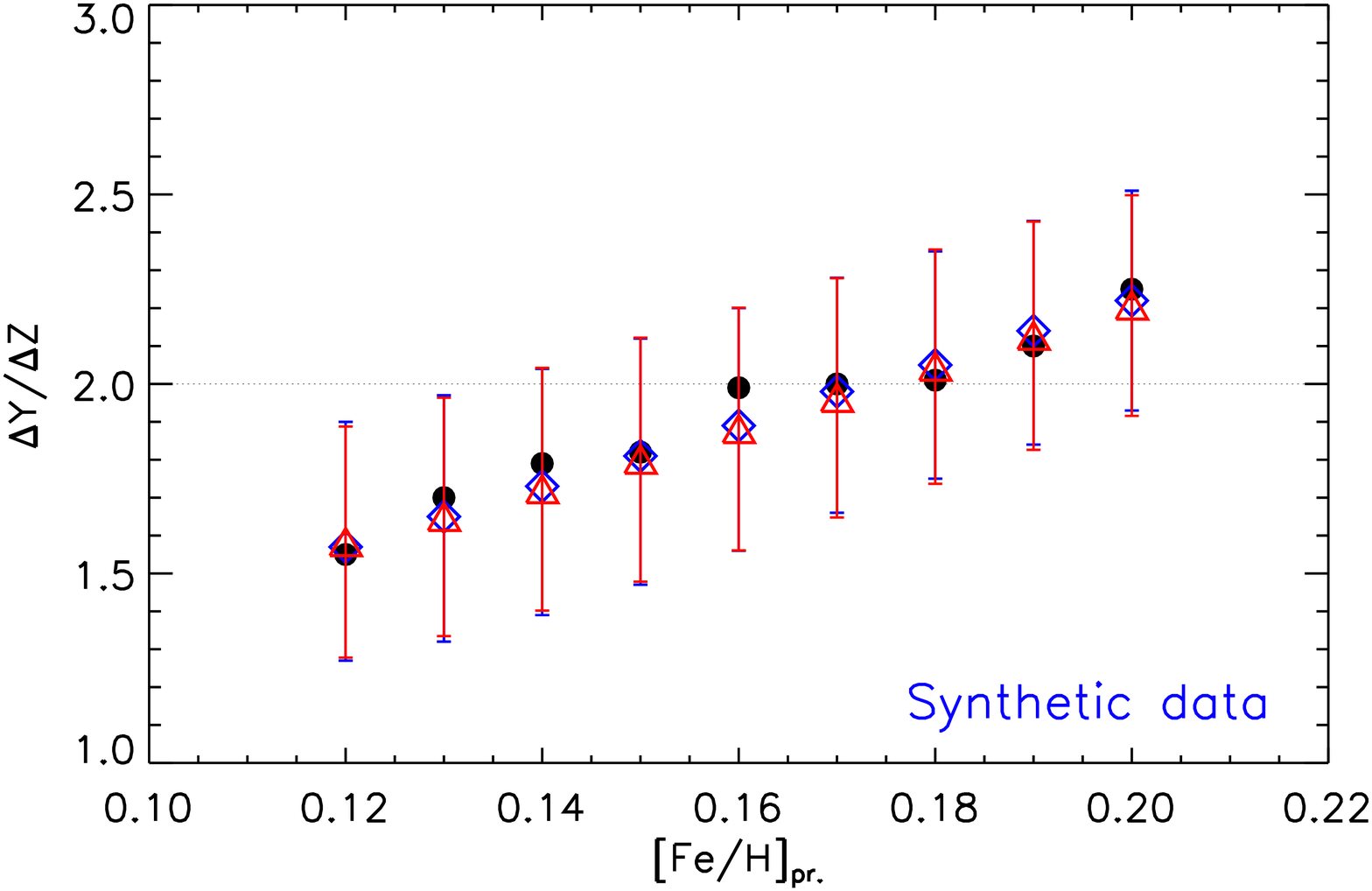}
\includegraphics[width=0.33\linewidth]{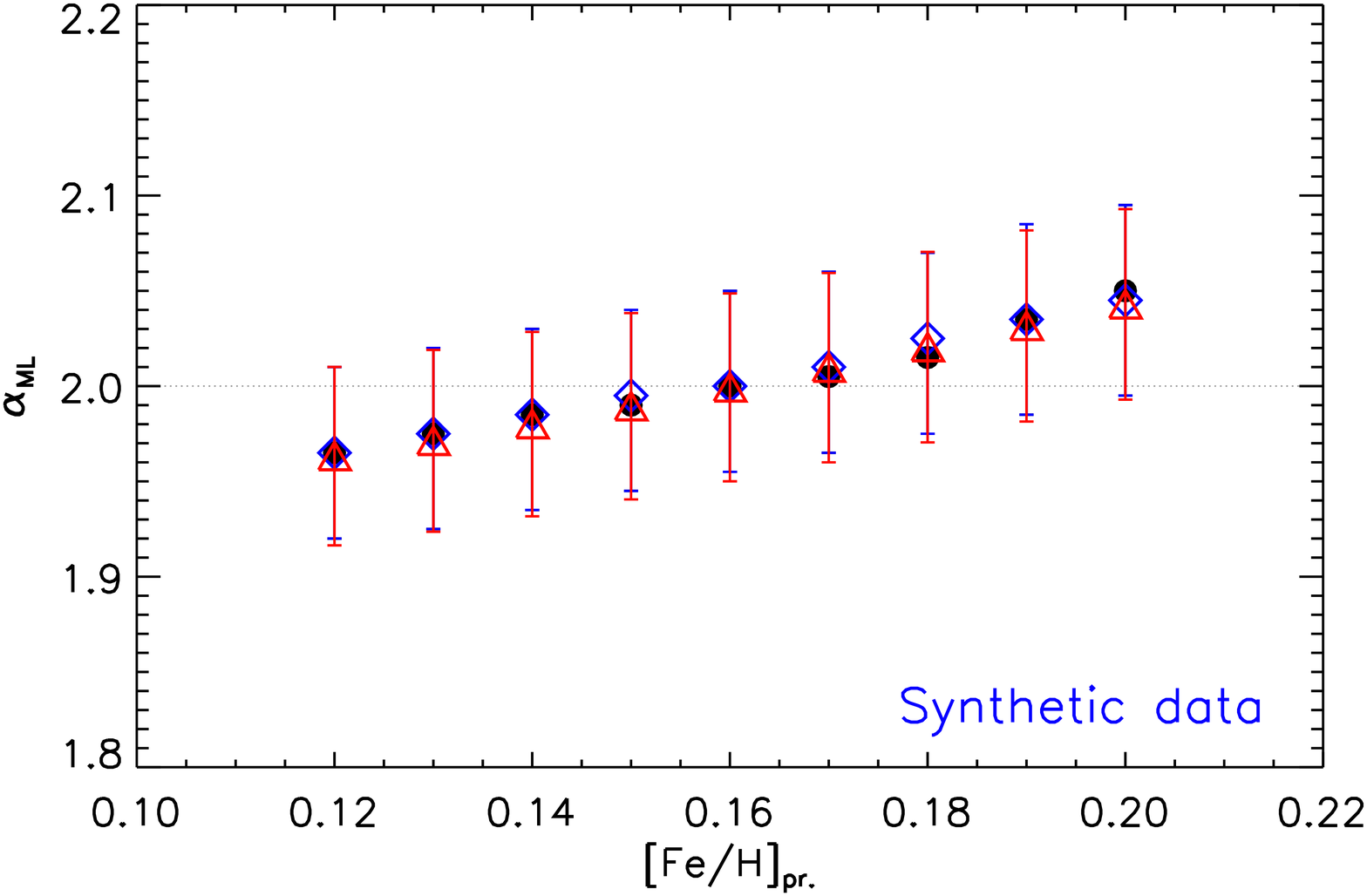}\\
\includegraphics[width=0.33\linewidth]{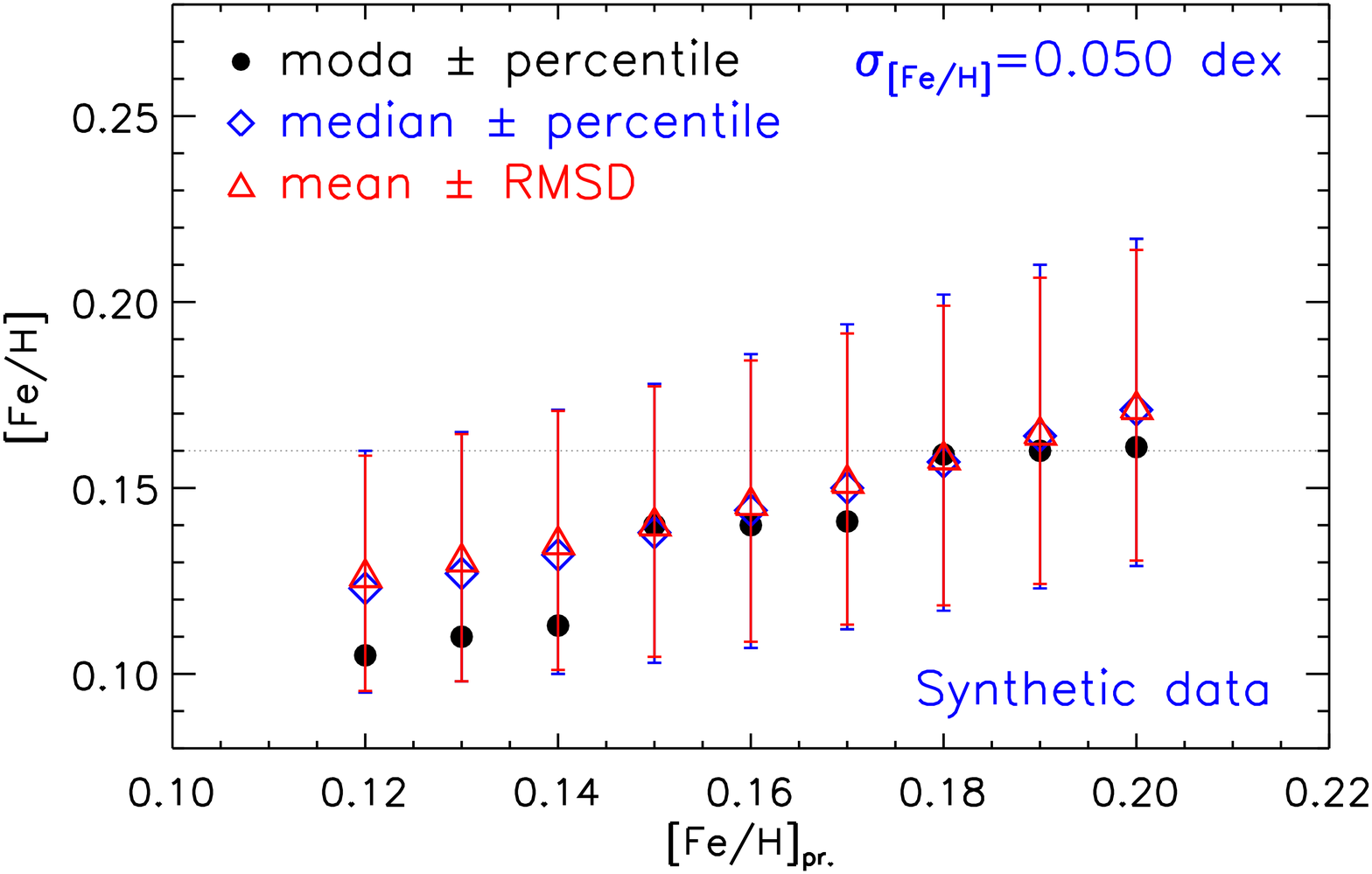}
\includegraphics[width=0.33\linewidth]{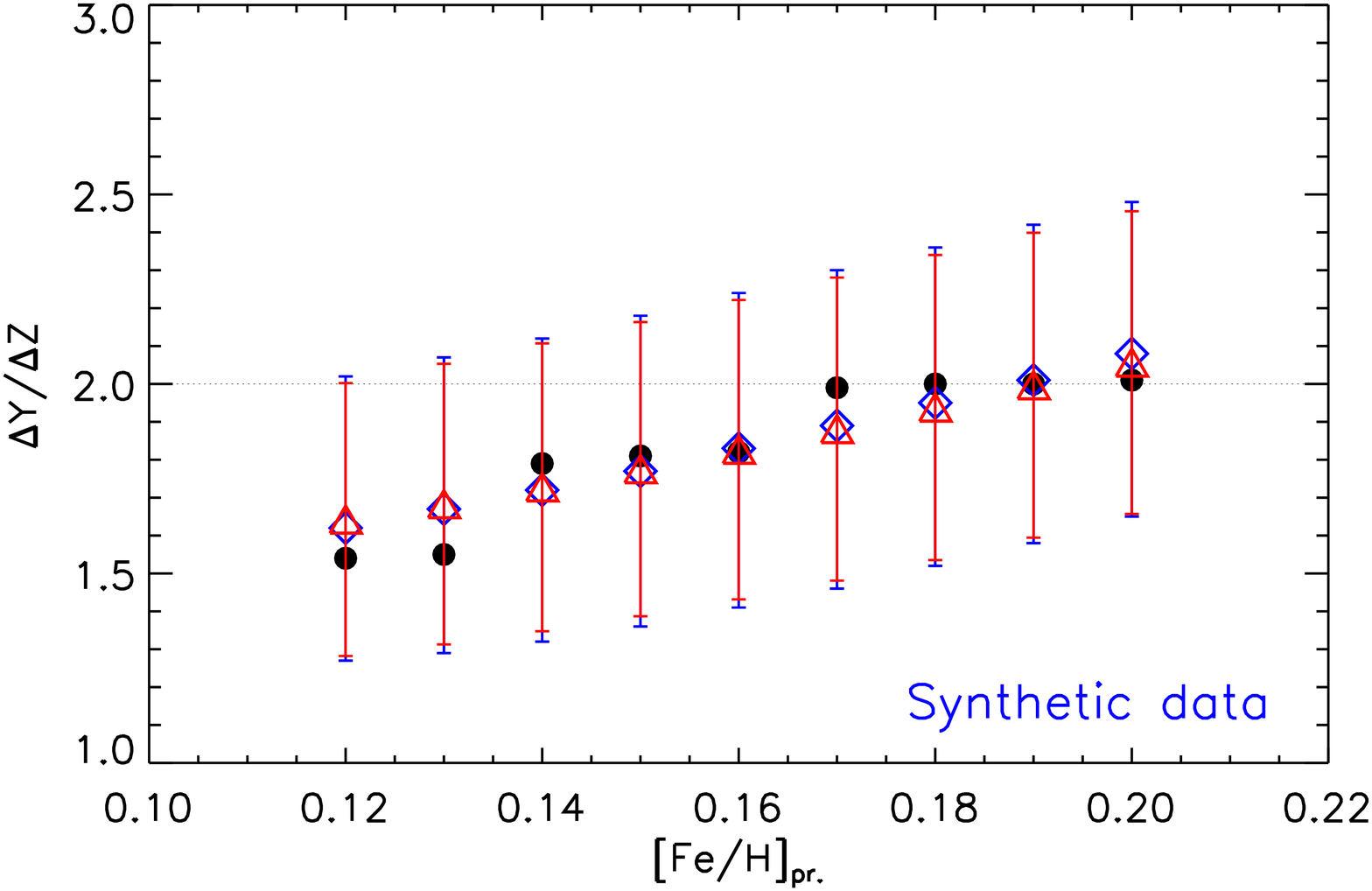}
\includegraphics[width=0.33\linewidth]{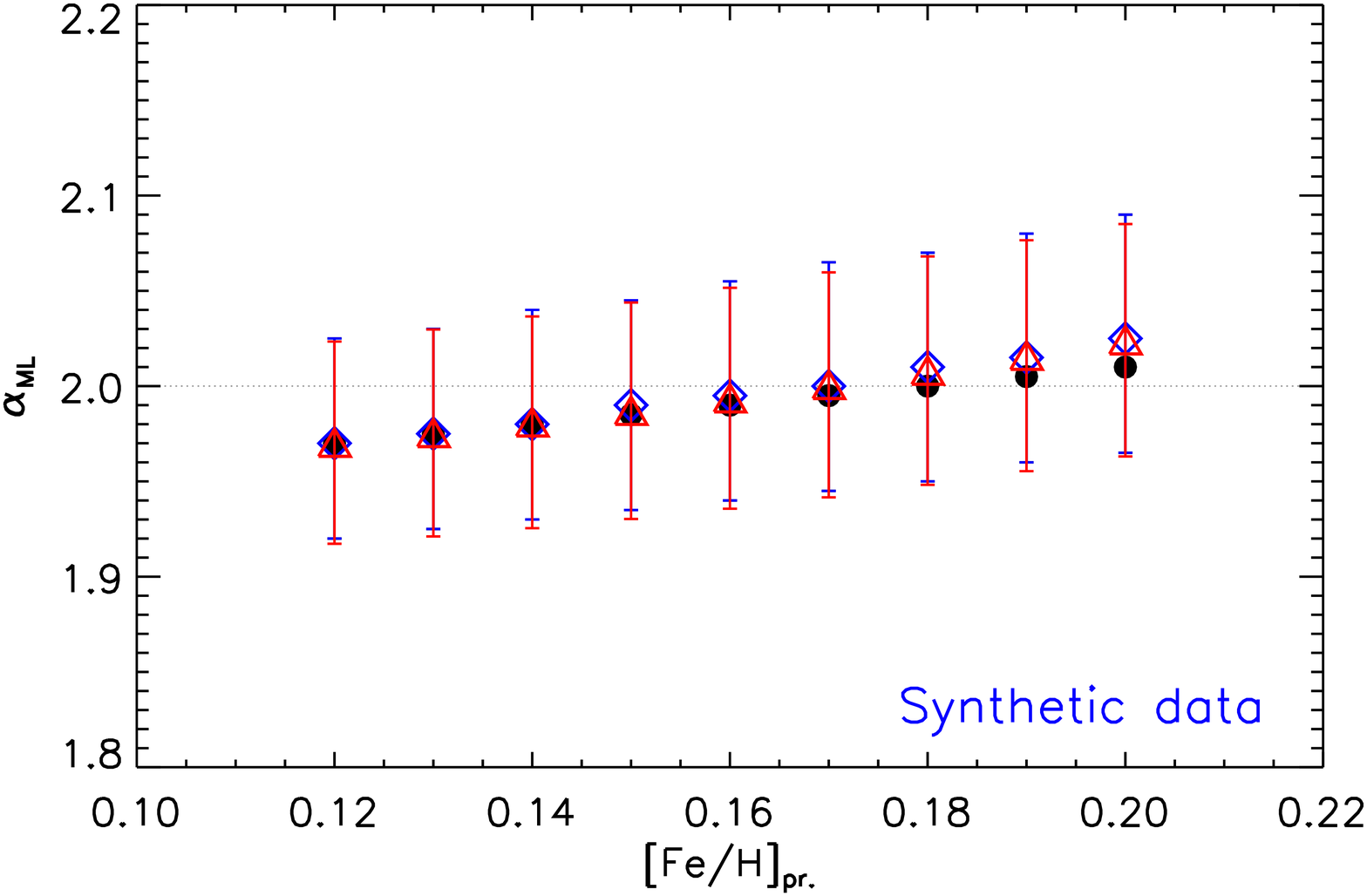}
\caption{Most probable value and corresponding CI for [Fe/H] (first column), \dydz{} (second column), and \ml{} (third column) for TYPE\_1 data set, as a function of \fehpr{} and for three values of \sigpr=0.02, 0.03, and 0.05~dex (first, second and third rows, respectively). The best value and the relative CI obtained using the median and the mean (see the text) are also shown.}
\label{fig:res_feh_type1}
\end{figure*}

To do this, we run the recovery using a grid of \fehpr, from +0.12 to +0.20 in step of +0.01, to cover a plausible range of \fehpr. Since the prior also depends on the width of the distribution \sigpr, we used for each value of \fehpr{} three values, \sigpr=0.02, 0.03 and 0.05$\equiv \{$\sigpr$_{,j}\}$. We used these values to construct a grid of Gaussian priors each with a couple (\fehpr$_{,j}$, \sigpr$_{,j}$). Then, we run the recovery over the TYPE\_1 data set, and we obtain for each choice of the Gaussian prior (\fehpr$_{,j}$, \sigpr$_{,j}$) the most probable value of [Fe/H], \dydz, and \ml{} with the corresponding CI. 
Figure~\ref{fig:res_feh_type1} shows the results of the recoveries, for a Gaussian prior with a width \sigpr=0.02 (first row), 0.03 (second row), and 0.05 (third row). Figure~\ref{fig:res_feh_type1} also shows the results obtained using different indicators for the most probable value and for the CI. In particular we used: (1) the mode (our reference) to define the most probable value and the 16th and 84th percentile for the CI, (2) the median for the most probable value and the 16th and 84th percentile for CI, and (3) the RMSD.
 
For what concerns [Fe/H] the recovered values of [Fe/H] resembles the value used in the prior depending on the value of \sigpr$_{,j}$. If the range of [Fe/H] values allowed by the prior distribution is well peaked, that is, \sigpr=0.02, then the recovery is forced to use [Fe/H] values close to \fehpr, and the derived [Fe/H] is very close to \fehpr. If the width of the prior is larger (i.e. 0.03 or 0.05), the best value of [Fe/H] can be different from the prior value. From Fig.~\ref{fig:res_feh_type1} (top panel) it is evident that by increasing \sigpr{} the derived CI on [Fe/H] increases too: in general the uncertainty on the derived [Fe/H] (i.e. the CI) is similar to \sigpr. In any case the derived [Fe/H] value is biased by the adoption of the prior.
 
The most probable value of \dydz{} is affected by (\fehpr$_{,j}$, \sigpr$_{,j}$). As a general comment, \dydz{} varies at maximum between 1.5 and 2.3, depending on \fehpr (about 15-25~percent), so it includes the true value (\dydz=2). As expected the CI of the recovered \dydz{} is affected by \sigpr, but not as much as for [Fe/H]:  the largest is \sigpr{} and the broader is \dydz{} CI, namely about 0.3 (about 15~percent) at \sigpr=0.02~dex and 0.4-0.5 (about 20-25~percent) at \sigpr=0.05~dex. This results show that also \dydz{} is affected by a bias if a prior on [Fe/H] is adopted, a bias that can reach 0.3-0.5.

Concerning \ml, as already noted, its value is less affected by the adopted prior on [Fe/H] and in all the cases, the recovered value is very close to the true one, and fully compatible with it within its CI. To be noted that \ml{} varies at maximum of about 2-3~percent considering all the possible values of (\fehpr$_{,j}$, \sigpr$_{,j}$). So, this parameter is recovered with a very high precision and accuracy if compared to \dydz{} and [Fe/H]. 

In any case, we can firmly state that in this particular data set, the adoption of a prior on [Fe/H] necessarily introduces a not negligible bias in all the recovered parameters but \ml.

Figure~\ref{fig:res_feh_type1} also shows that the three different indicators used to specify the most probable value and/or the CI generally agree, or, in any case, the difference among the results is smaller than the quoted CI. Noticeably, the CI obtained using the percentiles or the RMSD are essentially the same in all the cases.
 
\begin{figure*}
\centering
\includegraphics[width=0.33\linewidth]{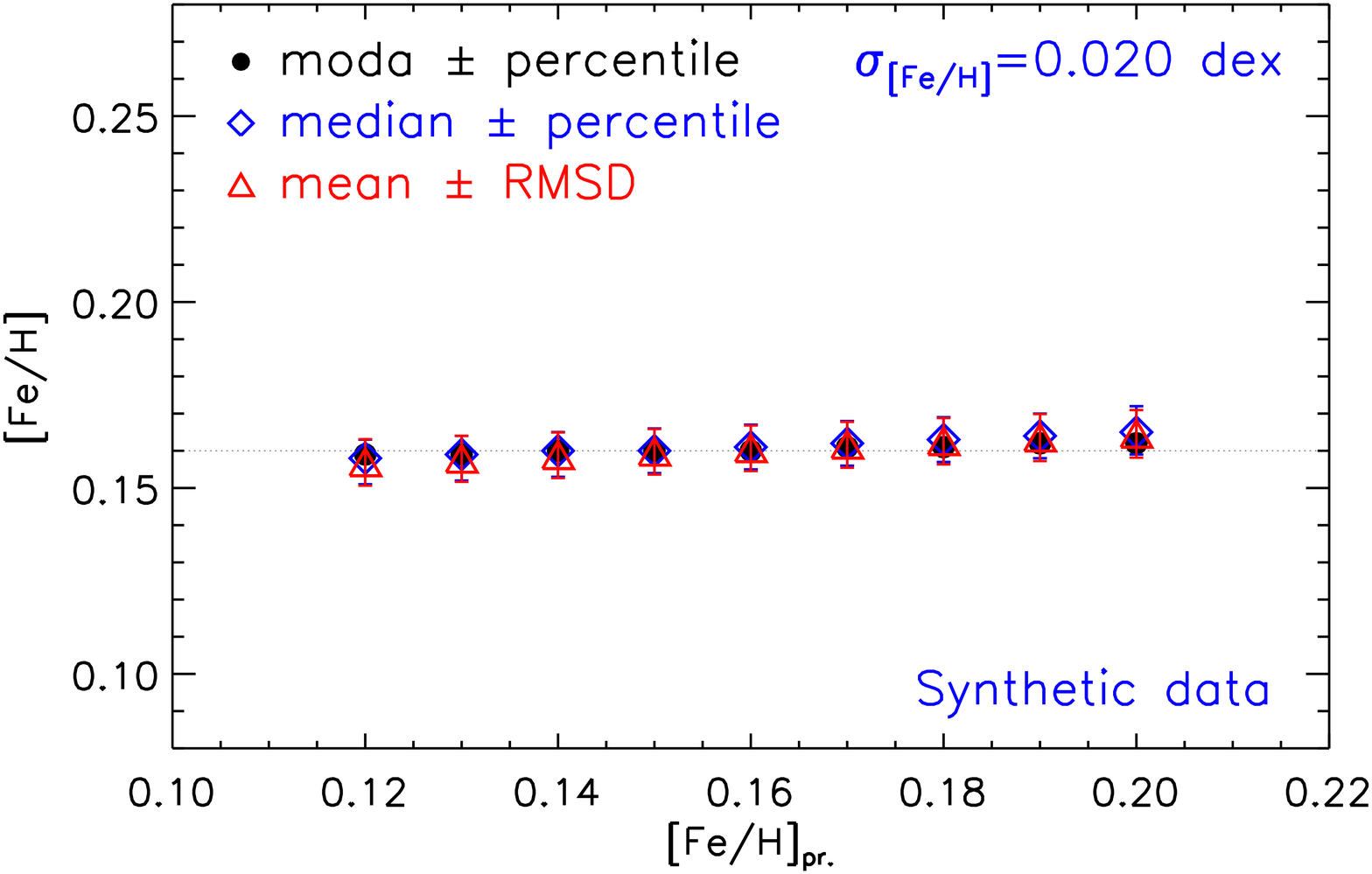}
\includegraphics[width=0.33\linewidth]{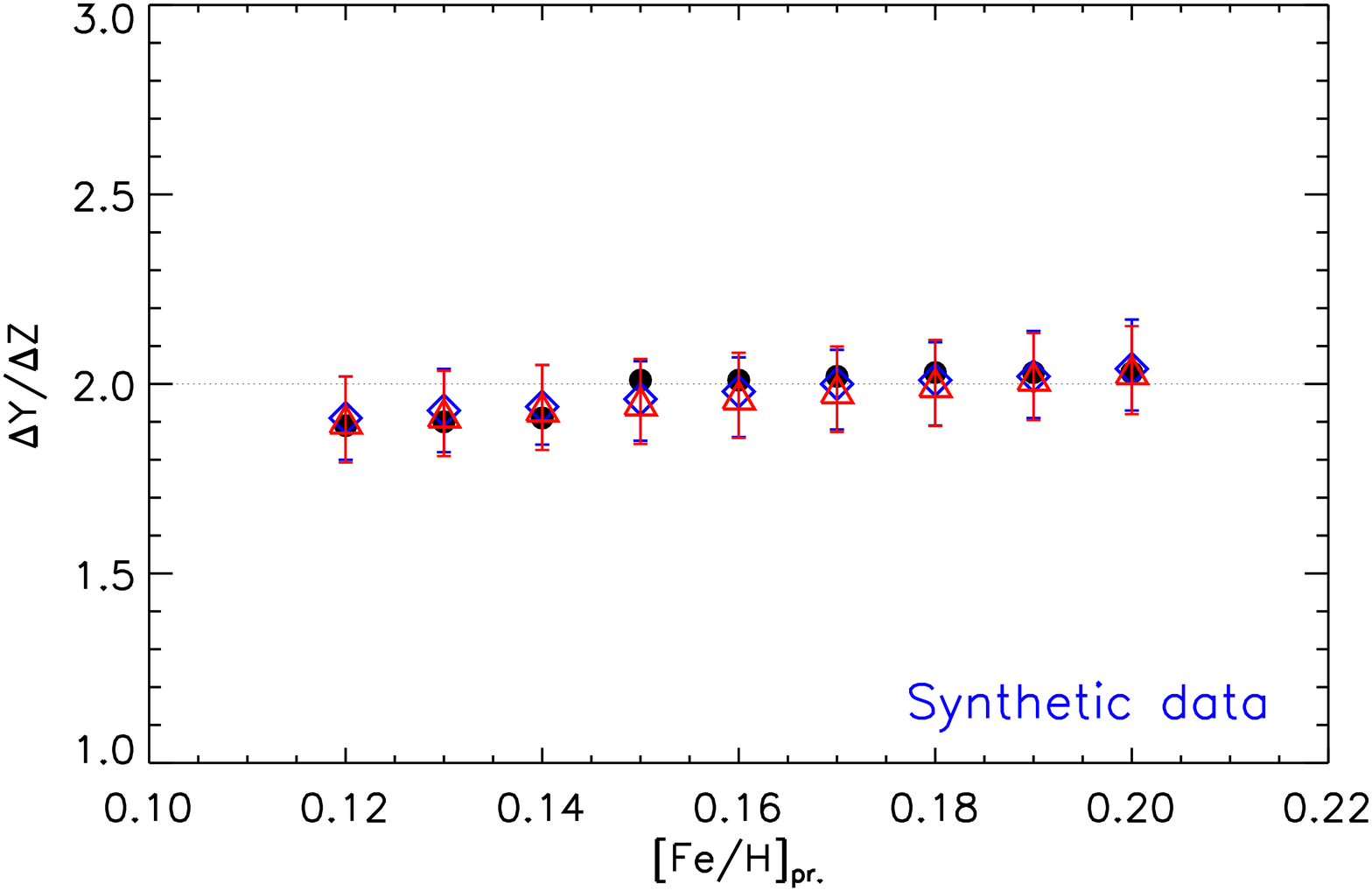}
\includegraphics[width=0.33\linewidth]{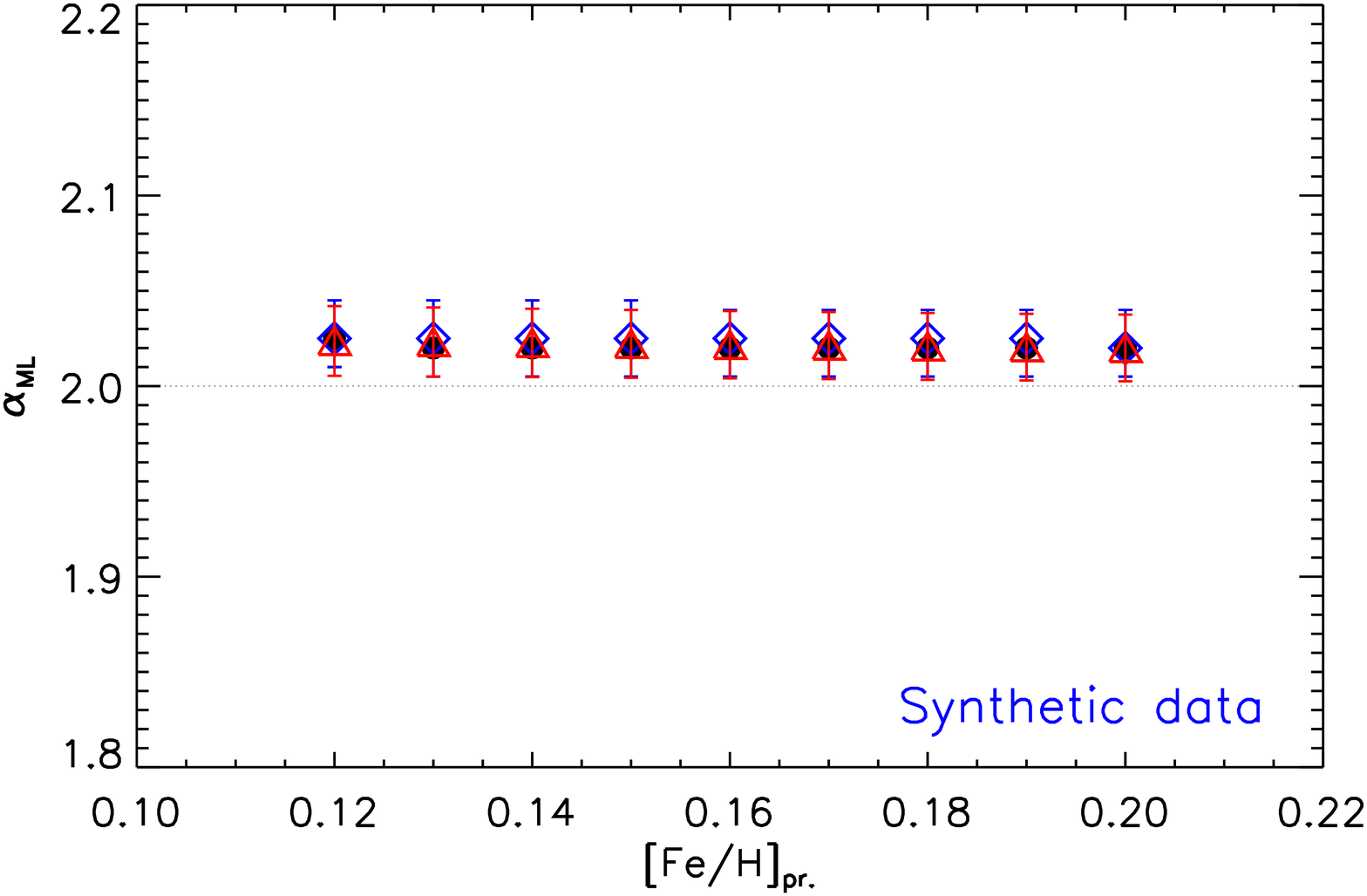}\\
\includegraphics[width=0.33\linewidth]{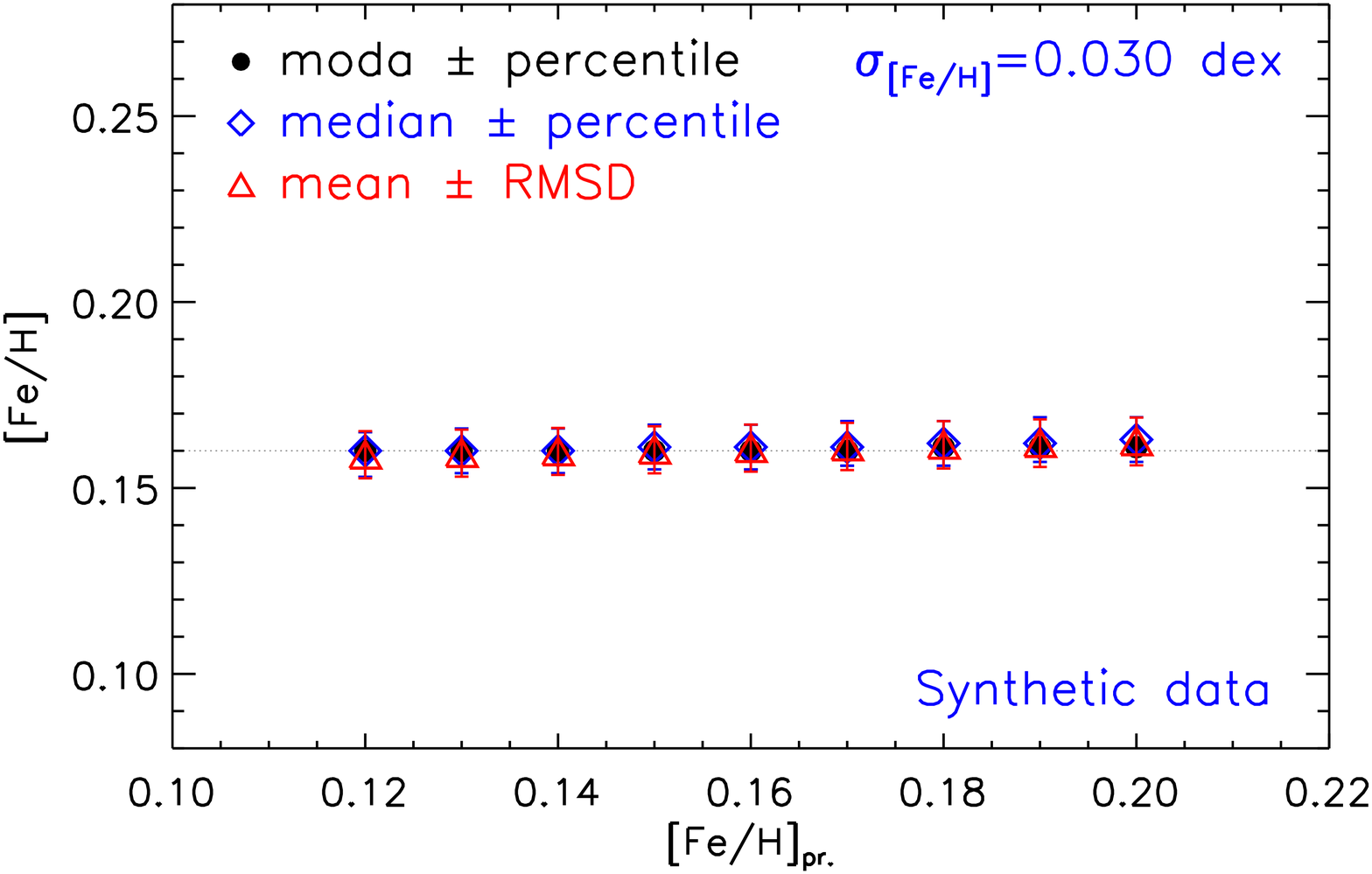}
\includegraphics[width=0.33\linewidth]{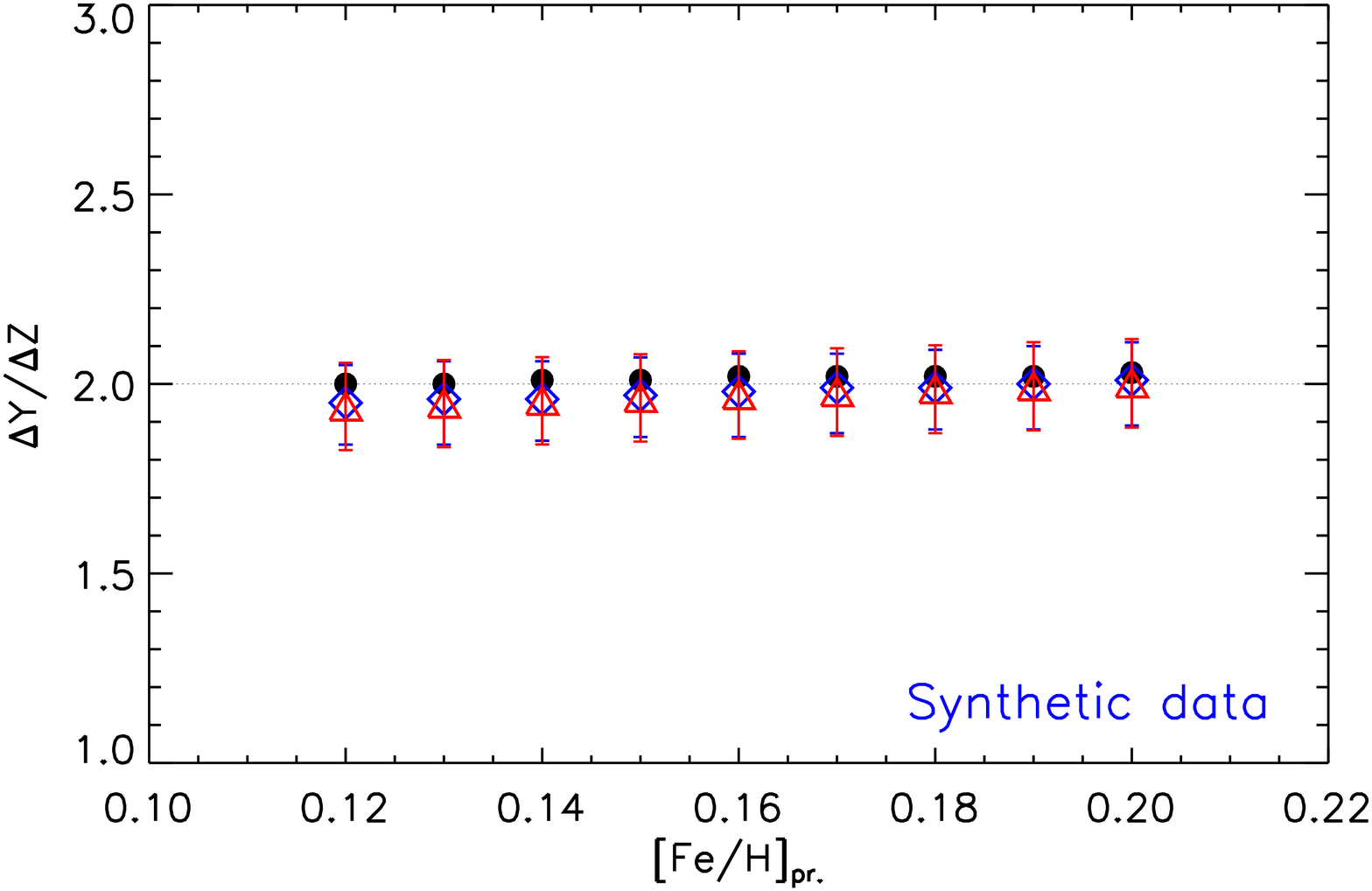}
\includegraphics[width=0.33\linewidth]{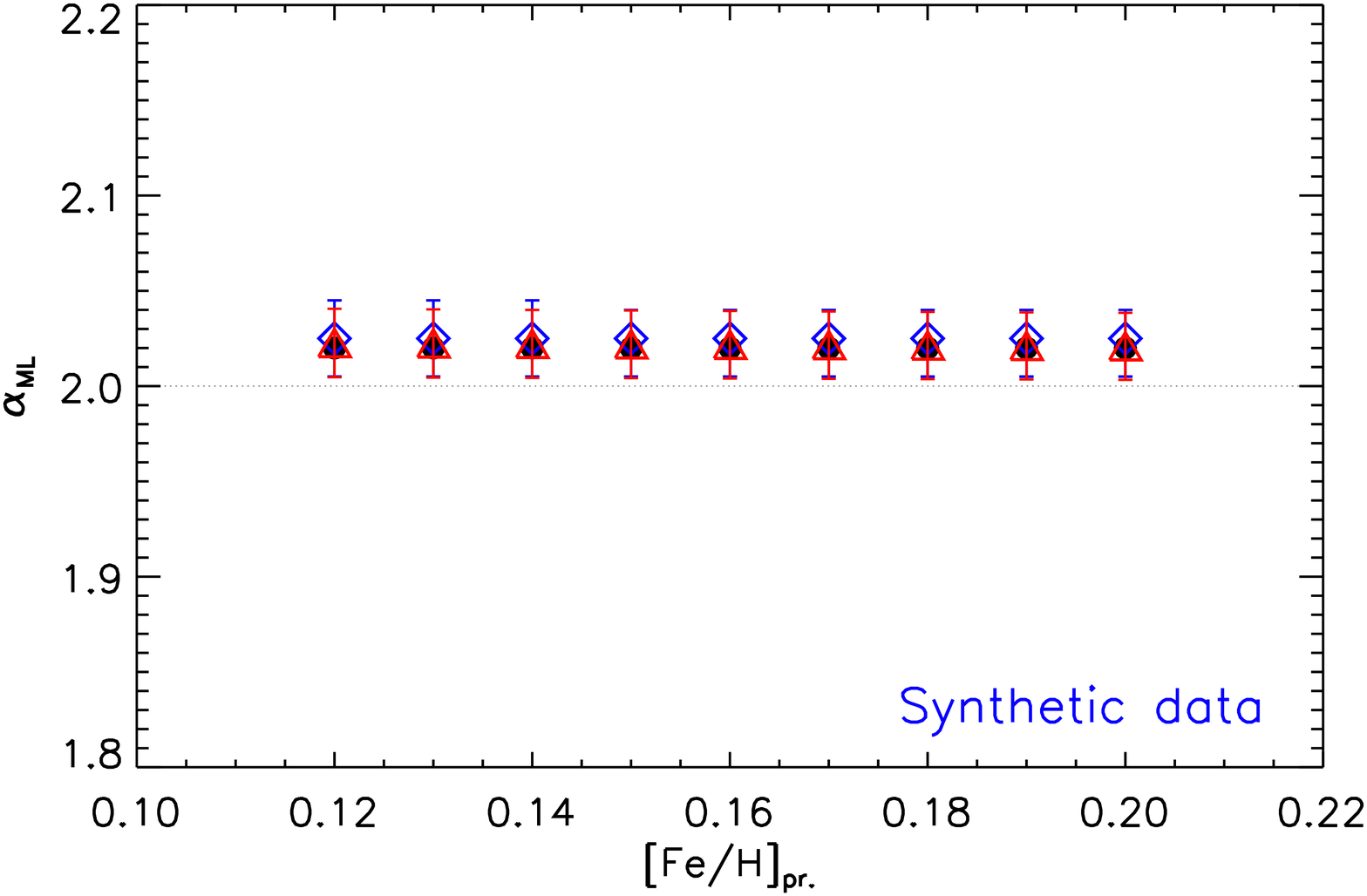}\\
\includegraphics[width=0.33\linewidth]{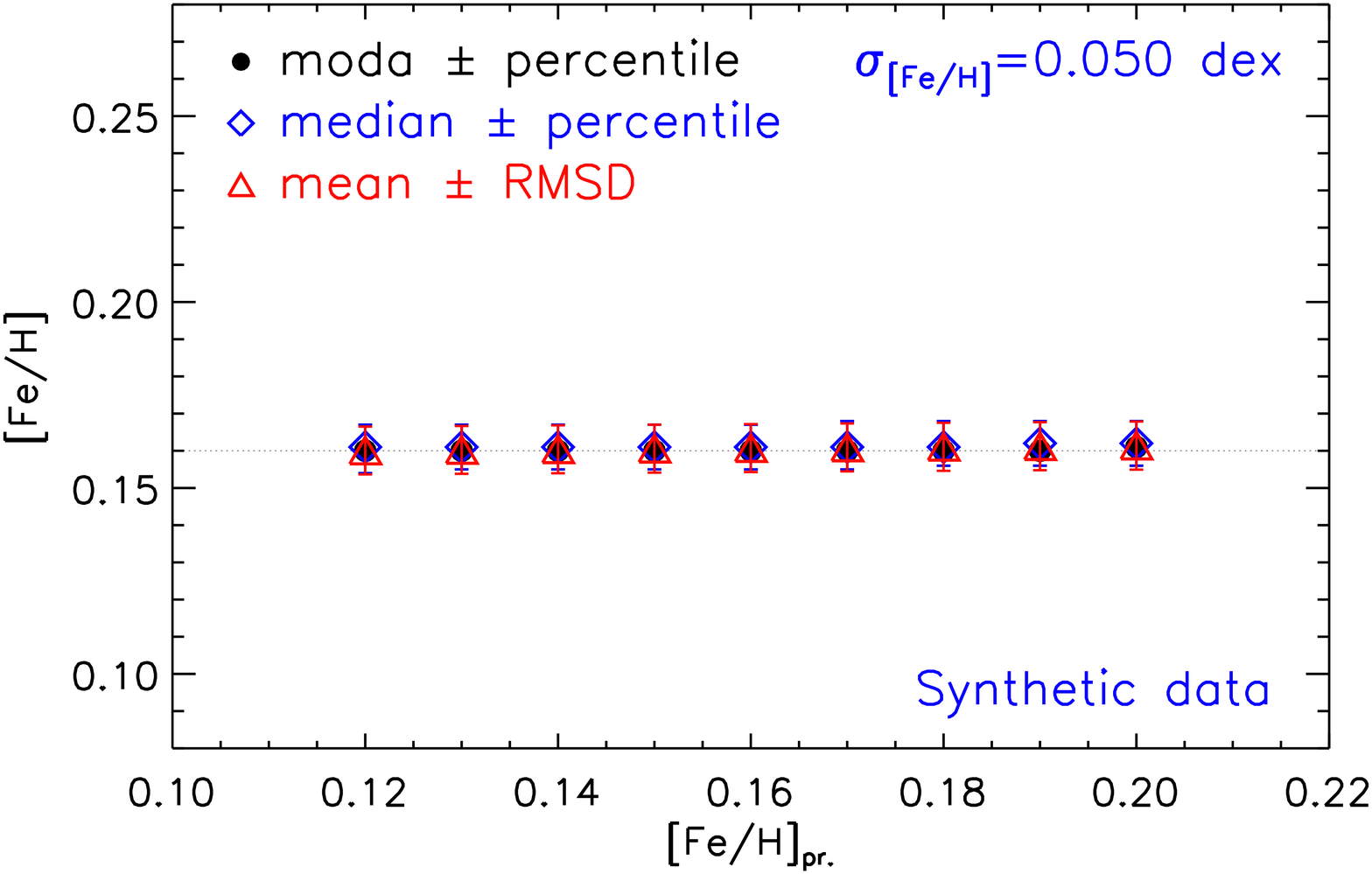}
\includegraphics[width=0.33\linewidth]{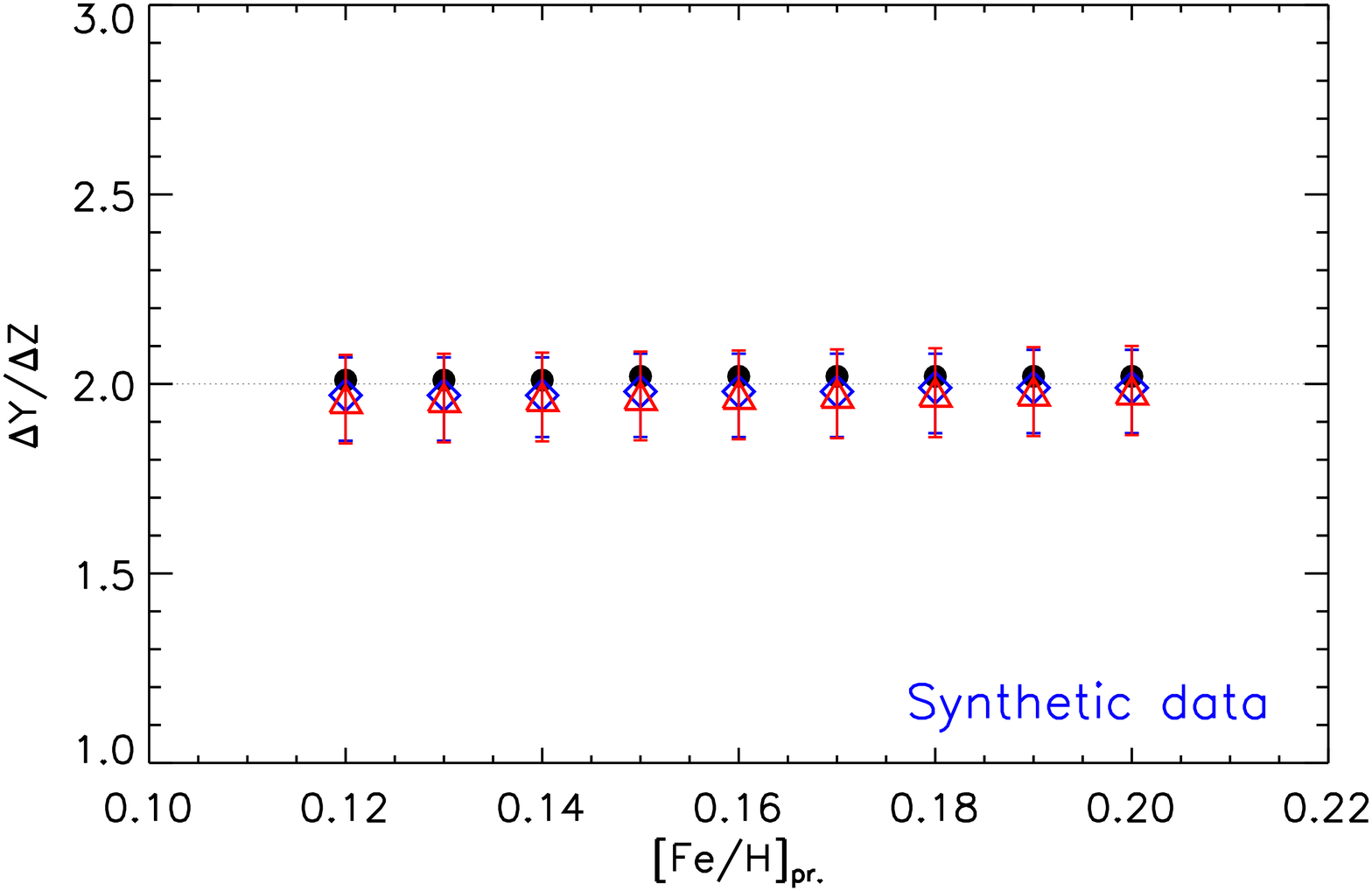}
\includegraphics[width=0.33\linewidth]{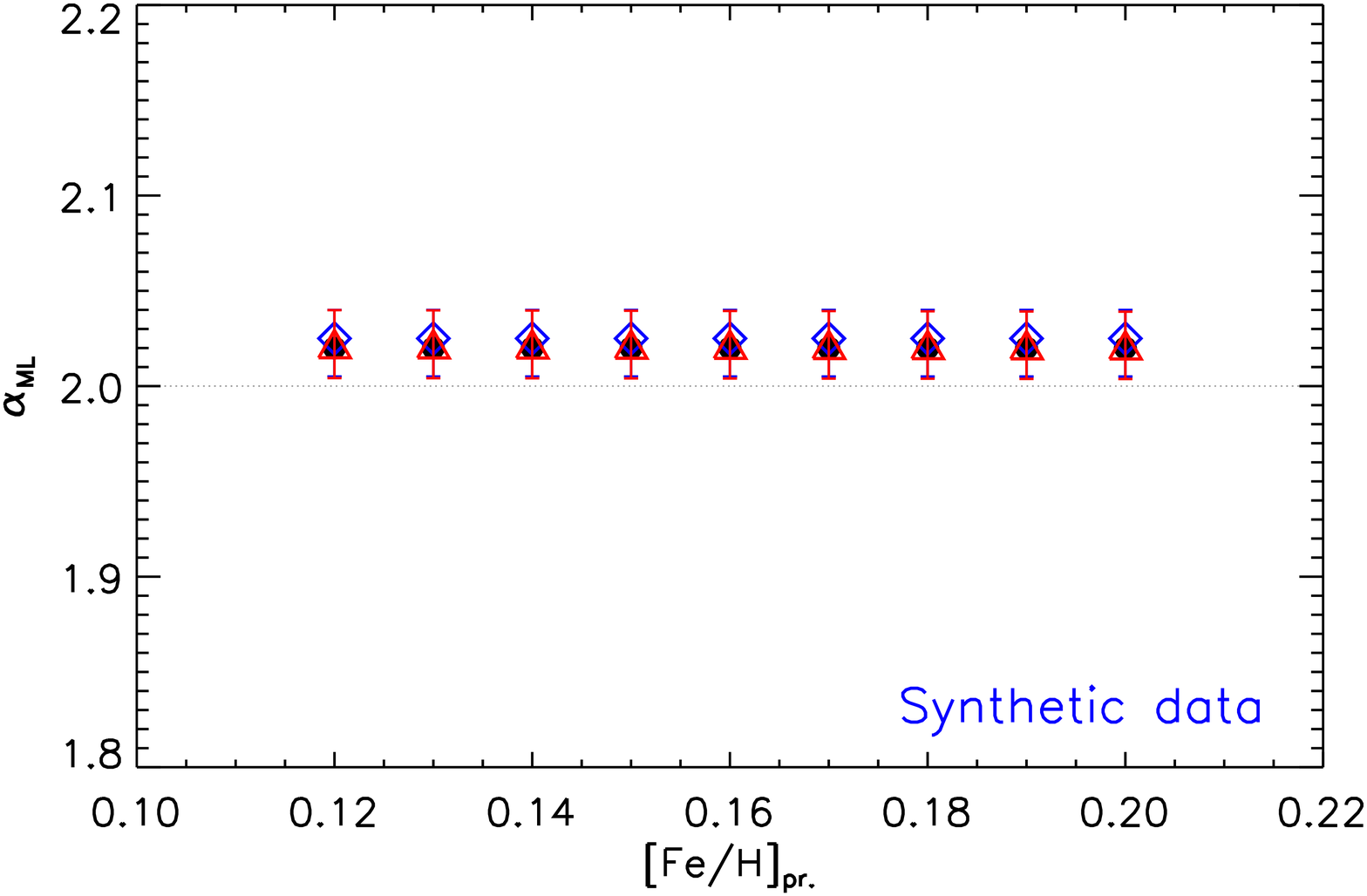}
\caption{As in Fig.~\ref{fig:res_feh_type1} but for TYPE\_3 data set.}
\label{fig:res_feh_type3}
\end{figure*}

The same analysis was repeated using the TYPE\_3 data set, that is, the set with the MS extended down to $B_P = 10$~mag. The results are shown in Figure~\ref{fig:res_feh_type3}.
This case is very close to the ideal one. Almost independently of the adopted \fehpr{} or \sigpr, the recovered parameters are affected by a negligible bias, they are in all the cases extremely close to the true ones, that is, those used to generate the synthetic data. Moreover the precision of the recovered parameters is good: namely within about 6-10~percent on [Fe/H], within about 10~percent on \dydz{} and 1-2~percent on \ml. This is the best precision that one can obtain for the selected parameters.

\section{Real data: Hyades cluster}
The method, tested in the previous sections, has been applied to the Hyades MS portion with $B_P \in [3.5,~6.5]$~mag with the isochrone age fixed to 500~Myr. The uncertainties on the photometry used for the recovery are similar to that discussed for TYPE\_1 data set: (1) the absolute $B_P$ magnitude errors take into account the magnitude error given in the original tables plus a systematic uncertainty of 0.01~mag (dominant contribution) due to a possible 0.1~mas zero-point error in the estimated parallax, and (2) for the $G-Rp$ colour index we used the errors in the original tables that range between 0.001 and 0.003~mag (with a mean value of about 0.002~mag). As previously shown, the adopted age is not crucial since the location in the CMD of the selected portion of MS is independent of the adopted age, at least within the range 300-700~Myr.  However, to be sure that the results are not affected at all by the adopted age, the recovery was performed also for two other different ages: 300 and 700 Myr. As expected the age variation does not affect the derived values of \dydz, \ml, and [Fe/H]. 
\begin{figure*}
\centering
\includegraphics[width=0.32\linewidth]{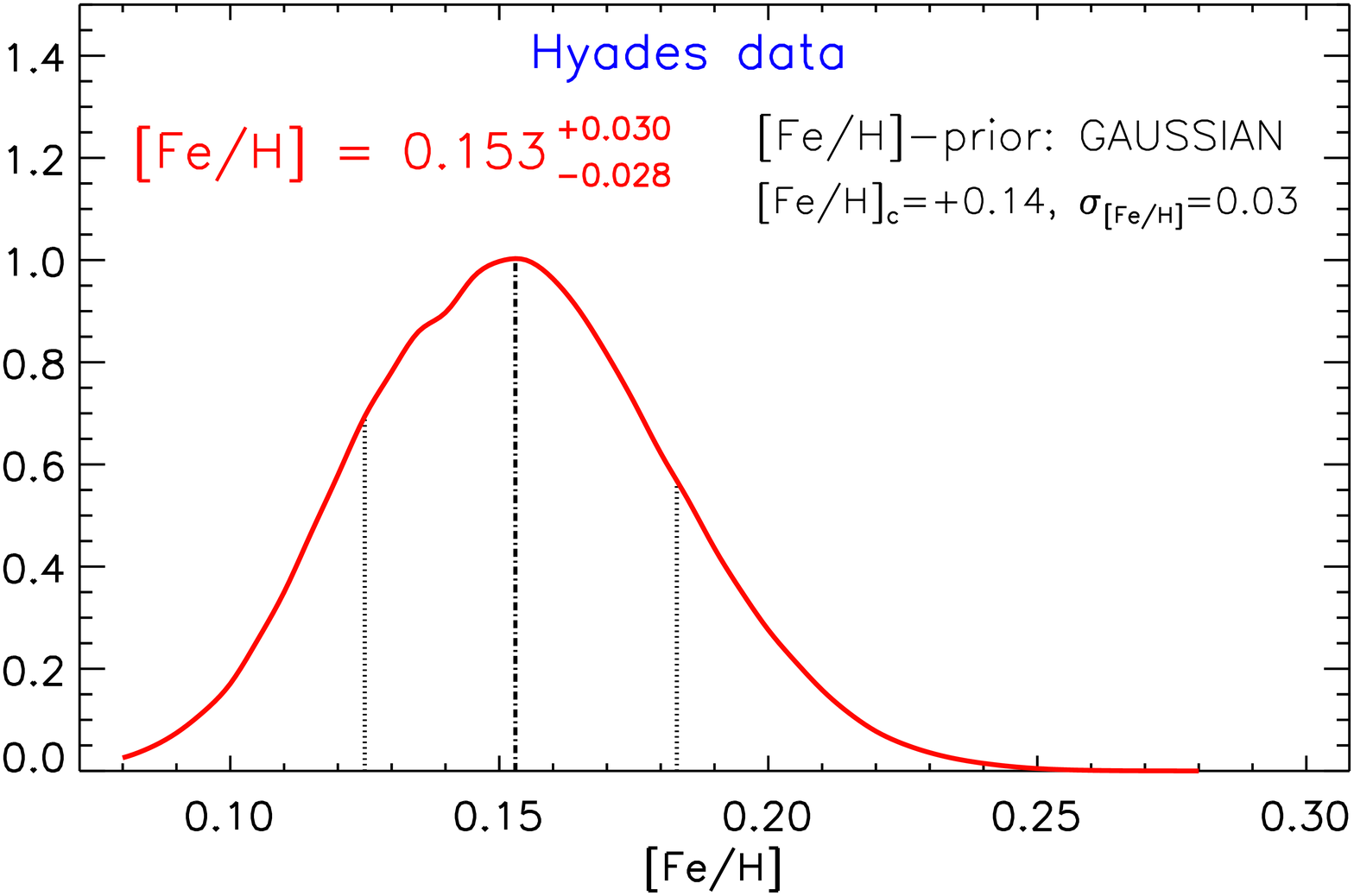}
\includegraphics[width=0.32\linewidth]{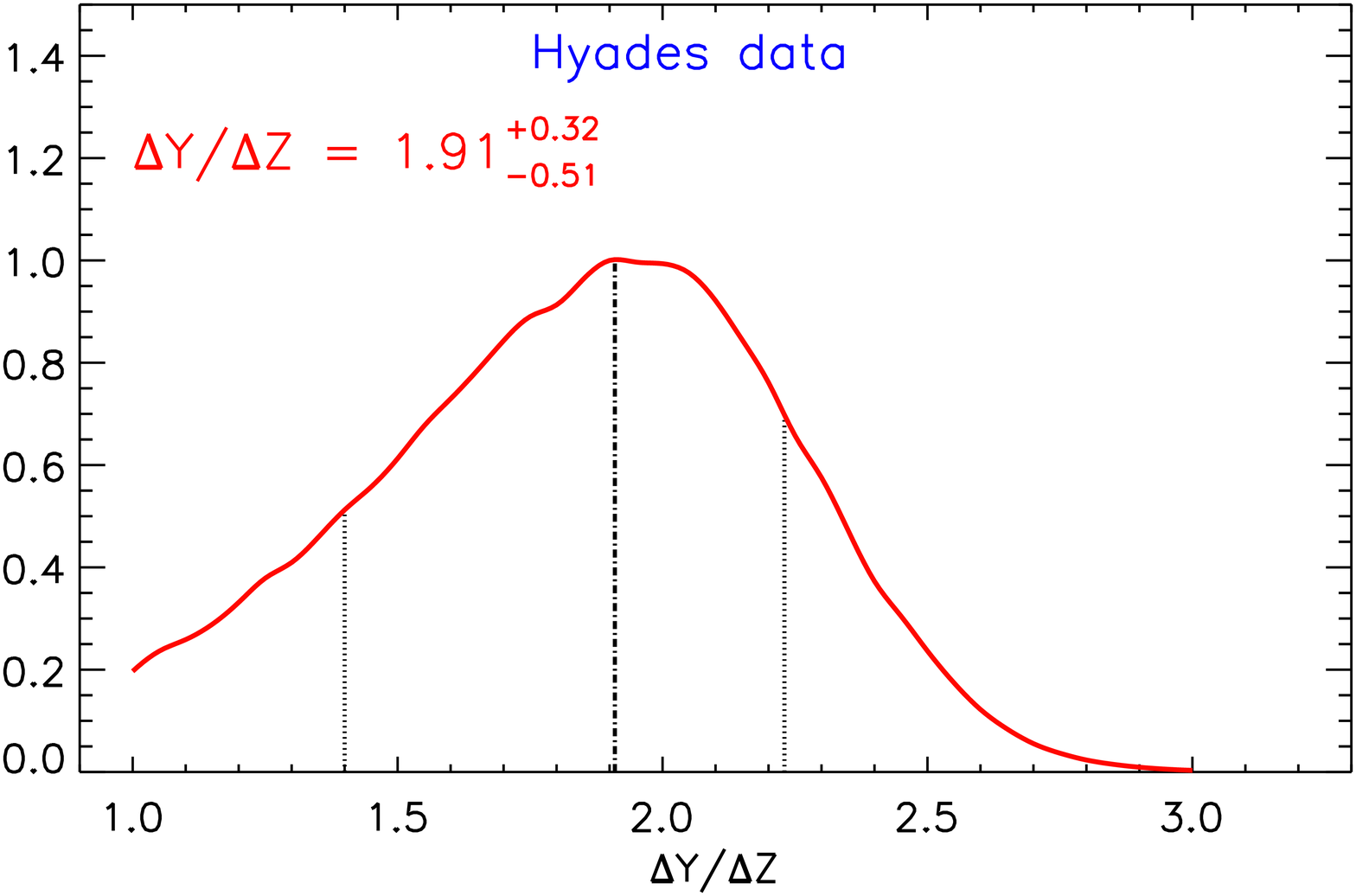}
\includegraphics[width=0.32\linewidth]{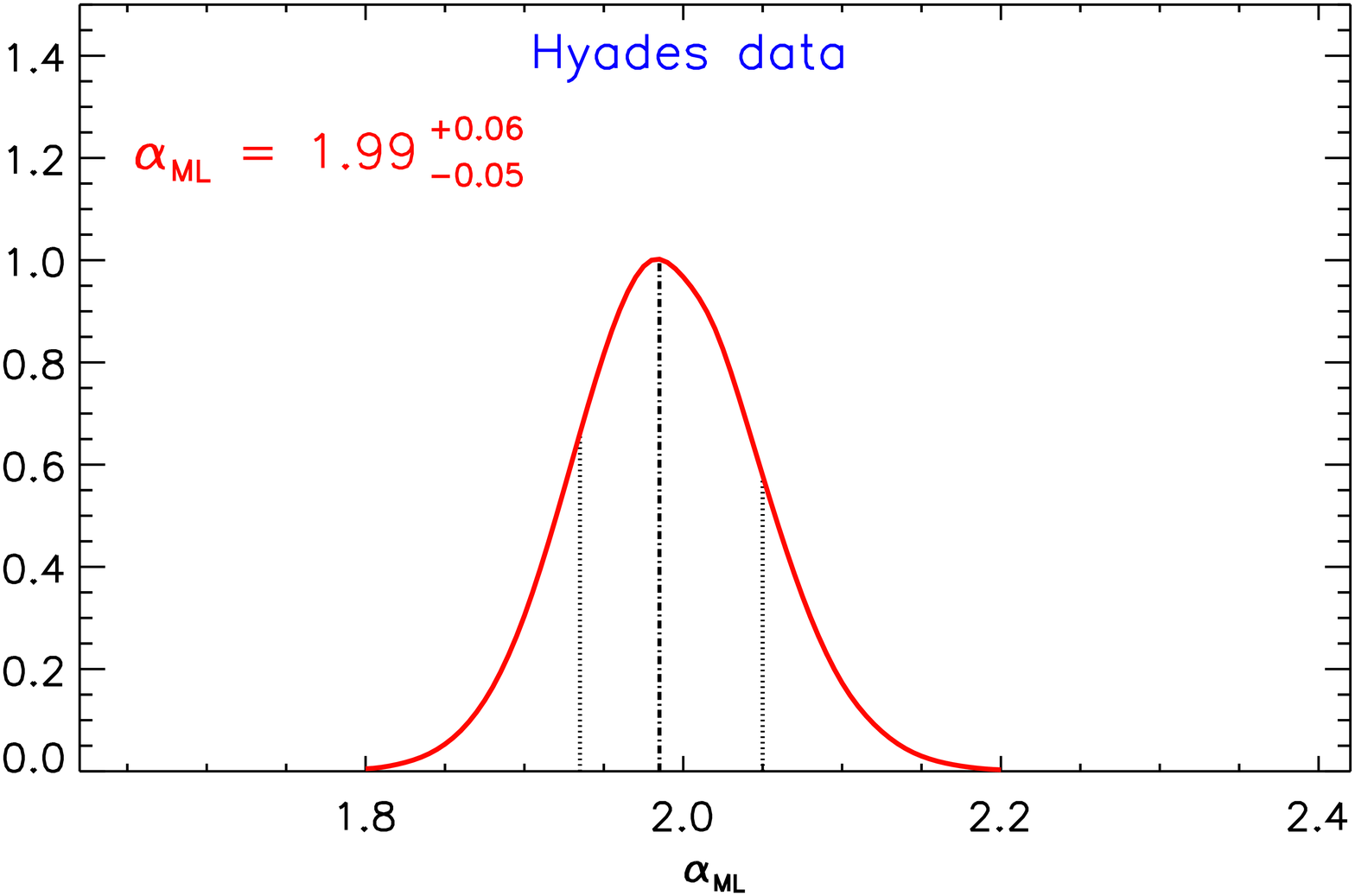}\\
\includegraphics[width=0.32\linewidth]{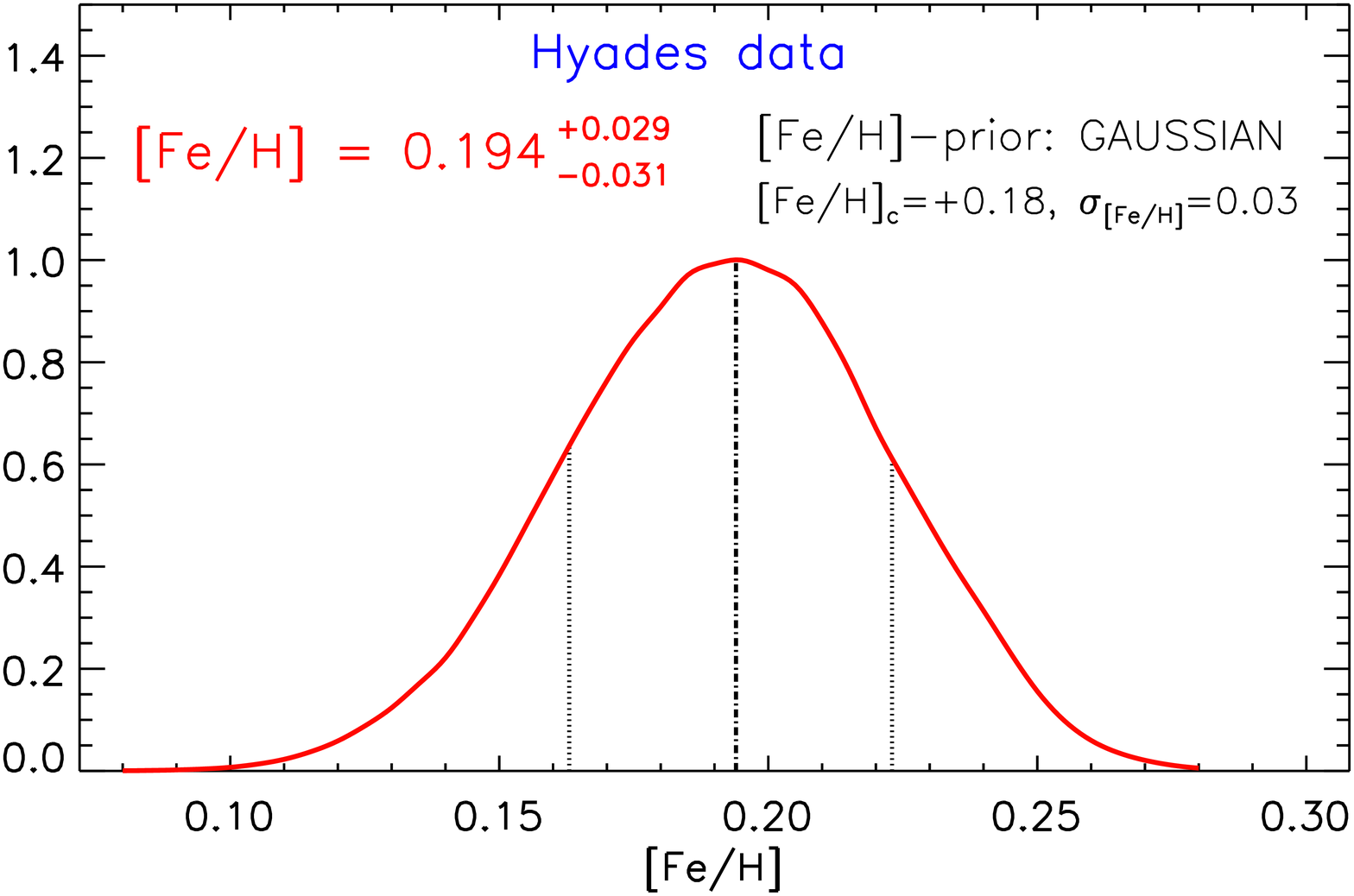}
\includegraphics[width=0.32\linewidth]{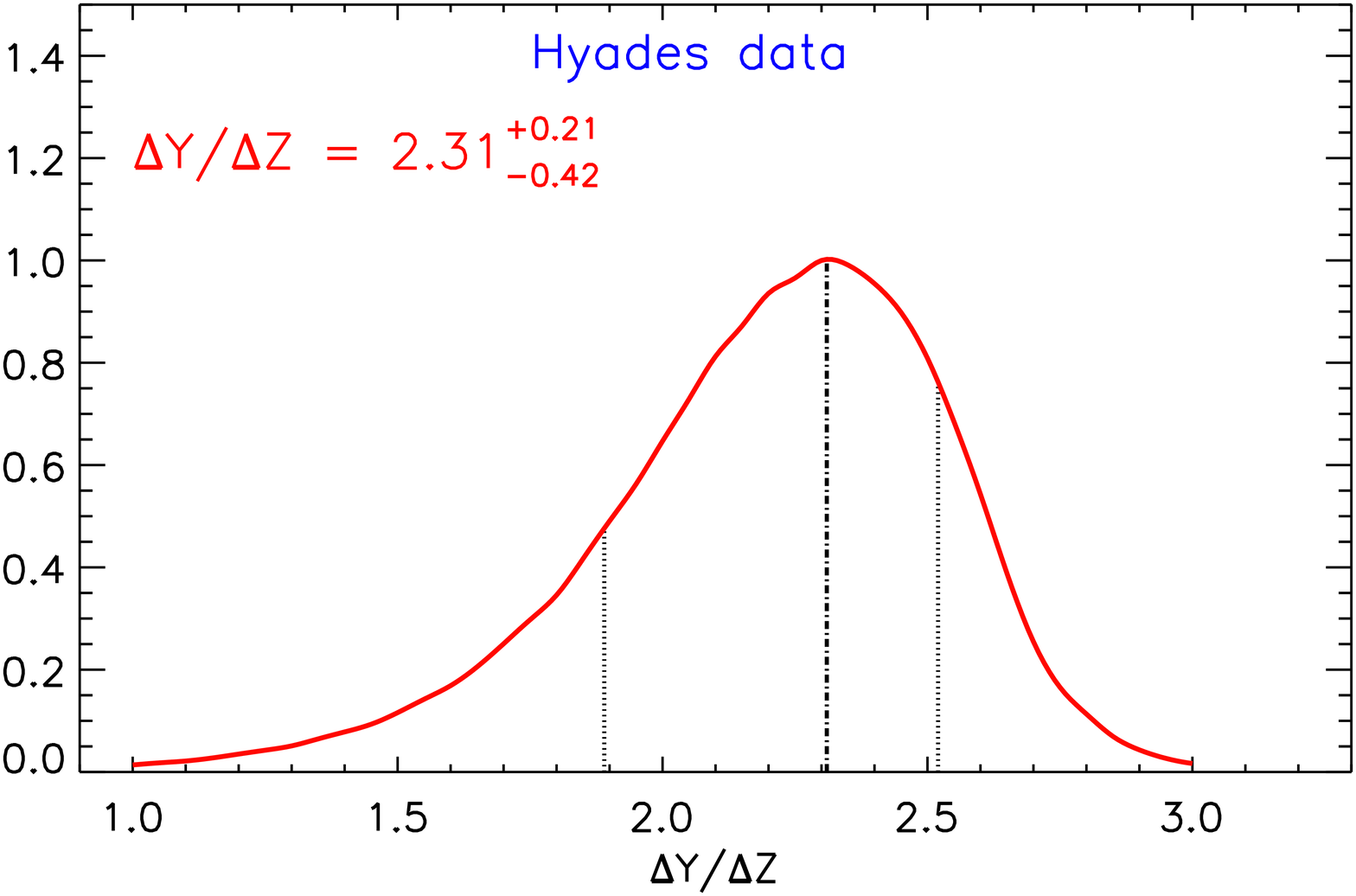}
\includegraphics[width=0.32\linewidth]{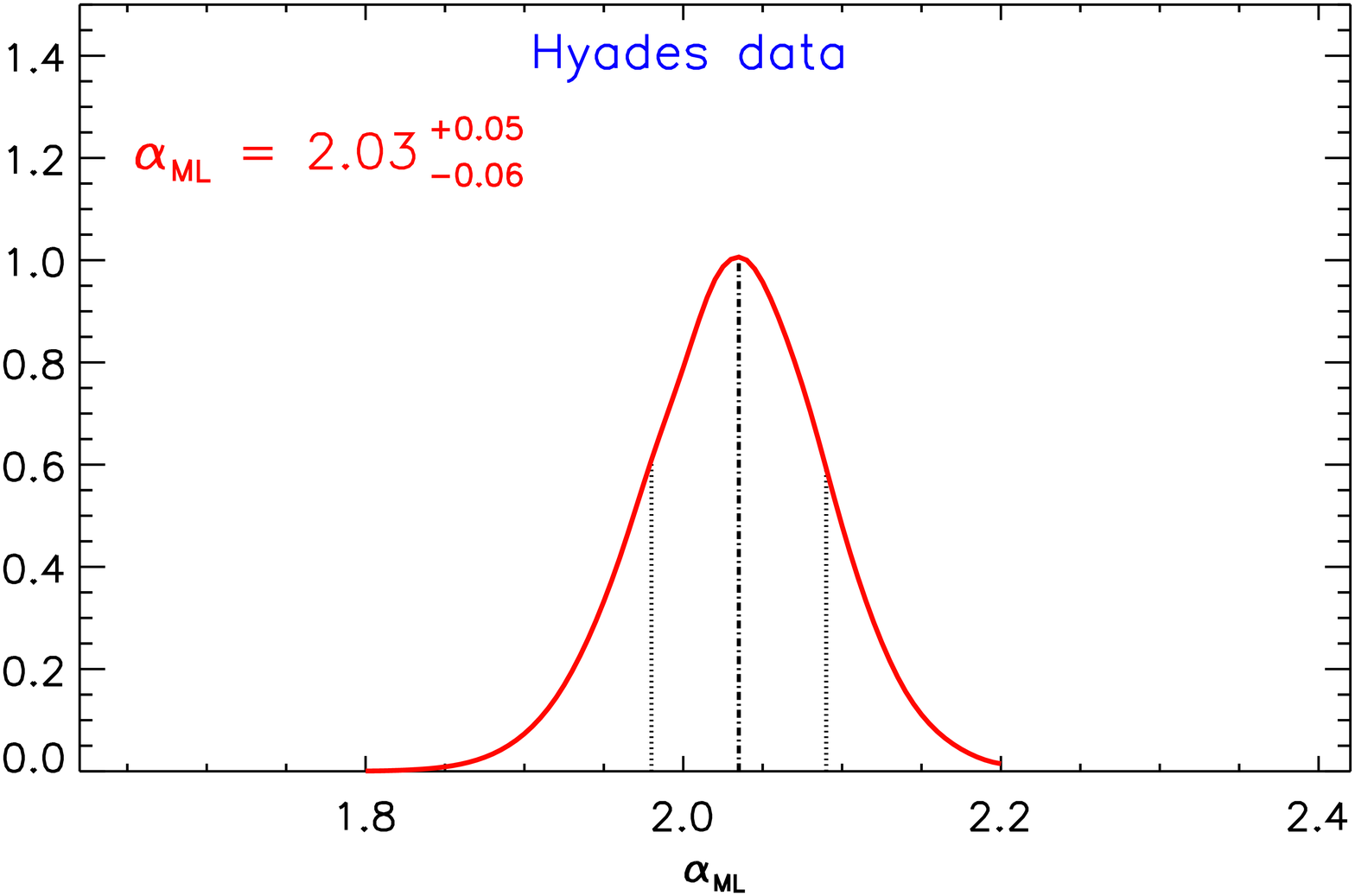}
\caption{Posterior marginalized distributions for Hyades data set, for [Fe/H] (left-hand panels), \dydz{} (mid panels), and \ml{} (right-hand panels), for two choices about the Gaussian metallicity prior, namely \fehpr=$+0.14$~dex and \sigpr=0.03~dex (top panel), and \fehpr=$+0.18$~dex and \sigpr=0.03~dex (bottom panel).}
\label{fig:hyades_lkl}
\end{figure*}

Figure~\ref{fig:hyades_lkl} shows the marginalized posterior distributions used to derive the most probable value of [Fe/H], \dydz, and \ml, for the cases where a Gaussian prior on [Fe/H] is adopted.  For the Hyades cluster, spectroscopic measurements of [Fe/H] are available, thus it is recommended to adopt a prior on [Fe/H]. The [Fe/H] value from the literature ranges between $+0.10$ and $+0.20$ \citep[see e.g.][]{cayrel85,boesgaard90,paulson03,schuler06,carrera11}. In particular recent measurements by \citet{dutra16}  give a best value of [Fe/H]=$+0.14\pm0.03$~dex  or $+0.18\pm0.03$~dex, depending on the setup used to derive [Fe/H] from the spectra. We adopted these two latter values, which are representative within the uncertainty of $\pm0.03$~dex of both low ($+0.13~\div~+0.14$) and high (up to $+0.20$) [Fe/H] values measurements present in the literature.

The top panels of Fig.~\ref{fig:hyades_lkl} refers to the case where we adopted a prior centred in \fehpr=$+0.14$ with \sigpr=0.03. The distributions of [Fe/H], \dydz, and \ml{} show a clear peak that can be used to define the most probable value of these parameters. We  obtain [Fe/H]=$+0.153$ with a CI of [$+0.125$,~$+0.183$], \dydz=1.91 with the corresponding CI of [1.40,~2.23], and \ml=1.99 with CI of [1.94,~2.05]. 

The results are affected by the adopted prior for [Fe/H]. If one chooses a Gaussian prior with \fehpr=$+0.18$ with \sigpr=0.03  one finds different values of the most probable [Fe/H], \dydz, and \ml, which however are consistent within their CI with the values obtained above. We obtain: [Fe/H]=$+0.194$ with CI of [$0.163$,~$+0.223$], \dydz=2.31 with a CI of [1.89,~2.52], and \ml=2.03 with a CI of [1.97,~2.08]. 

In both the cases, in agreement with the results from the synthetic data set, the parameter derived with the best precision is \ml: from the analysis we obtained an \ml{} value fully compatible with our solar calibrated mixing length, namely \ml$_{\sun}=2.0$. 

On the other hand \dydz{} is recovered with a relatively large CI, and more important, the most probable value depends on the adopted prior on [Fe/H]. For such parameter, the best value changes from \dydz$\approx 2$ at \fehpr$=+0.14$ to \dydz$\approx2.3$ at \fehpr$=+0.18$. Given such a dependence of \dydz{} value on the adopted prior for [Fe/H], we performed an analysis of the effect on the recovered quantities on \fehpr, as done in the previous section for the synthetic data sets.
\begin{figure*}
\centering
\includegraphics[width=0.33\linewidth]{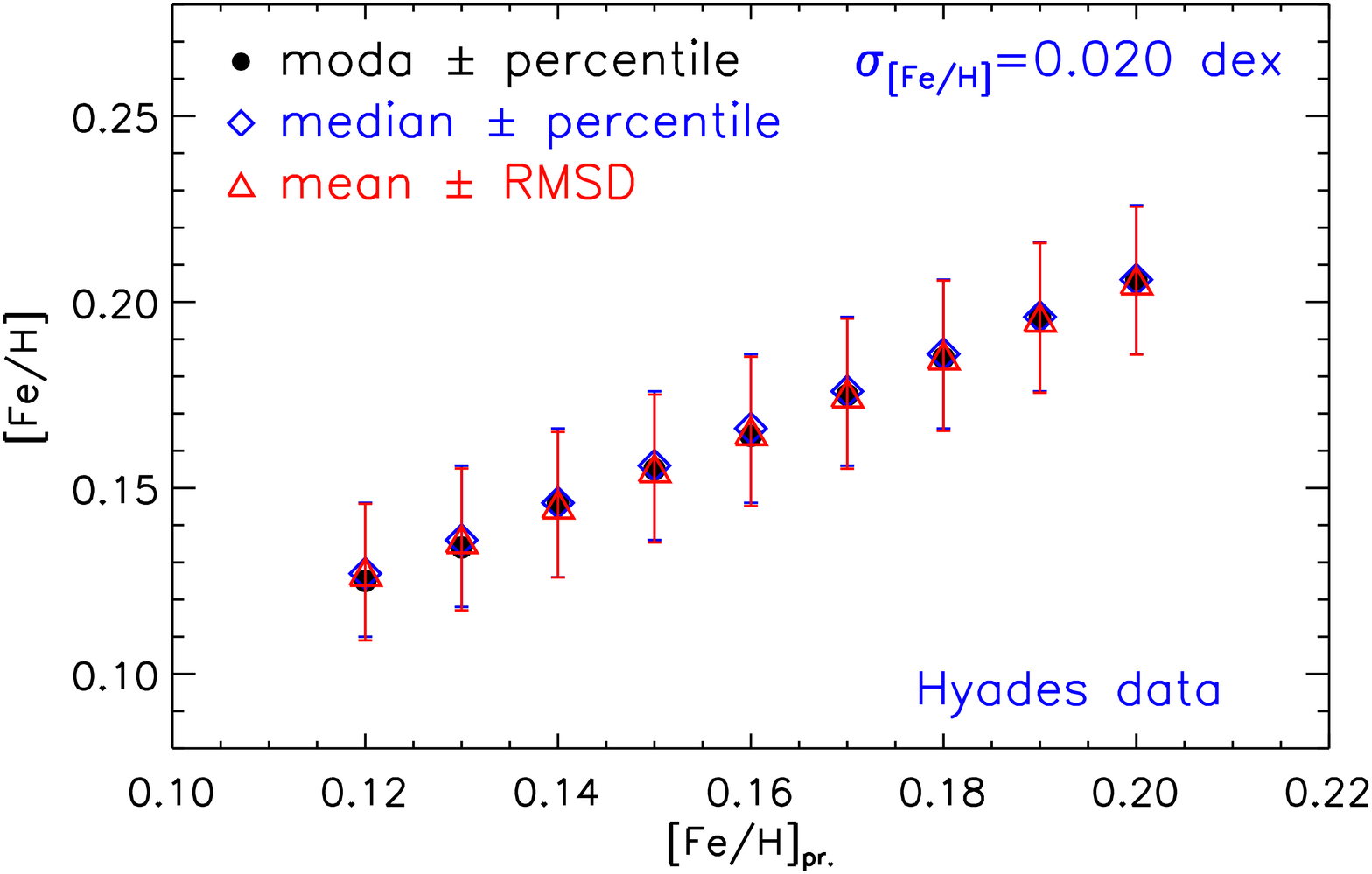}
\includegraphics[width=0.33\linewidth]{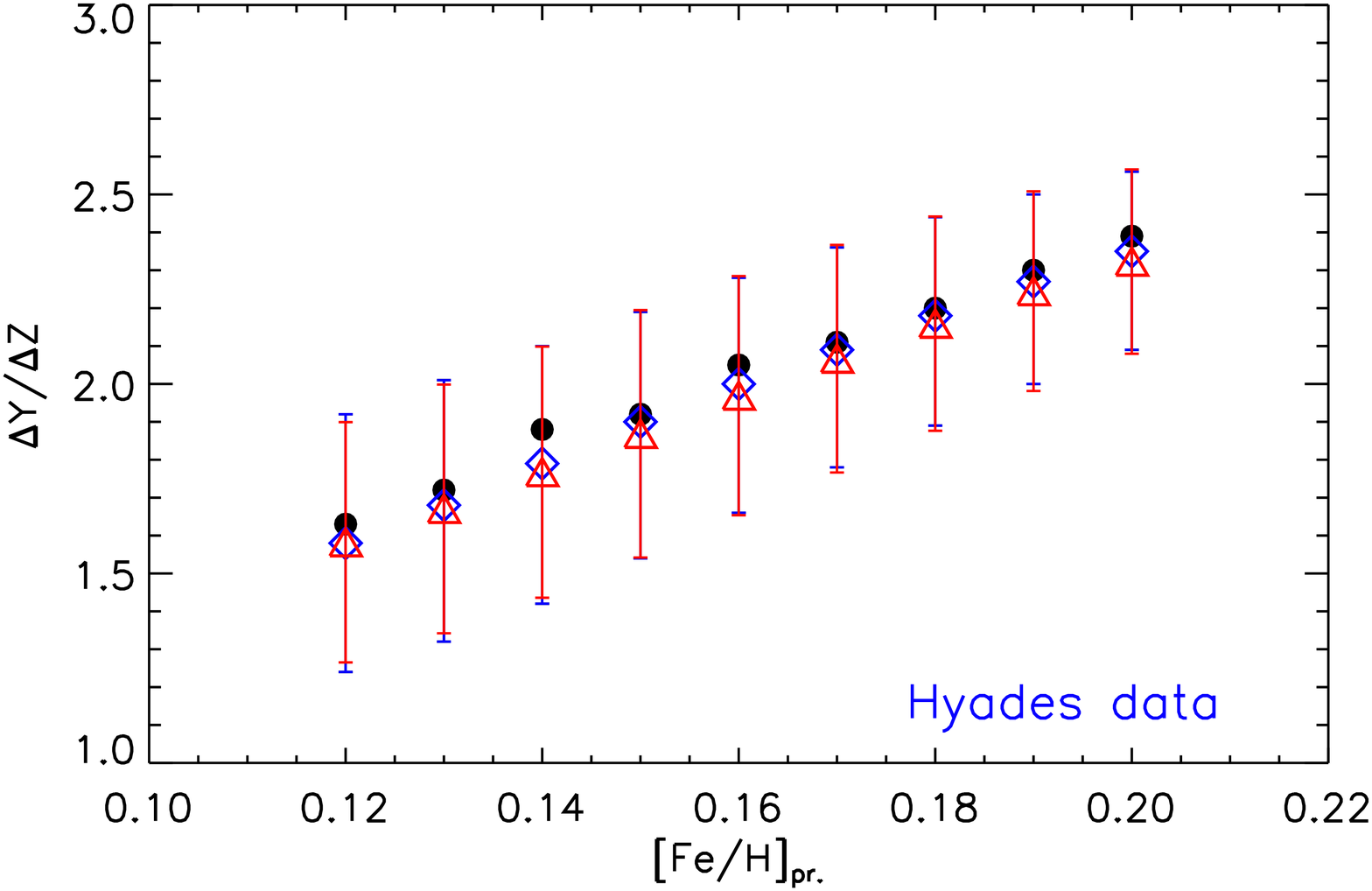}
\includegraphics[width=0.33\linewidth]{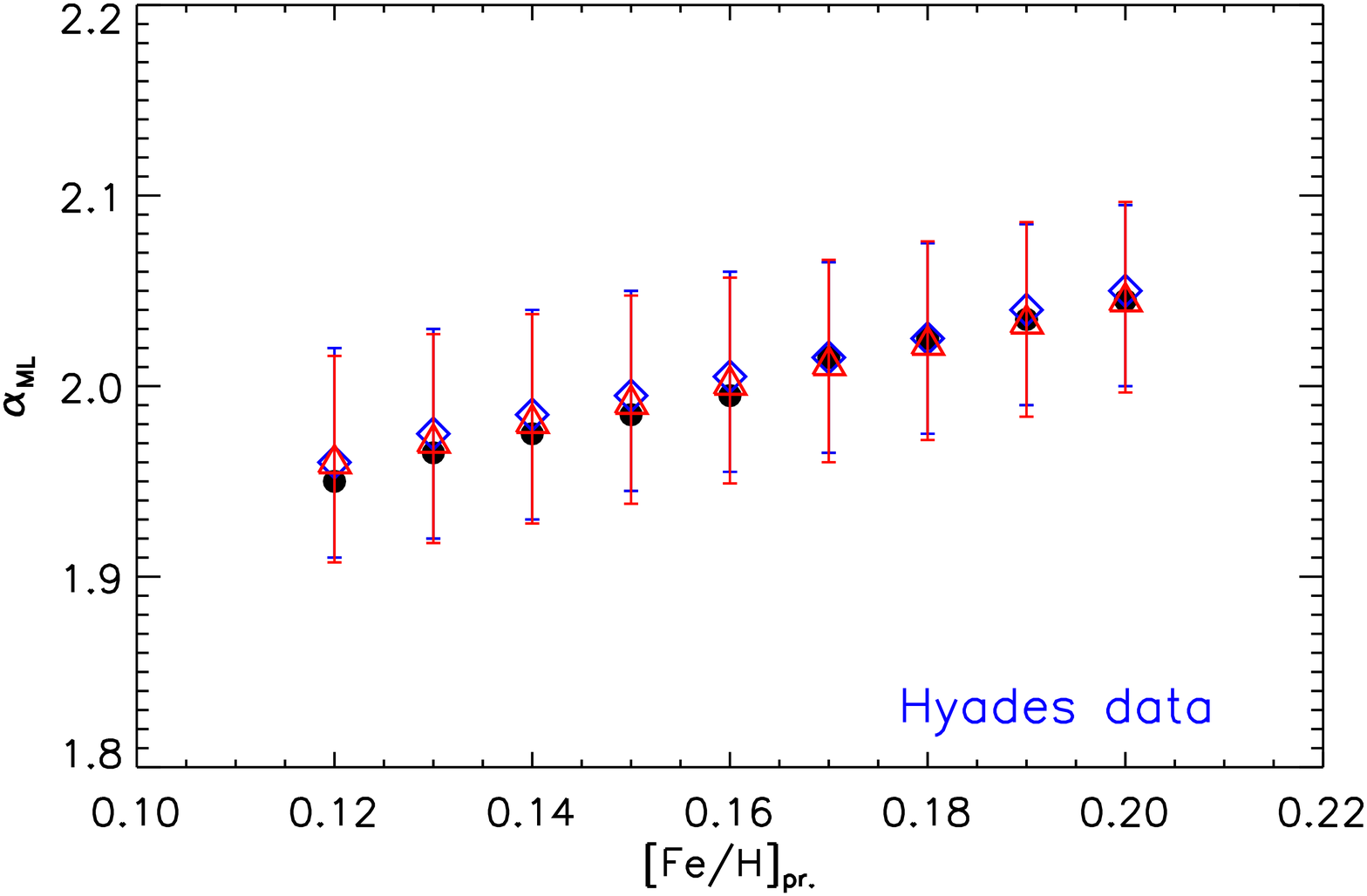}\\
\includegraphics[width=0.33\linewidth]{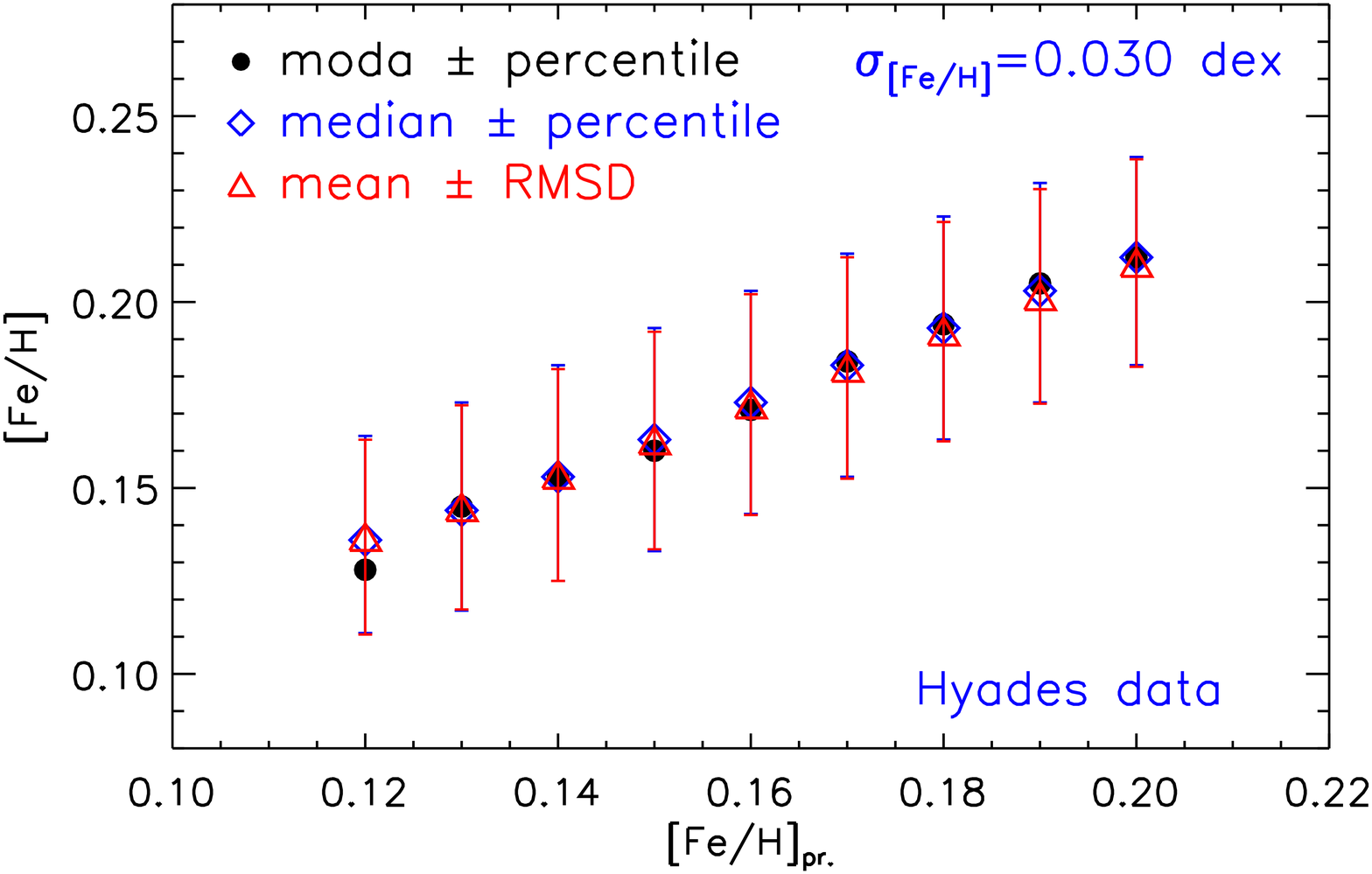}
\includegraphics[width=0.33\linewidth]{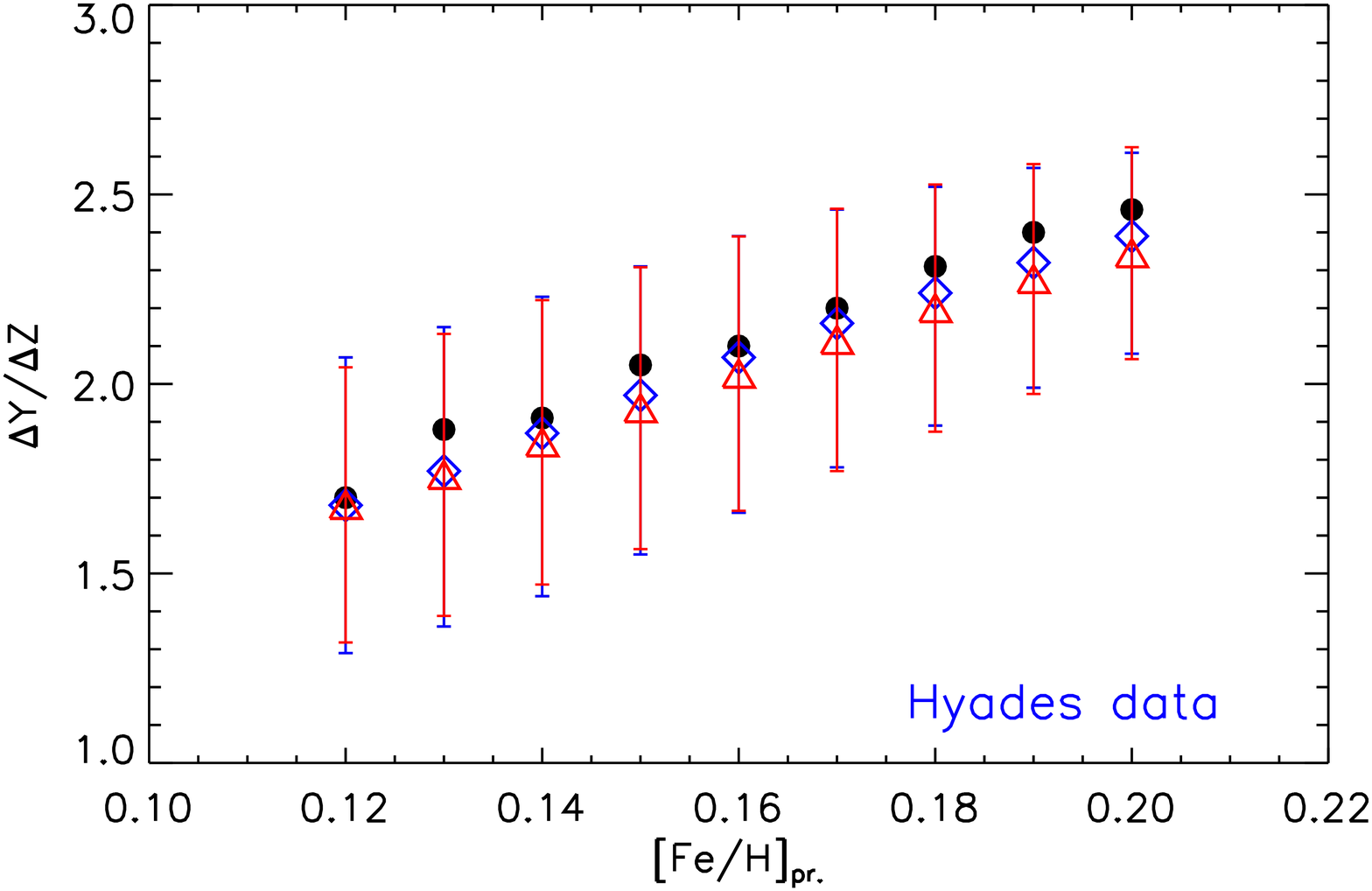}
\includegraphics[width=0.33\linewidth]{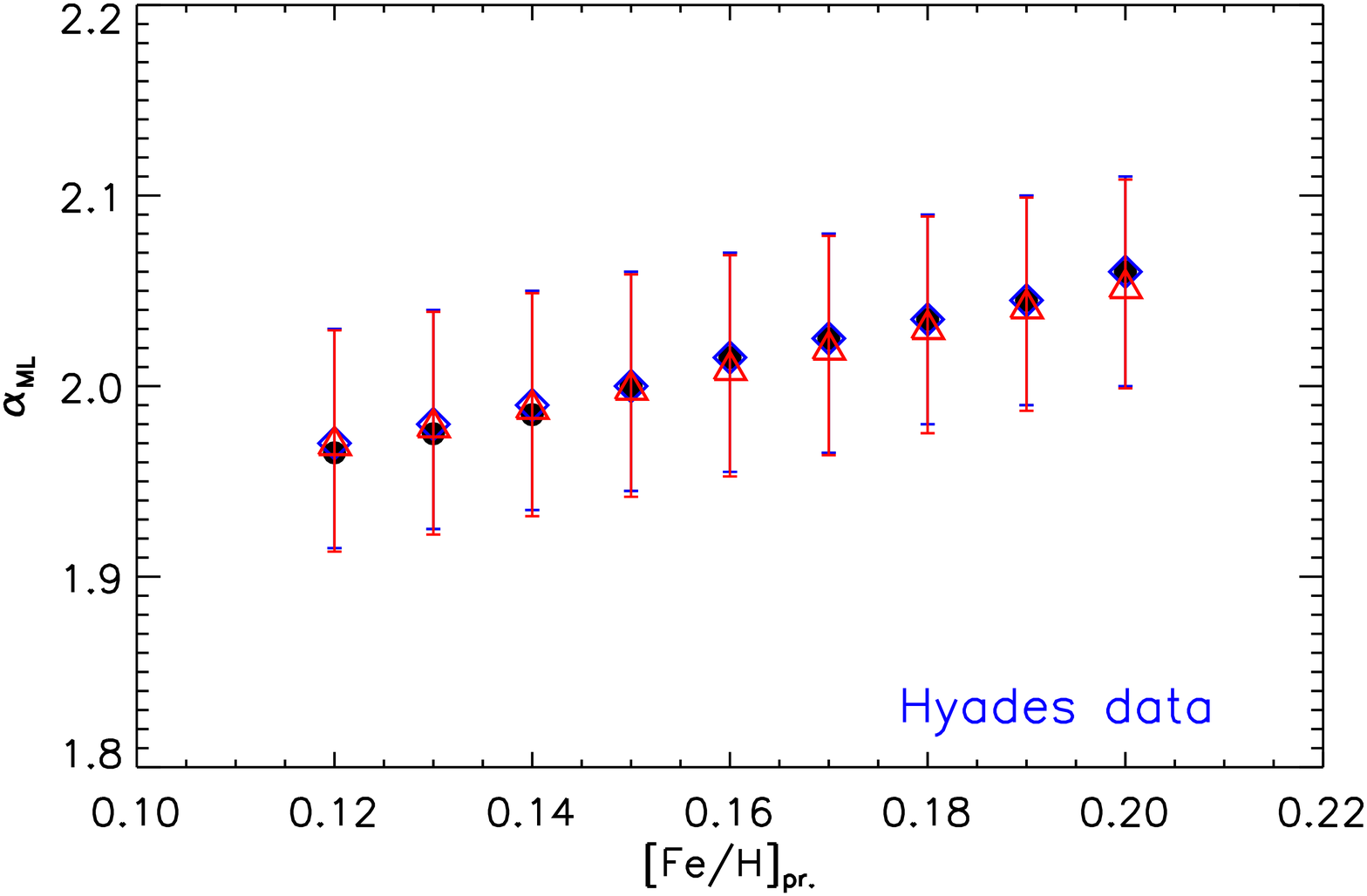}\\
\includegraphics[width=0.33\linewidth]{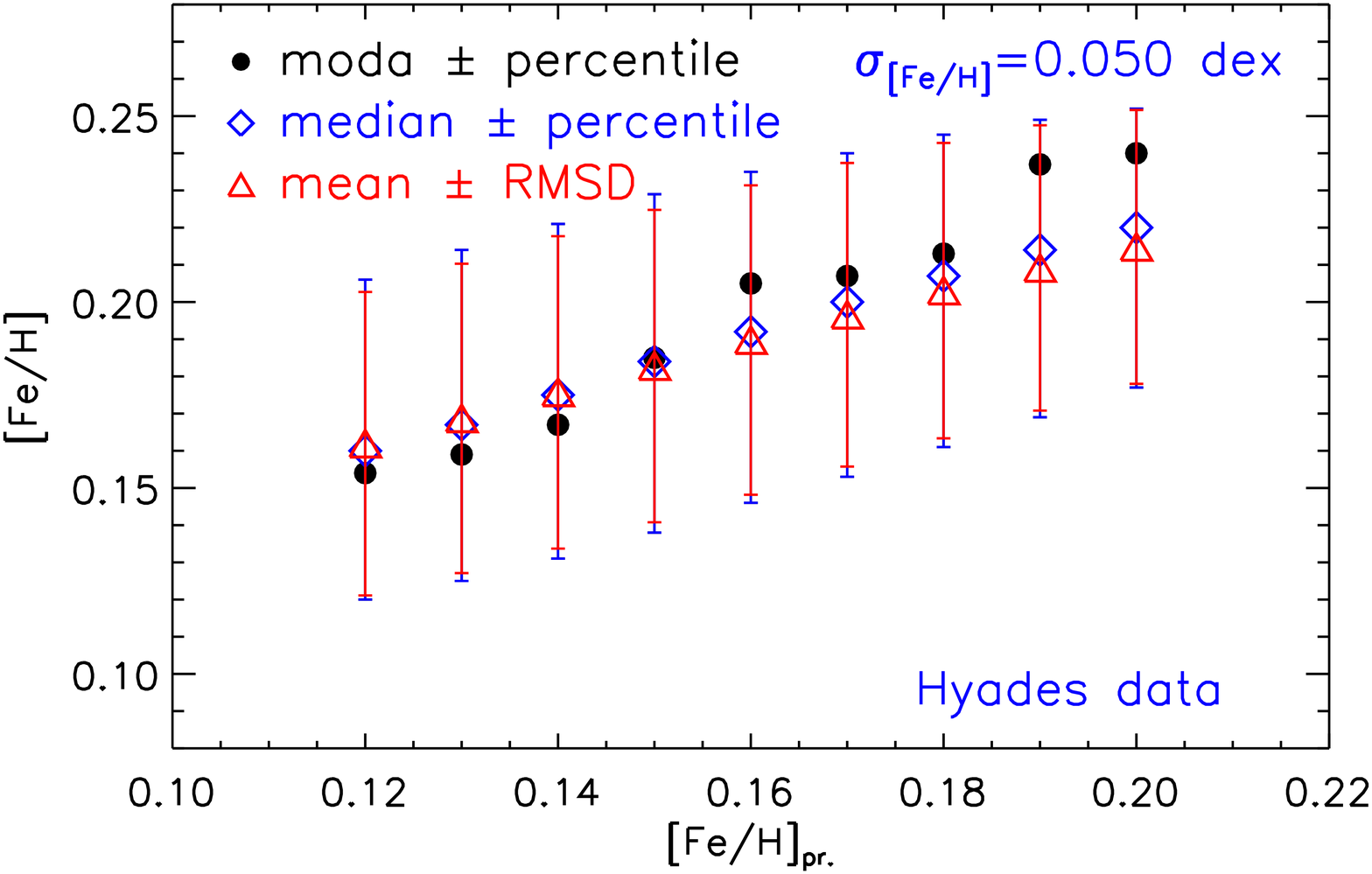}
\includegraphics[width=0.33\linewidth]{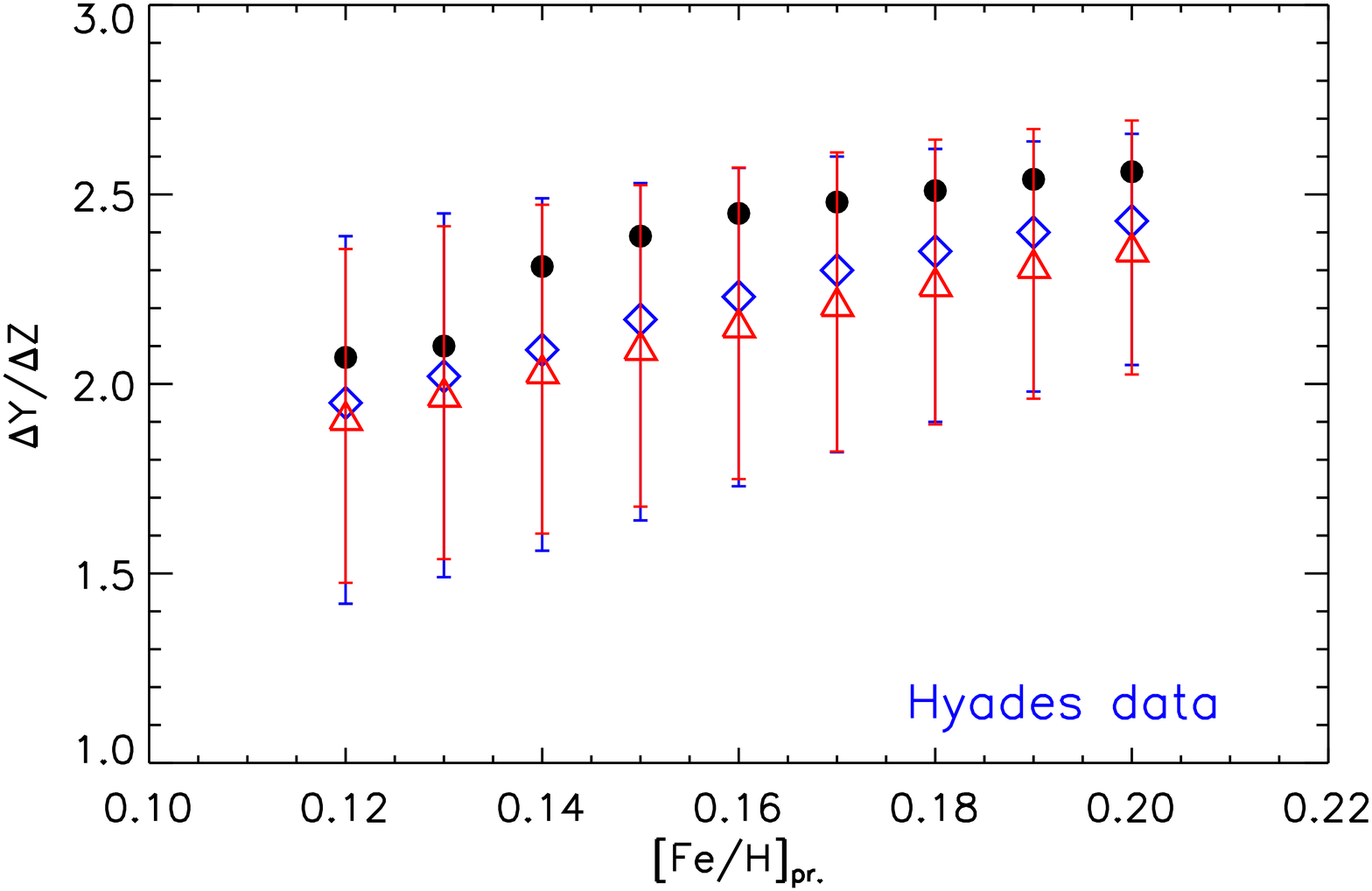}
\includegraphics[width=0.33\linewidth]{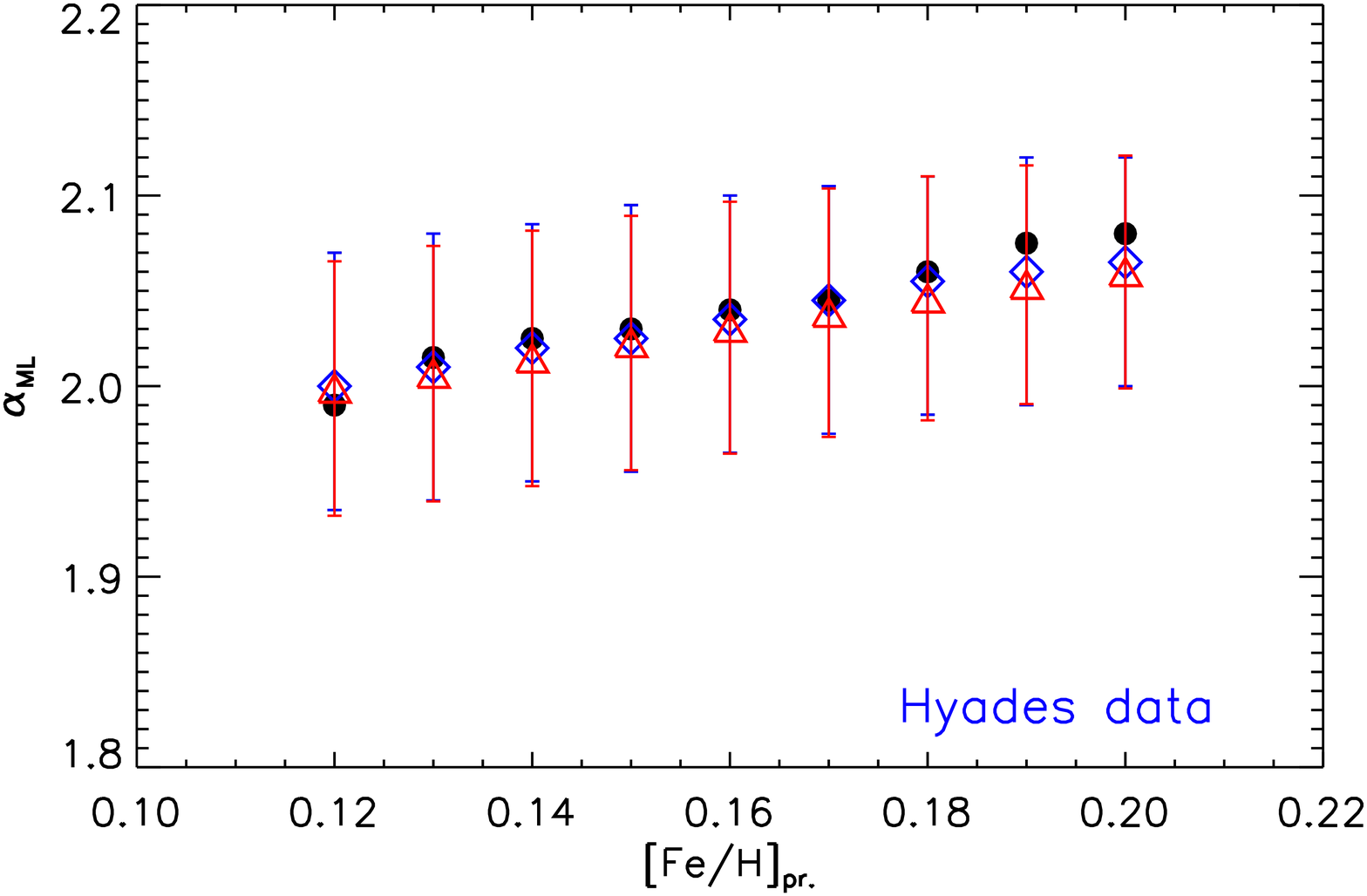}
\caption{Most probable value and corresponding CI for [Fe/H] (first column), \dydz{} (second column), and \ml{} (third column) for Hyades data set, as a function of \fehpr{} and for three values of \sigpr=0.02, 0.03, and 0.05~dex (first, second and third rows, respectively). In figure we also overplotted the best value and the relative CI obtained using the median and the mean (see the text).}
\label{fig:res_feh_data}
\end{figure*}

Figure~\ref{fig:res_feh_data} shows the values of [Fe/H], \dydz, and \ml{} obtained using the same grid of \fehpr{} and \sigpr{} defined in the previous Section. 
As already noted in the analysis of the mock data set, the most probable value of \dydz{} varies depending on \fehpr{} and \sigpr{} (and on the adopted estimator) in the interval [1.6,~2.6], being subjected to a bias that comes from the adopted prior on [Fe/H]. On the other hand, the mixing length parameter is, as expected, less affected by such bias. As in the synthetic data sets, the most probable value of \ml{} varies at maximum of about 2-3~percent for \sigpr=0.02 or 0.03~dex, while the maximum variation is a bit larger (about 4-5~percent) for \sigpr=0.05~dex.

We want to emphasize that not all the \fehpr{} values in figure are equally probable. To be very conservative we adopted a large interval of \fehpr, but as already noted, the recent determinations of [Fe/H] for the Hyades cluster suggest -- as prior -- values in the range [$+0.14$,~$+0.18$], with an uncertainty of 0.03~dex. If the analysis is restricted to this range of \fehpr{} and to \sigpr$\le 0.03$~dex, the most probable value of \dydz{} is in the range [1.7,~2.2] for \sigpr=0.02~dex or [1.8,~2.3] for \sigpr=0.03~dex. If one takes into account also the (large) CI of the \dydz{} estimates for these two cases, than the possible \dydz{} values vary in the interval [1.4,~2.4] for \sigpr=0.02~dex and [1.5,~2.5] for \sigpr=0.03~dex. 
\begin{figure}
\centering
\includegraphics[width=0.98\columnwidth]{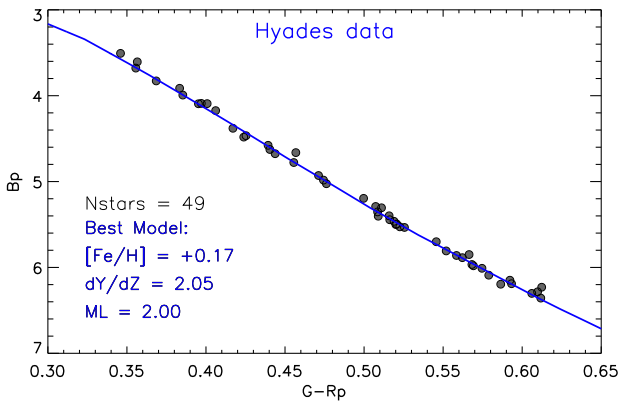}
\caption{Comparison between Hyades data and the ``best" isochrone obtained using a set of parameters among those available in the grid as close as possible to the average values $<$\dydz$>$, $<$\ml$>$ and $<$[Fe/H]$>$ obtained from the analysis.}
\label{fig:hyades_cmd}
\end{figure}

We can define an average value of \dydz{}, \ml{} and also [Fe/H] by considering the values obtained using (1) different estimators, (2) different values of \fehpr{} in $[+0.14,~+0.18]$, and (3) \sigpr{} equal to 0.02 and 0.03~dex. If we consider this sample, we obtain $<$\dydz$>$=$2.03 \pm0.33$ and $<$\ml$>$=$2.01\pm0.05$ and $<$[Fe/H]$>$=$+0.169\pm0.025$, which are our best estimate for these parameters using the current set of models. Figure~\ref{fig:hyades_cmd} shows as an example the comparison between the Hyades data and the isochrone computed with a set of parameters, among those available in the grid, as close as possible to the average values of $<$[Fe/H]$>$, $<$\dydz$>$, and $<$\ml$>$.  

From the estimated \dydz{} we obtained an initial helium content for the Hyades stars $Y=0.2867$ (for [Fe/H]$=+0.17$, $Z=0.01863$), which is larger than the solar one. As discussed in the introduction, in the literature one can find estimates of the Hyades helium content based on different techniques: 1) a binary system (vB22) or 2) MS stars. These two different methods resulted in different helium abundances \citep{perryman98,lebreton01,castellani02,pinsonneault03,gennaro10}. In particular, it seems to emerge that when the Hyades MS stars are used, then the derived helium content is compatible with a supersolar value, while the analysis of the vB22 binary systems suggests a lower helium content. The detailed comparison between these results goes beyond the aim of this paper. We limit to say that the results present in the literature were obtained using different methods and different stellar models with different input physics and parameters, which introduce differences and systematics in the derived helium content not easily quantifiable, as discussed in \citet{lebreton01} and \citet{pinsonneault03}. Nevertheless, it would be interesting, for future work, to analyse the difference in the derived helium contend when binary systems or MS Hyades stars are used, but relying on the same set/grid of stellar models.

As a final comment, we notice that the values of $<$[Fe/H]$>$, $<$\dydz$>$, and $<$\ml$>$ clearly depend on the adopted set of models, that is, on the input physics used to generate the tracks/isochrones. Thus the errors on the average values for the parameters derived here are probably an underestimation, because the systematic uncertainty deriving from the adoption of different input physics is not evaluated. Such an analysis, which would require a very huge computational effort, is beyond the scope of this paper, whose aim is to set up the method, test and analyse its applicability to synthetic data and then apply it to the real data using our reference set of models. 

\section{Conclusions}
The aim of the paper is to analyse the possibility to precisely calibrate the mixing length parameter \ml{} and the helium-to-metal enrichment ratio \dydz{} adopted in stellar models with the MS of open clusters. The essential precondition for a good performance is the availability of precise photometry, distance and reddening estimates. Relying on the recent {\it Gaia} DR2 data set, we selected the Hyades cluster, which being close, without reddening and having a very good photometry, satisfies all the needed requirements. We restricted our analysis to the Hyades MS in the $B_P$ magnitude range [3.5,~6.5]~mag because this region is insensitive to the (``a priori" unknown) cluster age and not affected by known problems in matching theoretical colours and observations in low MS. 

A Bayesian method is applied to simultaneously derive the set of parameters (namely \dydz, \ml, and [Fe/H]) that better reproduces the observed MS. We applied the method to a mock data set as close as possible to the Gaia DR2 Hyades data.

We proved that the Bayesian analysis is capable of recovering the most probable value of \ml{} with a very good precision, while [Fe/H] and \dydz{} are correctly determined only when a suitable prior on [Fe/H] is specified. This happens because of the well-known degeneracy [Fe/H]-\dydz{} which leads to a significant overlapping of the models with different combinations of [Fe/H] and \dydz, thus preventing a good determination of both these parameter when no prior information about [Fe/H] is available, especially when only part of the cluster MS is selected (i.e. in the range $B_P\in[3.5,~6.5]$~mag). In this range, we found a relatively large bias in the recovered [Fe/H] and \dydz. Since the sensitivity to a variation of the helium abundance and metallicity on the CMD location of low and high-mass MS stars is different, a significant improvement of both the accuracy and precision of the results can be achieved if the analysis is extended to a fainter portion of the MS, i.e. down to to $B_P\sim 10$~mag ($M\sim 0.5$~\msun). In this case, we proved that beside \ml, also the recovered \dydz and  [Fe/H] values are very close to the true ones used to generate the mock data set.

We tested whether the availability of an accurate spectroscopic [Fe/H] value could be used as a prior to restrict the interval of both [Fe/H] and \dydz{} values. We showed that adopting a Gaussian prior on [Fe/H] we can derive values for [Fe/H], \dydz, and \ml{} fully compatible with the true ones used to generate the data set. The best precision is achieved in the recovered value of \ml{} (with an uncertainty of 2-3~percent) without any significant bias. Using a reduced portion of the MS, the recovered \dydz{} value shows a small and negligible bias (\dydz{} is less than 1~percent smaller than the true one) with an uncertainty of about  10-15~percent. Also [Fe/H] show a negligible bias with an uncertainty that is very close to the dispersion adopted in the Gaussian prior. As in the case of no prior on [Fe/H], the accuracy and precision of the results are improved if the extended MS is adopted: in this case the bias in \dydz{} and [Fe/H] reduces to completely negligible values.

We also investigated the dependence of the recovered parameters on the [Fe/H] prior adopted (i.e. \fehpr{} and \sigpr), only when a portion of the MS is used. The derived [Fe/H] is strictly correlated to \fehpr{} and \sigpr: the smaller is \sigpr{} and greater is the correlation between [Fe/H] and \fehpr. Also \dydz{} is correlated to \fehpr, and the inferred \dydz{} values can differ by about 20-25~percent among each other depending on \fehpr: the lower is \fehpr{} and the lower is \dydz. This clearly show that the adoption of a not suitable prior on [Fe/H] leads to the introduction of a bias on the derived quantity. Both for [Fe/H] and \dydz{} the CI increases if \sigpr{} increases: for \dydz{} we found an uncertainty of about 0.3 (about 15~percent) for \sigpr=0.02~dex, which increases to about 0.4-0.5 (about 20-25~percent) for \sigpr=0.05~dex. For [Fe/H] the uncertainty is in almost all the cases equal to \sigpr. On the other hand, \ml{} is only marginally affected by the adopted prior on [Fe/H] and the values obtained using different \fehpr{} differs by about 2-3~percent, at maximum.

Finally, we applied the method to the Hyades stars, using a Gaussian prior on [Fe/H] based on recent observational determinations of this quantity. As already noticed, the parameter that is recovered with a very good precision is the mixing length, while \dydz{} and [Fe/H] are less precise and they are also affected by the adopted \fehpr. To obtain a robust estimate of the most probable value of [Fe/H], \dydz, and \ml, we evaluated these parameters adopting as prior a grid of values of \fehpr. Then, we took the average value for each estimated parameter, restricting the average to values of \fehpr$\in[+0.14,~+0.18]$, with \sigpr=0.02 and 0.03~dex, which are representative of the most recent determinations of [Fe/H] with the related uncertainty. The obtained results was: $<$[Fe/H]$>$=$+0.169\pm0.025$, $<$\dydz$>$=$2.03 \pm0.33$ and $<$\ml$>$=$2.01\pm0.05$. These tightly calibrated values of the helium-to-metal enrichment ratio \dydz and mixing length parameter \ml can be safely adopted to compute the grids of evolutionary tracks required to infer more precisely the stellar parameters of other clusters and field stars.

\section*{Acknowledgements}
The authors thank the anonymous referee for the useful comments that helped to improve the manuscript. ET acknowledges University of Pisa, Dipartimento di Fisica E.Fermi and INAF-OAAb for the fellowship "Analisi dell'influenza dell'evoluzione protostellare sull'abbondanza superficiale di elementi leggeri in stelle di piccola massa in fase di pre-sequenza principale". PGPM. and SD acknowledges INFN (iniziativa specifica TAsP).

\section*{Data Availability}
This work has made use of data from the European Space Agency (ESA) mission {\it Gaia} (\url{https://www.cosmos.esa.int/gaia}), processed by the {\it Gaia}
Data Processing and Analysis Consortium (DPAC, \url{https://www.cosmos.esa.int/web/gaia/dpac/consortium}). Funding for the DPAC has been provided by national institutions, in particular the institutions participating in the {\it Gaia} Multilateral Agreement.

The Hyades data underlying this article were derived from the GAIA DR2 catalogue available at the url: \url{https://gea.esac.esa.int/archive/}. 
\bibliographystyle{mn2e}
\bibliography{bibliography}
\label{lastpage}
\end{document}